\documentclass[twocolumn]{aastex631}
\usepackage{booktabs}
\usepackage{longtable}
\usepackage{tikz}
\usepackage{graphicx}

\newcommand{\niVI}{${}^{56}\textrm{Ni}$}

\newcommand{\CIV}{$\textrm{C\ IV}$}

\newcommand{\Harvard}{Center for Astrophysics \textbar{} Harvard \& Smithsonian, 60 Garden Street, Cambridge, MA 02138-1516, USA}
\newcommand{\JHU}{Physics and Astronomy Department, Johns Hopkins University, Baltimore, MD 21218, USA}
\newcommand{\ISEF}{ISEF International Fellowship}
\newcommand{\STScI}{Space Telescope Science Institute, Baltimore, MD 21218, USA}
\newcommand{\UCSC}{Department of Astronomy and Astrophysics, University of California, Santa Cruz, CA 95064, USA}
\newcommand{\UCBP}{Department of Physics, University of California, Berkeley, CA 94720, USA}
\newcommand{\UCBA}{Department of Astronomy, University of California, Berkeley, CA 94720-3411, USA}
\newcommand{\DARK}{DARK, Niels Bohr Institute, University of Copenhagen, Jagtvej 128, 2200 Copenhagen, Denmark}
\newcommand{\UHAWAII}{Institute for Astronomy, University of Hawaii, 2680 Woodlawn Drive, Honolulu, HI 96822, USA}
\newcommand{\CIERA}{Center for Interdisciplinary Exploration and Research in Astrophysics (CIERA), Northwestern University, Evanston, IL 60202, USA}
\newcommand{\UTAustin}{Department of Astronomy, University of Texas at Austin, 2515 Speedway Stop C1400, Austin, TX, 78712-1205, USA}
\newcommand{\Konkoly}{Konkoly Observatory, Research Centre for Astronomy and Earth Sciences, Budapest, Konkoly Thege Miklós út 15-17, 1121, Hungary}
\newcommand{\KonkolySecondary}{CSFK, MTA Centre of Excellence, Budapest, Konkoly Thege Miklós út 15-17, 1121, Hungary}
\newcommand{\Baja}{Baja Astronomical Observatory of University of Szeged, Szegedi {\'u}t Kt. 766, Baja, 6500, Hungary}
\newcommand{\ELTEHungary}{ELTE E\"otv\"os Lor\'and University, Institute of Physics and Astronomy, P\'azm\'any P\'eter s\'et\'any 1/A, Budapest, 1117 Hungary}
\newcommand{\EPSzeged}{Department of Experimental Physics, Institute of Physics, University of Szeged, D{\'o}m t{\'e}r 9, Szeged, 6720, Hungary }
\newcommand{\ELKH}{HUN-REN--SZTE Stellar Astrophysics Research Group, Szegedi {\'u}t Kt. 766, Baja, 6500, Hungary}
\newcommand{\LenduletMom}{MTA-ELTE Lend\"ulet "Momentum" Milky Way Research Group, Hungary}
\newcommand{\NWU}{Department of Physics and Astronomy, Northwestern University, Evanston, IL 60208, USA}
\newcommand{\IoACambridge}{Institute of Astronomy and Kavli Institute for Cosmology, Madingley Road, Cambridge, CB3 0HA, UK}
\newcommand{\GIANCU}{Graduate Institute of Astronomy, National Central University, 300 Zhongda Road, Zhongli, Taoyuan 32001, Taiwan}
\newcommand{\MTANFCosmology}{MTA CSFK Lendület Near-Field Cosmology Research Group, Konkoly Thege Miklós út 15-17, Budapest, 1121, Hungary}
\newcommand{\MTAMomentum}{MTA CSFK Lendület ``Momentum'' Milky Way Research Group, Hungary}
\newcommand{\Melbourne}{School of Physics, The University of Melbourne, VIC 3010, Australia}
\newcommand{\ELTEGothard}{ELTE E{\"o}tv{\"o}s Lor{\'a}nd University, Gothard Astrophysical Observatory, Szent Imre herceg \'ut 112, Szombathely, 9700, Hungary}
\newcommand{\WonderfulThacherHuman}{Thacher School, 5025 Thacher Rd., Ojai, CA 93203, USA}
\newcommand{\LCO}{Las Cumbres Observatory Global Telescope Network, Goleta, CA 93117, USA}
\newcommand{\UCSB}{Department of Physics, University of California, Santa Barbara, CA 93106, USA}

\def\carts{{CaRTs}}
\def\cart{{CaRT}}
\begin{document}

\title{SN~2022oqm: A Bright and Multi-peaked Calcium-rich Transient}

\author[0000-0002-0840-6940]{S.~Karthik~Yadavalli}
\affiliation{\Harvard}

\author[0000-0002-5814-4061]{V.~Ashley~Villar}
\affiliation{\Harvard}

\author[0000-0001-9695-8472]{Luca~Izzo}
\affiliation{\DARK}
\author[0000-0002-0632-8897]{Yossef~Zenati}
\altaffiliation{\ISEF}
\affiliation{\JHU}

\author[0000-0002-2445-5275]{Ryan~J.~Foley}
\affiliation{\UCSC}

\author[0000-0003-1349-6538]{J.~Craig~Wheeler}
\affiliation{\UTAustin}

\author[0000-0002-4269-7999]{Charlotte~R.~Angus}
\affiliation{\DARK}

\author{Dominik B{\'a}nhidi}
\affiliation{\Baja}

\author[0000-0002-4449-9152]{Katie~Auchettl}
\affiliation{\UCSC}
\affiliation{\Melbourne}

\author{Barna Imre B{\'i}r{\'o}}
\affiliation{\Baja}
\affiliation{\ELKH}

\author[0000-0002-8585-4544]{Attila B\'odi}
\affiliation{\Konkoly}
\affiliation{\KonkolySecondary}
\affiliation{\MTANFCosmology}

\author{Zs{\'o}fia Bodola}
\affiliation{\EPSzeged}

\author[0000-0001-5486-2747]{Thomas de Boer}
\affiliation{\UHAWAII}

\author[0000-0001-6965-7789]{Kenneth~C.~Chambers}
\affiliation{\UHAWAII}

\author[0000-0002-7706-5668]{Ryan~Chornock}
\affiliation{\UCBA}

\author[0000-0003-4263-2228]{David~A.~Coulter}
\affiliation{\UCSC}

\author{Istv{\'a}n Cs{\'a}nyi}
\affiliation{\Baja}

\author[0000-0002-6497-8863]{Borb\'ala Cseh}
\affiliation{\Konkoly}
\affiliation{\KonkolySecondary}
\affiliation{\MTAMomentum}

\author[0000-0003-4578-3216]{Srujan~Dandu}
\affiliation{\UCSC}

\author[0000-0002-5680-4660]{Kyle~W.~Davis}
\affiliation{\UCSC}

\author[0000-0001-9749-4200]{Connor~Braden~Dickinson}
\affiliation{\UCSC}

\author[0000-0001-7576-5428]{Diego~Farias}
\affiliation{\DARK}

\author[0000-0003-4914-5625]{Joseph~Farah}
\affiliation{\LCO}
\affiliation{\UCSB}

\author[0000-0002-8526-3963]{Christa~Gall}
\affiliation{\DARK}

\author[0000-0003-1015-5367]{Hua~Gao}
\affiliation{\UHAWAII}

\author[0000-0003-4253-656X]{D.~Andrew~Howell}
\affiliation{\LCO}
\affiliation{\UCSB}

\author[0000-0002-3934-2644]{Wynn~V.~Jacobson-Galan}
\affiliation{\UCBA}

\author[0000-0003-2720-8904]{Nandita~Khetan}
\affiliation{\DARK}
\affiliation{School of Mathematics and Physics, University of Queensland, 4101, Australia}

\author[0000-0002-5740-7747]{Charles~D.~Kilpatrick}
\affiliation{\CIERA}

\author[0000-0002-8770-6764]{R\'eka K\"onyves-T\'oth}
\affiliation{\Konkoly}
\affiliation{\KonkolySecondary}
\affiliation{\EPSzeged}
\affiliation{\ELTEGothard}

\author[0000-0002-1792-546X]{Levente Kriskovics}
\affiliation{\Konkoly}
\affiliation{\KonkolySecondary}

\author[0000-0002-2249-0595]{Natalie~LeBaron}
\affiliation{\UCBA}

\author{Kayla~Loertscher}
\affiliation{\UCSC}

\author[0009-0004-3242-282X]{X. K. Le Saux}
\affiliation{\UCSC}

\author[0000-0003-4768-7586]{Raffaella~Margutti}
\affiliation{\UCBA}
\affiliation{\UCBP}

\author[0000-0002-7965-2815]{Eugene~A.~Magnier}
\affiliation{\UHAWAII}

\author[0000-0001-5807-7893]{Curtis~McCully}
\affiliation{\LCO}

\author[0000-0002-1052-6749]{Peter~McGill}
\affiliation{\UCSC}

\author[0000-0003-2736-5977]{Hao-Yu~Miao}
\affiliation{\GIANCU}

\author[0000-0001-9570-0584]{Megan~Newsome}
\affiliation{\LCO}
\affiliation{\UCSB}

\author[0000-0003-0209-9246]{Estefania~Padilla~Gonzalez}
\affiliation{\LCO}
\affiliation{\UCSB}

\author[0000-0001-5449-2467]{Andr\'as P\'al}
\affiliation{\Konkoly}
\affiliation{\KonkolySecondary}
\affiliation{\ELTEHungary}

\author{Bor{\'o}ka H. P{\'a}l}
\affiliation{\EPSzeged}

\author[0000-0001-8415-6720]{Yen-Chen~Pan}
\affiliation{\GIANCU}

\author[0000-0003-3727-9167]{Collin~A.~Politsch}
\affiliation{\IoACambridge}

\author[0000-0003-4175-4960]{Conor~L.~Ransome}
\affiliation{\Harvard}

\author[0000-0003-2558-3102]{Enrico~Ramirez-Ruiz}
\affiliation{\UCSC}

\author[0000-0002-4410-5387]{Armin~Rest}
\affiliation{\JHU}
\affiliation{\STScI}

\author[0000-0002-3825-0553]{Sofia~Rest}
\affiliation{\JHU}
\affiliation{\STScI}

\author{Olivia~Robinson}
\affiliation{\UCSC}

\author[0000-0001-8023-4912]{Huei~Sears}
\affiliation{\CIERA}
\affiliation{\NWU}

\author[0000-0002-3522-6312]{Jackson~Scheer}
\affiliation{\WonderfulThacherHuman}

\author[0000-0001-7806-2883]{\'Ad\'am S\'odor}
\affiliation{\Konkoly}
\affiliation{\KonkolySecondary}
\affiliation{\MTANFCosmology}

\author[0000-0002-9486-818X]{Jonathan~Swift}
\affiliation{\WonderfulThacherHuman}

\author{P{\'e}ter Sz{\'e}kely}
\affiliation{\EPSzeged}
\affiliation{\ELKH}

\author[0000-0002-1698-605X]{R\'obert~Szak\'ats}
\affiliation{\Konkoly}
\affiliation{\KonkolySecondary}

\author[0000-0003-4610-1117]{Tam{\'a}s Szalai}
\affiliation{\EPSzeged}
\affiliation{\ELKH}
 \affiliation{\LenduletMom}

\author[0000-0002-5748-4558]{Kirsty~Taggart}
\affiliation{\UCSC}

\author[0000-0003-0794-5982]{Giacomo~Terreran}
\affiliation{\LCO}
\affiliation{\UCSB}

\author[0000-0001-8638-2780]{Padma~Venkatraman}
\affiliation{\UCBA}

\author[0000-0001-8764-7832]{J\'ozsef~Vink\'o}
\affiliation{\UTAustin}
\affiliation{\Konkoly}
\affiliation{\KonkolySecondary}
\affiliation{\EPSzeged}
\affiliation{\ELTEHungary}

\author[0000-0001-7823-2627]{Grace~Yang}
\affiliation{\WonderfulThacherHuman}

\author[0000-0002-2093-6960]{Henry~Zhou}
\affiliation{\WonderfulThacherHuman}

\begin{abstract}

We present the photometric and spectroscopic evolution of SN~2022oqm, a nearby multi-peaked hydrogen- and helium-weak calcium-rich transient (\cart{}). SN~2022oqm was detected $13.1$~kpc from its host galaxy, the face-on spiral galaxy NGC~5875. Extensive spectroscopic coverage reveals an early hot (T $\geq$ 40,000~K) continuum and carbon features observed $\sim$1~day after discovery, SN~Ic-like photospheric-phase spectra, and strong forbidden calcium emission starting 38~days after discovery. SN~2022oqm has a relatively high peak luminosity ($M_B = -17$~mag) for \carts{}, making it an outlier in the population. We determine that three power sources are necessary to explain the light curve (LC), with each corresponding to a distinct peak. The first peak is powered by an expanding blackbody with a power law luminosity, suggesting shock cooling by circumstellar material (CSM). Subsequent LC evolution is powered by a double radioactive decay model, consistent with two sources of photons diffusing through optically thick ejecta. From the LC, we derive an ejecta mass and $\rm{^{56}Ni}$ mass of $\sim$0.6~M$_{\odot}$ and $\sim$0.09~M${_\odot}$. Spectroscopic modeling  $\sim$0.6~M$_{\odot}$ of ejecta, and with well-mixed Fe-peak elements throughout. We discuss several physical origins for SN 2022oqm and find either a surprisingly massive white dwarf progenitor or a peculiar stripped envelope model could explain SN 2022oqm. A stripped envelope explosion inside a dense, hydrogen- and helium-poor CSM, akin to SNe Icn, but with a large 56Ni mass and small CSM mass could explain SN 2022oqm. Alternatively, helium detonation on an unexpectedly massive white dwarf could also explain SN 2022oqm.

\end{abstract}

\section{Introduction}

Calcium-rich transients (\carts{}) are observationally rare. They tend to be fainter than most supernovae (typically $-16\lesssim \text{M}_\mathrm{B}\lesssim-15$~mag), show strong forbidden calcium and weaker forbidden oxygen emission during the nebular phase ([\ion{Ca}{2}]/[\ion{O}{1}] $> 2$), and have a fast light curve rise time ($\rm t_{rise} \lesssim 15$~days) \citep{Filippenko2003, Perets10, Waldman+11, Kasliwal+12, MengHan15, de2020zwicky, Wynn2022}. In total, only 38 such transients have been identified in the literature. Although the rates of \carts{} are likely in the range of 5-20$\%$ of the normal SNe~Ia rates, the relatively dim peak magnitudes and quick light curve evolution times make these supernovae (SNe) more difficult to detect than other types \citep{Perets10, Kasliwal+12, MengHan15, de2020zwicky, Zenati+23_carich}. As the set of well-observed \carts{} has grown, a rich diversity of observational characteristics has emerged.

\carts{} are typically found offset from their host galaxy, suggesting a predominantly older progenitor population (with some notable exceptions; see \citealt{de2020zwicky} for a review). \carts{} are typically discovered in stellar clusters in elliptical galaxies \citep{Perets10, Kasliwal+12,Foley+15,de2020zwicky,PeretsBeniamini21,Jacobson-Galan2019ehk,Zenati+23_carich}, consistent with this older population. However, some of the \carts{}, namely iPTF15eqv \citep{milisavljevic2017}, iPTF16hgs \citep{De2018}, SN 2016hnk \citep{Galbany2019, jacobson-galan2016hnk}, SN 2019ehk \citep{Jacobson-Galan2019ehk}, and SN 2021gno \citep{Wynn2022, 2023Ertini} are located in spiral star-forming galaxies, with significant offsets from the closest star-forming regions within their host galaxies \citep{Wynn2022}. Nevertheless, population studies of CaRTs (e.g., \citealt{Shen2019}) find that \cart{} locations are generally consistent with populations of old ($\geq 5$ Gyr) and low-metallicity ($Z/Z_\odot ~\leq$  0.3) stars. 

\carts{} are not the only class of SNe that have strong forbidden nebular calcium emission. Particularly, SNe Iax are known to have strong forbidden calcium emission \citep{Silverman12:bsnip, Siebert20b}. However, these are likely distinct classes of transients as SNe Iax have longer-lived light curves ($\sim$~years compared to $\sim$~months) and bluer photospheric-phase spectra \citep{Foley13_Iax, Kawabata+21_Iax}. In addition, \carts{} themselves have large spectroscopic diversity. Compared to other classes of SNe, \cite{de2020zwicky} find that some \carts{} are spectroscopically more similar to Type Ia SNe (Ia-like \carts{}; \carts{}-Ia), while others are more similar to Type Ibc SNe (Ibc-like \carts{}; \carts{}-Ibc). Furthermore, \cite{Das2022} presented a population of \carts{}-IIb, although they predicted a distinct progenitor for these transients. 
Such spectroscopic diversity points to the possibility that several distinct progenitor systems give rise to the observed population of \carts{}.

Although the progenitor nature of \carts{} remains elusive, \cite{Perets10} proposed a theory to explain the prototypical \cart{} SN~2005E: the detonation of a helium shell on the surface of a WD in a binary WD (BWD) system. Helium can be accreted onto a WD without detonating the WD from either a He-rich WD or from a non-degenerate He star \citep{Holcomb+13_SHedet, Shen2019}. As He is accreted, the He shell can reach the critical density to ignite \citep{Guillochon+10, DanM+12, Dessart2015}. However, the burning of the He shell should not trigger a WD core detonation, which constrains the WD mass to $\leq${}$0.8~\mathrm{M_\odot}$. Therefore, a BWD system within this narrow region of parameter space could be a \cart{} progenitor \citep{Shen2010, Waldman+11, DanM+11, DanM+12, Shen2019, Jacobson-Galan2019ehk, de2020zwicky, Zenati+23_carich}. Indeed,  \cite{Zenati+23_carich} presented a similar scenario involving a BWD system consisting of a secondary carbon-oxygen WD (C/O-WD) and a primary He+C/O hybrid WD, where the He-envelope contains $\sim2-20\%$ of the primary WD's mass. Before the merger, the secondary WD is fully disrupted by the primary WD, causing CO from the secondary to accrete onto the primary and heat the He shell. A He-enriched detonation occurs in the primary, leading to a weak explosion and an intact core.

Here, we present observations of a recent \cart{}-Ic, SN~2022oqm, in the nearby ($z = 0.012$) star-forming spiral galaxy NGC~5875. Upon discovery, SN~2022oqm was initially classified as a SN~I \citep{2022oqm_original_classification}, given the absence of hydrogen and helium, and the lack of  obvious silicon in its earliest spectrum. In addition, the peak magnitude of SN~2022oqm ($\rm{M_B}=-17$) is typical of SNe~Ic, further supporting the SN~Ic classification. However, nebular phase spectroscopy reveals that SN~2022oqm is a \cart{}, the 39th of its class (see Table~\ref{table:all_CaRTs}), and among the brightest \carts{}-Ibc detected so far \citep{Jacobson-Galan2019ehk, de2020zwicky, Das2022}. Therefore, an important question emerges: Was SN~2022oqm the result of the core collapse of a massive star (core-collapse supernova, CCSN), the standard progenitor of SNe~Ic, or was its progenitor a binary WD system? In this study, we address this and other peculiarities in the evolution of SN~2022oqm. \cite{irani2022} also presents a detailed analysis of this event, focusing primarily on the first peak, and constraining the mass of the circumstellar material. In this article, we present a largely independent and complementary analysis of SN~2022oqm and find broad agreement with \cite{irani2022}. Where possible, we present comparisons to their work and show potential disagreements between the two analyses.

In Section~\ref{sec:observations}, we present our photometric and spectroscopic observations. In Section~ \ref{sec:photometry}, we present the photometric analysis of SN~2022oqm. In Section~\ref{sec:spectroscopy}, we present the full spectral sequence of SN~2022oqm, along with a detailed analysis of the spectra of SN~2022oqm. In Section~\ref{sec:mosfit}, we present modeling of the light curve of the SN~2022oqm. In Section~\ref{sec:ca_rich_comparisons}, we compare SN~2022oqm to the broader \cart{} population. In Section~\ref{sec:allmodels} we explore possible progenitor scenarios of SN~2022oqm. We discuss our results in Section~\ref{sec:discussion} and conclude in Section~\ref{sec:conclusions}. We assume a standard $\Lambda$CDM cosmology throughout ($H_0 = 67.8$ km~s$^{-1}$~Mpc$^{-1}$, $\Omega_M$ = 0.31, $\Omega_\Lambda ~=~0.69$; \citealt{Planck15Cosmology}).

\begin{deluxetable}{lcr}
\tabletypesize{\footnotesize}
\tablecolumns{3}
\tablecaption{SN~2022oqm and NGC~5875 (host galaxy) Properties. \label{table:discovery}}
\tablehead{
\colhead{Parameter} & \colhead{Value} & \colhead{Reference}}
 \startdata
     R.A. & $15^h09^m08.21^s$ & [2]\\
     \\
     Declination & $+52^\circ32'05''.14$ & [2]\\
     \\
     Redshift & 0.012 & [4]\\
     \\
     Distance Modulus & 33.575 & [4]\\
     \\
     Milky Way E[B-V] & 0.016 & [3]\\
     \\
     Explosion Time & MJD = 59770-59771 & This Work \\
      & 2022 July 10 - 11 & \\
     Time of Peak 1 & MJD = 59771  & This Work \\
      & 2022 July 11 & \\
     Time of Peak 2 & MJD = 59775 & This Work \\
      & 2022 July 13 - 17 & \\
     Time of Peak 3 & MJD = 59785& This Work \\
      &  2022 July 18 - 25 &\\
    \\
     Host RA & $+15^h09^m13.16^s$ & [1]\\
     \\
     Host Dec & $52^\circ31'42''.40$ & [1]\\
     \\
     Host-SN Offset & $50.6''$ ($13.1$ kpc) & This Work \\
\enddata
\tablecomments{[1] \cite{SDSS_DR3_data}, [2] \cite{2022oqm_original_discovery}, [3] \cite{MW_extinction_2022oqm}, [4] \cite{van2016}}
\end{deluxetable}

\section{Observations}
\label{sec:observations}
SN~2022oqm was reported to the Transient Name Server with a discovery date of 2022-07-11 04:33 UT (MJD = 59771.69) and the last non-detection a day earlier on 2022-07-10 06:14 UT (MJD = 59770.75) by \cite{2022oqm_original_discovery}, and initially classified as a SN~I \citep{2022oqm_original_classification}. Immediately after detection, a clear shock-like UV excess had become apparent, and the community began to follow the evolution of SN~2022oqm. After discovery, SN~2022oqm was next classified as a SN~Ic \citep{2022oqm_class_Ic_1, 2022oqm_class_Ic_2}, and finally as a SN~Ic-pec \citep{2022oqm_class_Icpec}. Its observed properties are summarized in Table~\ref{table:discovery}. SN~2022oqm was found offset by $50.6''$ ($13.1$ kpc) from the center of NGC~5875, an extended spiral galaxy at $53.5 \pm 1$ Mpc \citep{Tully13}.

\subsection{Photometry}

\subsubsection{Pan-STARRS}
SN~2022oqm was observed with the Pan-STARRS telescope \citep[PS1/2;][]{Kaiser2002, Chambers2017} on 8 September 2022 in $rz$-bands through the Young Supernova Experiment (YSE) \citep{yse_ref_3}. Data storage/visualization and follow-up coordination was done through the YSE-PZ Target and Observation Manager \citep{Coulter22, Coulter23}. The YSE photometric pipeline is based on {\tt photpipe} \citep{Rest+05}, which relies on calibrations from \citet{Magnier20a} and \citet{waters20}. Each image template was taken from stacked PS1 exposures, with most of the input data from the PS1 3$\pi$ survey. All images and templates were resampled and astrometrically aligned to match a skycell in the PS1 sky tessellation. An image zero-point is determined by comparing PSF photometry of the stars to updated stellar catalogs of PS1 observations \citep{flewelling20}. The PS1 templates are convolved with a three-Gaussian kernel to match the PSF of the nightly images, and the convolved templates are subtracted from the nightly images with {\tt HOTPANTS} \citep{becker15}. Finally, a flux-weighted centroid is found for the position of the SN in each image and PSF photometry is performed using ``forced photometry": the centroid of the PSF is forced to be at the SN position. The nightly zero-point is applied to the photometry to determine the brightness of the SN for that epoch.

\subsubsection{Las Cumbres Observatory - Global Supernova Experiment}
The Las Cumbres Observatory (LCO) triggered observations through the Global Supernova Experiment on SN~2022oqm within two weeks prior to the peak. Observations were conducted by the Sinistro 1m telescopes from Las Cumbres Observatory. Data were recorded in the $B$, $g$, $V$, $r$, and $i$ bands covering 11 days prepeak and 52 days postpeak. We reduced the photometry in-house using the \texttt{lcogtsnpipe}\footnote{https://github.com/LCOGT/lcogtsnpipe} infrastructure \citep{valenti2016}, which uses the point-spread-function (PSF) fitting procedure to extract target magnitudes. We calibrated photometry in the $B$, and $V$ bands using Vega magnitudes in the Landolt catalog \citep{Landolt1992}. We calibrated photometry in the $g$, $r$, and $i$ bands to AB magnitudes using the Sloan Digital Sky Survey (SDSS) catalog \citep{Smith2002}.

\subsubsection{Thacher, Lulin, LCO, Nickel}
We observed SN\,2022oqm with the Thacher 0.7\,m telescope \citep{Swift21},  in $griz$ bands from 12 July to 9 September 2022, with the Lulin 1\,m telescope in $griz$ bands from 9--30 August 2022, and with the LCO 1\,m telescopes and Sinistro imagers in $ugri$ bands from 2--10 August 2022.  All images were reduced in {\tt photpipe} \citep{Rest+05} with bias, flat, and dark frames obtained in the same instrumental configuration as our science images.  We regridded each frame to a common pixel scale and field center with {\tt SWarp} \citep{swarp} and performed point-spread function photometry with a custom version of {\tt DoPhot} \citep{Schechter93}.  All photometry was calibrated using standard stars from the PS1 3$\pi$ DR2 catalog \citep{flewelling20} observed in the same field as SN\,2022oqm.  We subtracted pre-explosion $griz$ template images from PS1 using {\tt hotpants} \citep{hotpants} and performed forced photometry at the site of SN\,2022oqm in the subtracted images, which is the final photometry presented here. 

\subsubsection{Konkoly, Baja Observatories}
Photometric observations of SN2022oqm were collected from Piszkesteto Station of Konkoly Observatory and from Baja Observatory of University of Szeged, Hungary. Both sites are equipped with a robotic 0.8m Ritchey-Chretien-Nasmyth telescope, manufactured by ASA AstroSysteme GmbH, Austria.  Photometry was performed by applying a back-illuminated, liquid-cooled, $2048 \times 2048$ FLI ProLine PL230 CCD camera through Johnson $B, V$, and Sloan $g', r', i'$ and $z'$ bands. Image reductions were done by custom-made IRAF\footnote{https://iraf-community.github.io/} and {\tt  fitsh}\footnote{https://fitsh.net/} scripts. Photometry of the SN was calibrated via local point sources within the CCD field-of-view using their PS1 photometry\footnote{https://catalogs.mast.stsci.edu/panstarrs/}.

\subsubsection{Asteroid Terrestrial-impact Last Alert System (ATLAS)}
SN~2022oqm was detected in the $c$ and $o$ bands by ATLAS between $-20$ and $70$ days relative to the phase of r-band peak. Using the ATClean toolkit \citep{Rest2023}, we searched for any explosion activity by running a Gaussian-weighted rolling sum on the flux/dflux of the pre-SN light curve to compare to the control lightcurves close to the SN position. The pre-SN rolling sum was within the noise of the control lightcurve sums suggesting no evidence for any pre-SN bumps in the ATLAS light curve data. We used the ATLAS forced photometry server \citep{Tonry2018, Smith2020, Shingles2021} to recover the difference-image photometry for SN~2022oqm. To remove erroneous measurements and have significant SN flux detection at the location of SN~2022oqm, we applied several cuts on the total number of individual data points and nightly averaged data. Our first cut used the $\chi^{2}$ and uncertainty values of the point-spread-function (PSF) fitting to remove discrepant data. We then obtained forced photometry of eight control light curves located in a circular pattern around the location of the SN with a radius of 17\arcsec. The flux of these control light curves is expected to be consistent with zero, and any significant deviation from zero would indicate that there are either unaccounted systematic biases or underestimated uncertainties. We searched for such deviations by calculating the weighted mean of the set of control light-curve measurements for a given epoch after removing any $>3\sigma$ outliers (for a more detailed discussion, \cite{Rest2023}.\footnote{\url{https://github.com/srest2021/atlaslc}}). If the weighted mean of these photometric measurements was inconsistent with zero, we flagged and removed those epochs from the SN light curve. This method allows us to identify potentially incorrect measurements without using the SN light curve itself. We then binned the SN~2022oqm light curve by calculating a 3$\sigma$-cut weighted mean for each night (ATLAS typically has four epochs per night), excluding the flagged measurements from the previous step.

\subsubsection{ZTF}

We retrieved photometry of SN\,2022oqm from the Zwicky Transient Facility (ZTF) public survey in ZTF-$g$ and ZTF-$r$ bands \citep[see][for details]{Bellm19}.  Following the methodology in \citet{Aleo22}, we used forced photometry of SN\,2022oqm in ZTF difference imaging from the Las Cumbres Observatory's Make Alerts Really Simple (MARS) public alerts broker Fink \citep{Moller21}.

\subsubsection{\textit{Swift}}

SN~2022oqm was observed with the Ultraviolet Optical Telescope (UVOT; \citealt{Roming05}) onboard the Neil Gehrels \emph{Swift} Observatory \citep{Gehrels04} from 11 July 2022 until 29 July 2022. We performed aperture photometry with a 5$\arcsec$ region with \texttt{uvotsource} within HEAsoft v6.26\footnote{We used the calibration database (CALDB) version 20201008.}, following the standard guidelines from \cite{Brown14}. In order to remove contamination from the host galaxy, we employed images acquired at $t\approx122$~days after the explosion, assuming that the SN contribution is negligible at this phase. This is supported by visual inspection in which we found no flux associated with SN~2022oqm. We subtracted the measured background count rate at the location of the SN from the count rates in the SN images following the prescriptions of \cite{Brown14}. Consequently, we detect bright UV emission from the SN directly after the explosion (Figure~\ref{fig:reduced_photometry}) until maximum bolometric light. Subsequent non-detections in $w1, m2, w2$ bands indicate significant cooling of the photosphere and/or Fe-group line blanketing. 

Observed contemporaneously with the UVOT, \textit{Swift} also observed SN2022oqm using the X-Ray Telescope (XRT; \citealt{burrows05}) in photon-counting mode. Using the most up to date calibrations and the standard filters and screenings, all level one XRT observations were processed using the XRTPIPELINE version 0.13.7. Using a region with a radius of 47'' centered on the location of SN2022oqm and a source free background region, we found no X-ray emission coincident with the location of SN2022oqm. To place the deepest constraints on the presence of X-ray emission, we merged all available \textit{Swift} observations of  SN2022oqm using the HEASOFT tool \textsc{xselect} version 2.5b. We then derived a 3$\sigma$ upperlimit to the absorbed flux in the 0.3-10.0 keV energy range of $6.79\times10^{-14}$ erg cm s$^{-1}$, assuming an absorbed powerlaw  with a column density of 1.73$\times10^{20}$ cm$^{-2}$ \citep{HI4PI16} and a photon index of 2 which is redshifted to the location of SN2022oqm’s host galaxy.

\subsection{Spectroscopy}
\subsubsection{Hobby Eberly Telescope}
Six optical spectra were taken through the Low-Resolution Spectrograph 2 (LRS2) instrument on the Hobby Eberly Telescope (HET) on the 13th, 16th, 17th, 24th, 30th of July and the 2nd of August 2022. The LRS2 data were processed with \texttt{Panacea}\footnote{\url{https://github.com/grzeimann/Panacea}}, the HET automated reduction pipeline for LRS2.  The initial processing includes bias correction, wavelength calibration, fiber-trace evaluation, fiber normalization, and fiber extraction; moreover, there is an initial flux calibration from default response curves, an estimation of the mirror illumination, as well as the exposure throughput from guider images.  After the initial reduction, we used an advanced code designed for crowded IFU fields to perform a careful sky subtraction and host-galaxy subtraction.  Finally, we modeled the target SNe with a \citet{Moffat69} PSF model and performed a weighted spectral extraction.

\subsubsection{Kast \& LRIS}
Seven optical spectra were taken with the Kast dual-beam spectrograph \citep{KAST} on the Shane 3~m telescope at Lick Observatory,  on the 19th, 24th, 28th of July 2022, the 2nd, 7th, 13th, and 17th of August 2022. One spectrum was taken with the Low-Resolution Imaging Spectrograph (LRIS; \citealt{LRIS}) on the 10 m Keck I telescope on the 25th of September 2022. To reduce the Kast and LRIS spectral data, we used the {\tt UCSC Spectral Pipeline}\footnote{\url{https://github.com/msiebert1/UCSC\_spectral\_pipeline}} \citep{siebert19}, a custom data-reduction pipeline based on procedures outlined by \citet{Foley03}, \citet{silverman12}, and references therein.  The two-dimensional (2D) spectra were bias-corrected, flat-field corrected, adjusted for varying gains across different chips and amplifiers, and trimmed.  Cosmic-ray rejection was applied using the {\tt pzapspec} algorithm to individual frames.  Multiple frames were then combined with appropriate masking.  One-dimensional (1D) spectra were extracted using the optimal algorithm \citep{Horne86}.  The spectra were wavelength-calibrated using internal comparison-lamp spectra with linear shifts applied by cross-correlating the observed night-sky lines in each spectrum to a master night-sky spectrum. Flux calibration was performed using standard stars at a similar airmass to that of the science exposures, with ``blue'' (hot subdwarfs; i.e., sdO) and ``red'' (low-metallicity G/F) standard stars.  We correct for atmospheric extinction. By fitting the continuum of the flux-calibrated standard stars, we determine the telluric absorption in those stars and apply a correction, adopting the relative airmass between the standard star and the science image to determine the relative strength of the absorption.  We allow for slight shifts in the telluric A and B bands, which we determine through cross-correlation.  For dual-beam spectrographs, we combine the sides by scaling one spectrum to match the flux of the other in the overlap region and use their error spectra to correctly weight the spectra when combined.  More details of this process are discussed elsewhere \citep{Foley03, silverman12, siebert19}.

\subsubsection{NIRES}
We obtained a Near Infrared (NIR; 0.94--2.45~$\mu$m) spectrum of SN~2022oqm using the Near-Infrared Echellette Spectrometer \citep[NIRES;][]{NIRES} on the 10~m Keck~II telescope as part of the Keck Infrared Transient Survey (KITS), a NASA Keck Key Strategic Mission Support program (PI R. Foley). We observed the SN at two positions along the slit (AB pairs) to perform background subtraction. An A0V star was observed immediately before or after the science observation. We reduced the NIRES data using \texttt{spextool v.5.0.2} \citep{cushing2004}; the pipeline performs flat-field corrections using observations of a standard lamp and wavelength calibration based on night-sky lines in the science data. We performed telluric correction using \texttt{xtellcor} \citep{vacca2003}. The NIR spectrum is shown in Figure~\ref{fig:nir_series}.

\subsubsection{ALFOSC}
One spectrum was taken with the Alhambra Faint Object Spectrograph and Camera (ALFOSC) on the Nordic Optical Telescope (NOT) at La Palma on the 15th of July 2022. This spectrum was taken using grism 4 and a 1.000 slit, aligned along the parallactic angle, and under clear observing conditions and good seeing. It was reduced with a custom pipeline running standard {\tt pyraf} procedures.

\subsubsection{Las Cumbres Observatory - Global Supernova Experiment}
Two spectra were obtained by the Faulkes Telescope North with the FLOYDS low resolution spectrograph on the 12th and 19th of August 2022. These data were reduced via the pipeline as detailed in \cite{Valenti2012hn}.

\section{Photometric Analysis}
\label{sec:photometry}

\begin{figure*}
    \centering
    \includegraphics[width = 0.8\linewidth]{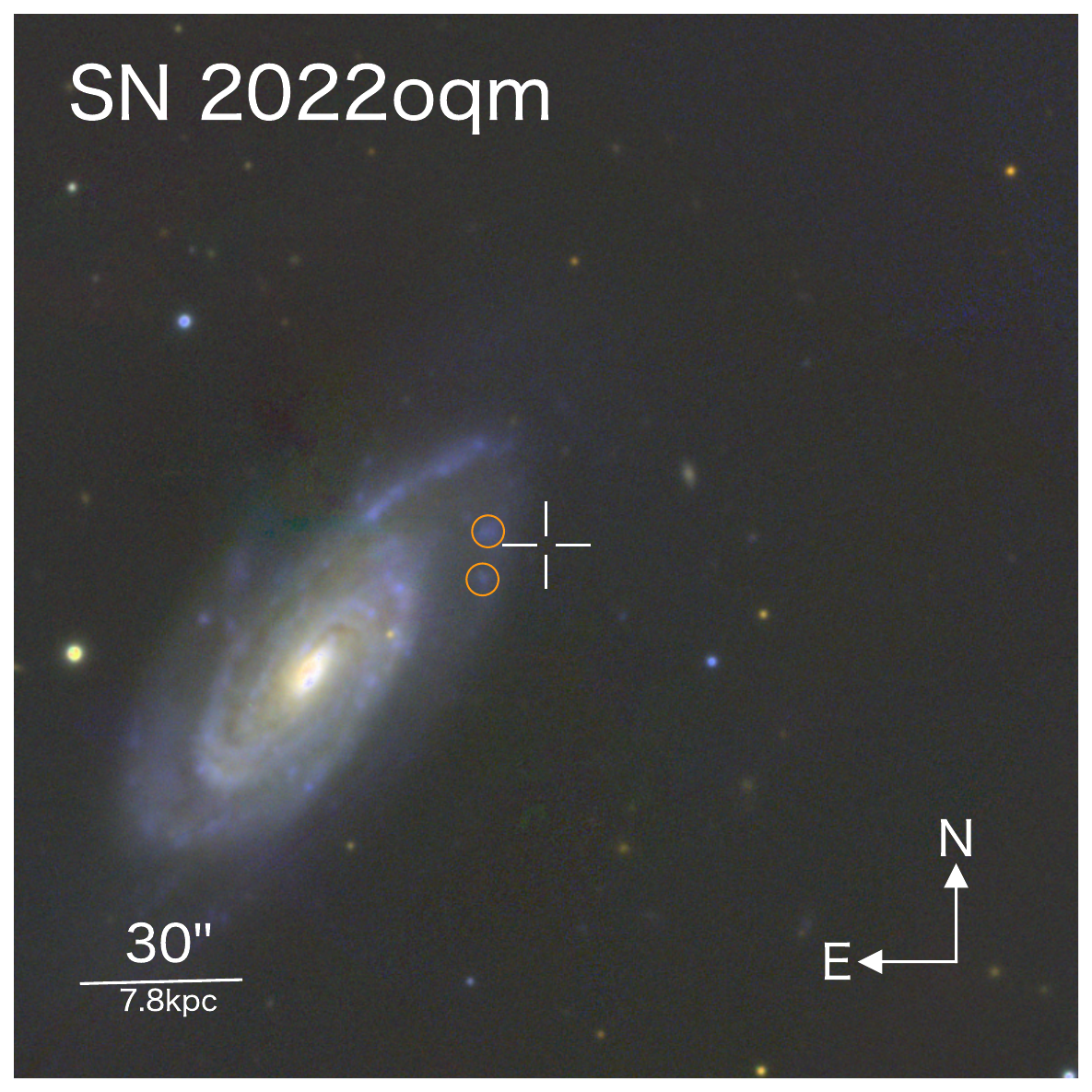}
    \caption{Pan-STARRS image the host galaxy NGC5875, at ($15^h09^m10.31^s$, $52^\circ32'19''.72$). We indicate the location of SN~ 2022oqm in the white crosshairs. The closest knots of star formation (marked in orange) are $2.9$ and $3.5$~kpc away.}
    \label{fig:host_galaxy}
\end{figure*}

SN~2022oqm was discovered in the outskirts of a face-on, spiral galaxy, NGC~5875, shown in Figure~\ref{fig:host_galaxy}. Like most \carts{}, SN~2022oqm is highly offset from the center of the galaxy. The host redshift is consistent with SN~2022oqm, showing SN~2022oqm is not a chance coincidence with the galaxy's location. SN~2022oqm is visibly offset from any ongoing star formation. To quantify the radial offset of SN\,2022oqm in terms of the host light, we adopt a fractional light method previously employed in SN environmental studies \citep[e.g.,][]{Fruchter2006, Habergham_2014, Ransome_2022}. In the case of SN\,2022oqm, we use $g$-band imaging from a Pan-STARRS PS1 pre-explosion image of the galaxy. This technique uses ellipses in the same `aspect ratio' as the galaxy (position angle and axis ratio from NED\footnote{\url{https://ned.ipac.caltech.edu/}} with an axis ratio of $\sim$\,2.5). An ellipse that is gradually expanded until the difference between iterations becomes small (i.e. consistent with reaching the background) is the host ellipse. Another ellipse with the same aspect ratio that intercepts the SN location is the SN ellipse. The total light emitted inside the SN ellipse is divided by the total light emitted inside the host ellipse, giving the fractional light enclosed by the SN, which is $96\%$ for SN~2022oqm. We also calculate that the location of SN~2022oqm is $\sim2.4$ half-light radii from the center of the galaxy. Using Pan-Starrs PS1 pre-explosion imaging of the galaxy, we find no indication of pre-explosion flux at the position of the transient, to a limiting surface brightness $24.5$ mag arcsec$^{-2}$. The two closest clusters of optical light are 11.2'' (2.9 kpc) and 13.6'' (3.5 kpc) away.

\begin{figure*}
    \centering
    \includegraphics[width = \textwidth]{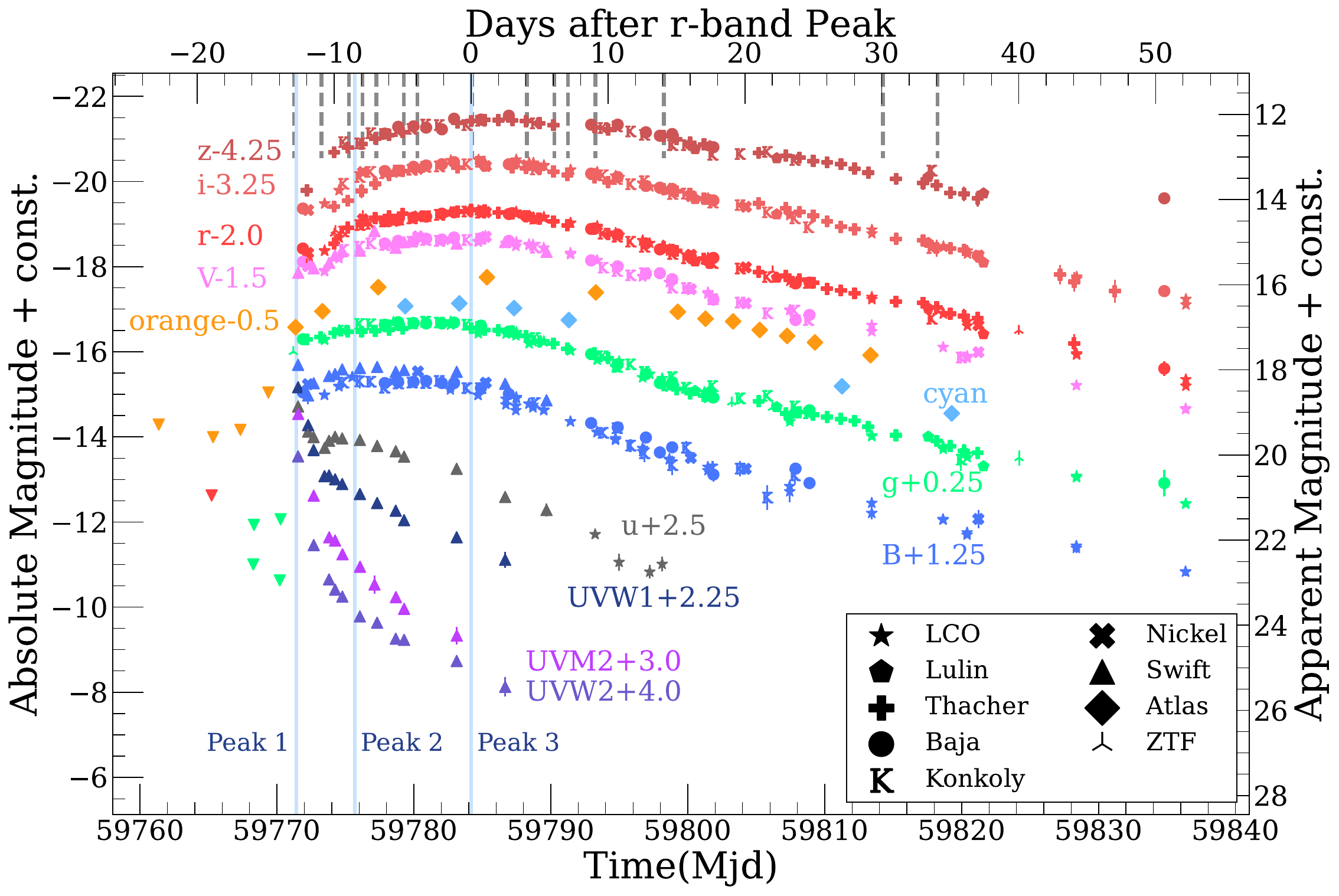}
    \caption{Multiband light curve of SN~2022oqm. Dotted grey vertical lines represent epochs where we obtain spectroscopy. Phases of peaks 1, 2, and 3 are labeled with blue vertical lines. Upper limits ($3\sigma$) are identified as downward-facing triangles. Photometry in the $B$-filter appears to be scattered because the \textit{swift}~$B$-filter bandpass is slightly different from the bandpass of the $B$-filter on other instruments. The precise shape of the bandpass is accounted for in photometric analysis presented throughout.}
    \label{fig:reduced_photometry}
\end{figure*}

Photometry reveals SN~2022oqm is a multi-peaked SN. Our near-infrared to ultraviolet light curve of SN~2022oqm is shown in Figure~\ref{fig:reduced_photometry}. We detected an early peak that is more pronounced in shorter wavelengths at the time of detection (MJD $59771.19$). There is another, broader peak between MJD $\sim 59781$ and $\sim 59788$. We also note a weak local maximum of the $U$-band flux at MJD $59774.24$ (see Section~\ref{subsec:pow_double_arnett} for a detailed discussion on the confirmation and origin of these three peaks). We note the approximate phases of these three as peaks 1, 2, and 3 in Figure~\ref{fig:reduced_photometry}. We report phase relative to the time of maximum $r$-band flux (MJD $59784.17$) and present epochs with respect to this phase henceforth in this article. The peak luminosity of SN~2022oqm is in the typical range of peak SN~Ibc luminosities ($-17$ to $-18$~mag; \citealt{Drout2011}), but is $\sim1.5$~mag brighter than the population of \carts{}-Ibc at peak \citep{de2020zwicky, Zenati+23_carich}.

\begin{figure}
    \centering
    \includegraphics[width = \linewidth]{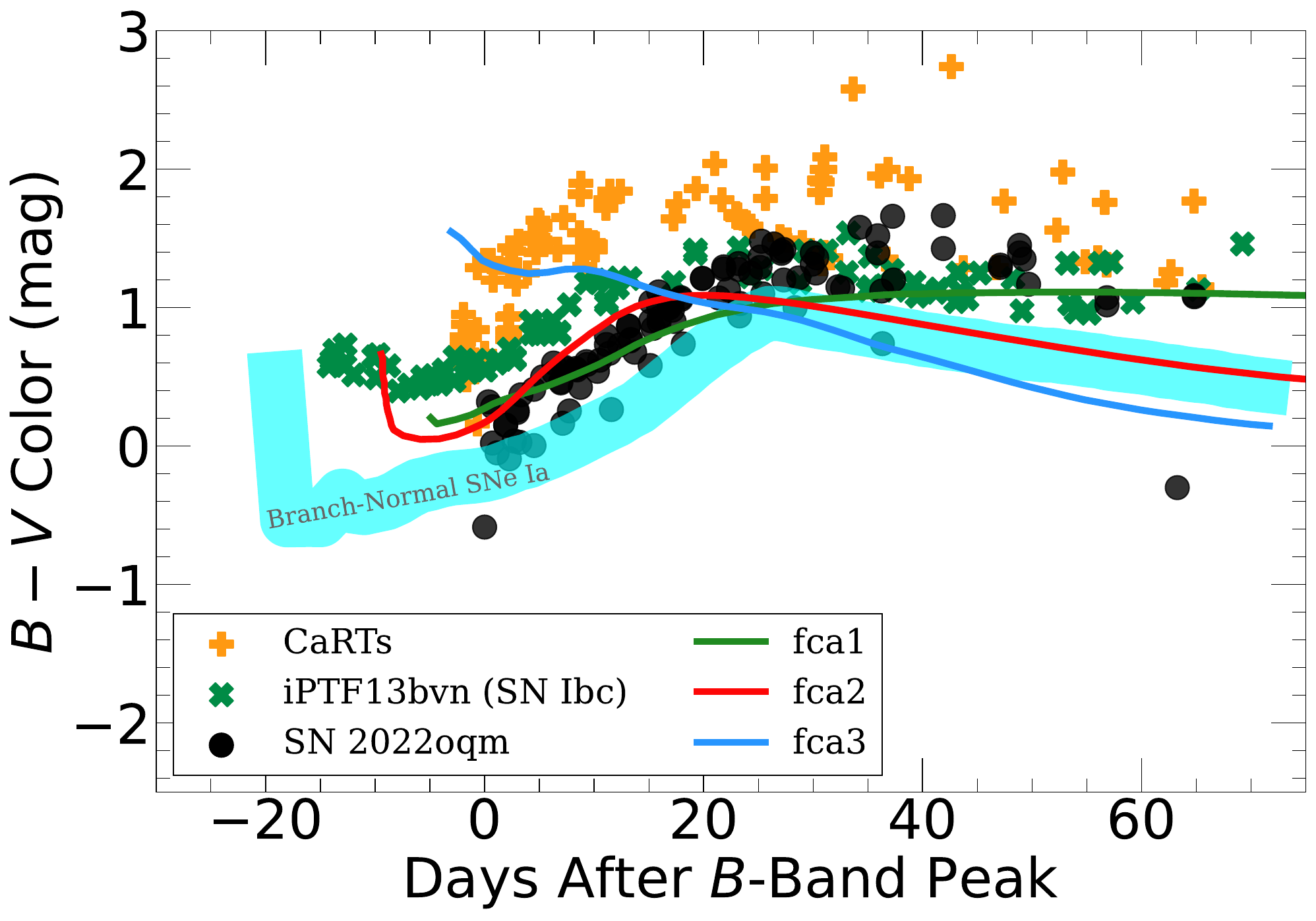}
    \includegraphics[width = \linewidth]{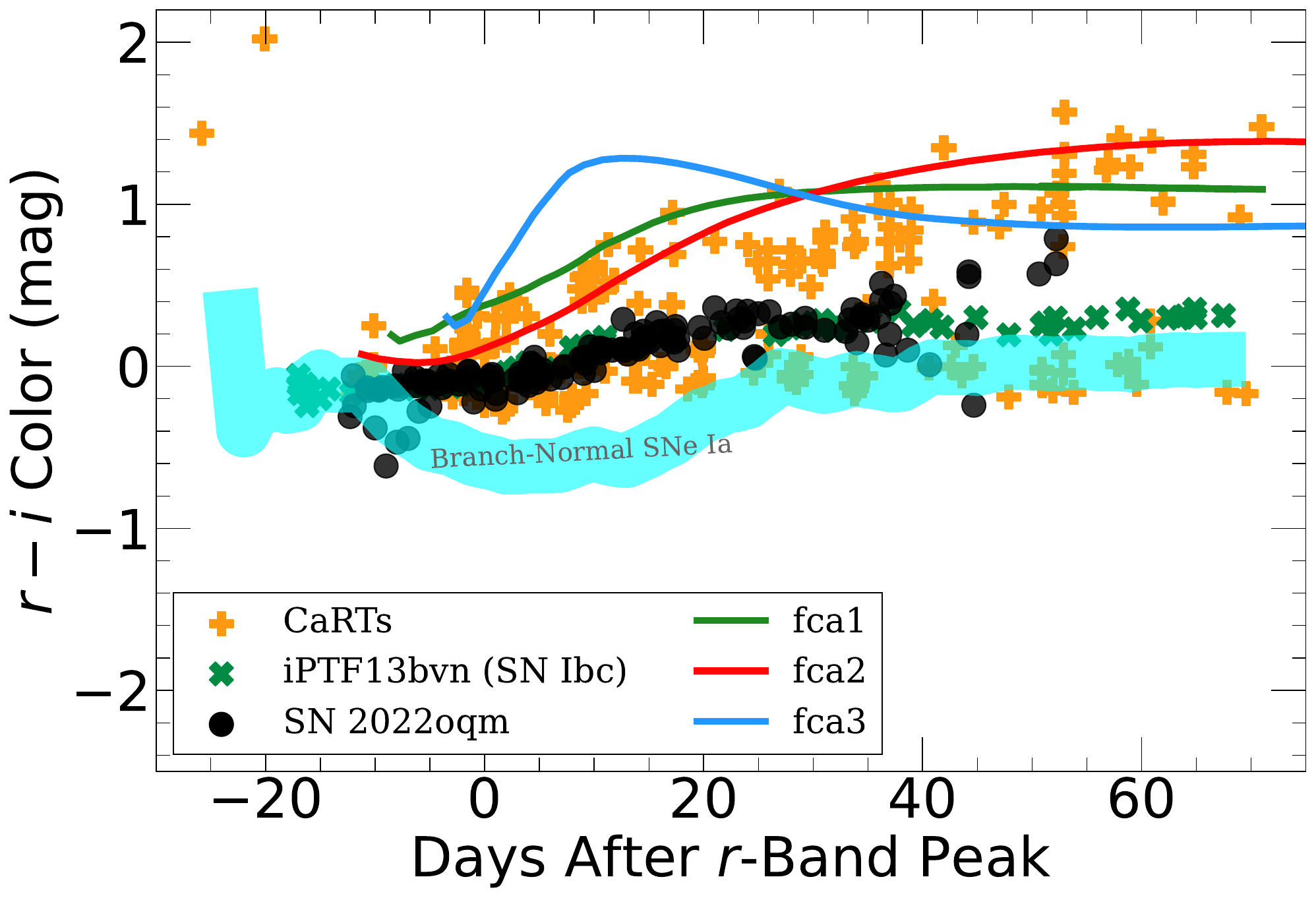}
    \caption{$\textit{B-V}$ (top) and $\textit{r-i}$ (bottom) color evolution of SN~2022oqm, compared to that of all \carts{} with well-sampled light curves, iPTF13bvn, the well-studied SN~Ib with evidence of a binary progenitor, and the template of Branch-normal SNe~Ia. SN~2022oqm is bluer than most \carts{} in both $r-i$ color and $B-V$ color. The three BWD models simulated by \cite{Zenati+23_carich} are shown as $fca1$, $fca2$, and $fca3$.}
    \label{fig:color_evolution}
\end{figure}

In Figure~\ref{fig:color_evolution}, we show $B-V$ (top) and $r-i$ (bottom) color evolution of SN~2022oqm with known \carts{},  the well-studied SN~Ib with evidence of a binary progenitor iPTF13bvn \citep{iPTF13bvn1, iPTF13bvn2}, and a template SN~Ia light curve. We use the canonical $s=1$ model from \cite{Nugent2002} to find a template SN~Ia light curve (see Section~\ref{sec:ca_rich_comparisons} for a more detailed description of this model). We also show the three theoretical models of differing progenitor scenarios that result in \carts{} from \cite{Zenati+23_carich}: fca1, fca2, and fca3. The $r-i$ color evolution of SN~2022oqm matches that of other \carts{} but the $B-V$ color evolution for SN~2022oqm matches that of SNe~Ia and SNe~Ibc. The $B-V$ color of SN~2022oqm is bluer than that of most \carts{}. The tracks of \emph{fac1, fca2,} and \emph{fca3} predict redder colors than what is seen in SN~2022oqm (see Section~\ref{sec:allmodels} for a detailed description of this progenitor system).

\subsection{Bolometric Properties}
\label{subsec:bol_facts}

\begin{deluxetable*}{cccc}
\tabletypesize{\footnotesize}
\tablecolumns{4}
\tablecaption{Summary of blackbody properties of all three peaks as derived from fitting a single blackbody evolution to the SN~2022oqm light curve with \texttt{extrabol} \label{table:extrabol_info}. Note that reported errors only account for statistical uncertainties.}
\tablehead{
\colhead{Parameter} & \colhead{Peak 1} & \colhead{Peak 2} & \colhead{Peak 3} }
 \startdata
     Temperature (K) & $32,000 \pm 3200$ & $7,600 \pm 170$ & $5,700 \pm 170$\\
     Photosphere Radius (cm) & $(1.5\pm~0.14)\times10^{14}$ & $(9.2\pm~0.49)\times10^{14}$ & $(1.7\pm~0.12)\times10^{15}$\\
     Bol. Luminosity (erg/s) & $(1.5\pm~0.39)\times10^{43}$ & $(2.0\pm~0.08)\times10^{42}$ & $(2.2\pm~0.1)\times10^{42}$ \\
     Bol. Magnitude & $-19.4\pm~0.56$ & $-17.1\pm~0.09$ & $-17.2\pm~0.11$\\
 \enddata
\end{deluxetable*}

\begin{figure}
    \centering
    \includegraphics[width = 0.95\linewidth]{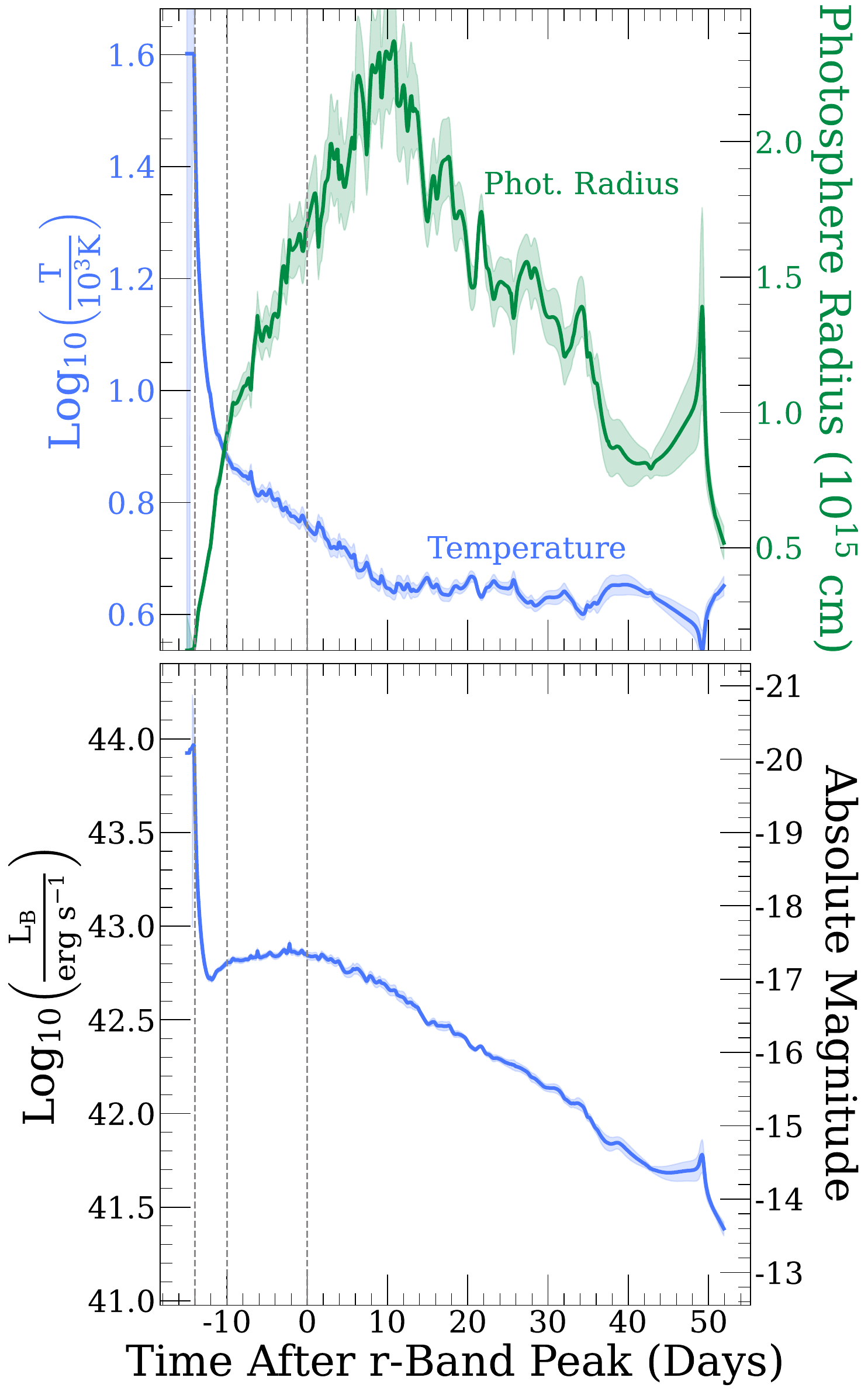}
    \caption{Blackbody fits to the photometry of SN~2022oqm, showing temperature and photosphere radius (top); and bolometric luminosity (bottom) evolution. Derived values at peaks 1, 2, and 3 are shown in grey. Numerical values are tabulated in text (see subsection \ref{subsec:bol_facts}).}
    \label{fig:extrabol_output}
\end{figure}

We use the \texttt{extrabol} \citep{extrabol_cite} package to estimate the bolometric luminosity ($L_{\rm bol}$), blackbody temperature ($T$), and photosphere radius ($R_{\rm phot}$) of SN~2022oqm over time. The {\tt extrabol} package interpolates the light curve in each band using a Gaussian process with a 2D 3/2-Matern kernel, accounting for correlation in both time and wavelength. A blackbody spectral energy distribution (SED) is then fit to each observed epoch, inferring bolometric luminosities, blackbody radii, and blackbody temperatures with time. The fitted $L_{\rm bol}$, $T$, and $R_{\rm phot}$ are shown in Figure~\ref{fig:extrabol_output}, while the bolometric luminosity, temperature, and radius during peaks 1, 2, and 3 are listed in Table~\ref{table:extrabol_info}. 

At the time of detection, SN~2022oqm is very hot ($T \approx 40,000$~K) rapidly expanding and cooling between peak 1 (first detection) to peak 2, where the temperature decreases by a factor of $5$ and the bolometric luminosity decreases by a factor of $10$. After peak 2, the object continues to cool down and expand, but the bolometric luminosity increases by a factor $1.5$ until peak 3. By integrating the bolometric luminosity over all time, we find at $1\sigma$ significance, the total energy emitted ranges from $48.6 \leq \log_{10}\left[E_{\rm{tot}}/\rm{erg}\right] \leq 49.1$. We find our derived blackbody luminosity and temperatures are within $1\sigma$ of the derived values in \cite{irani2022}.

\section{Spectroscopic Analysis}
\label{sec:spectroscopy}

\begin{figure*}
    \centering
    \includegraphics[width = 0.95\linewidth]{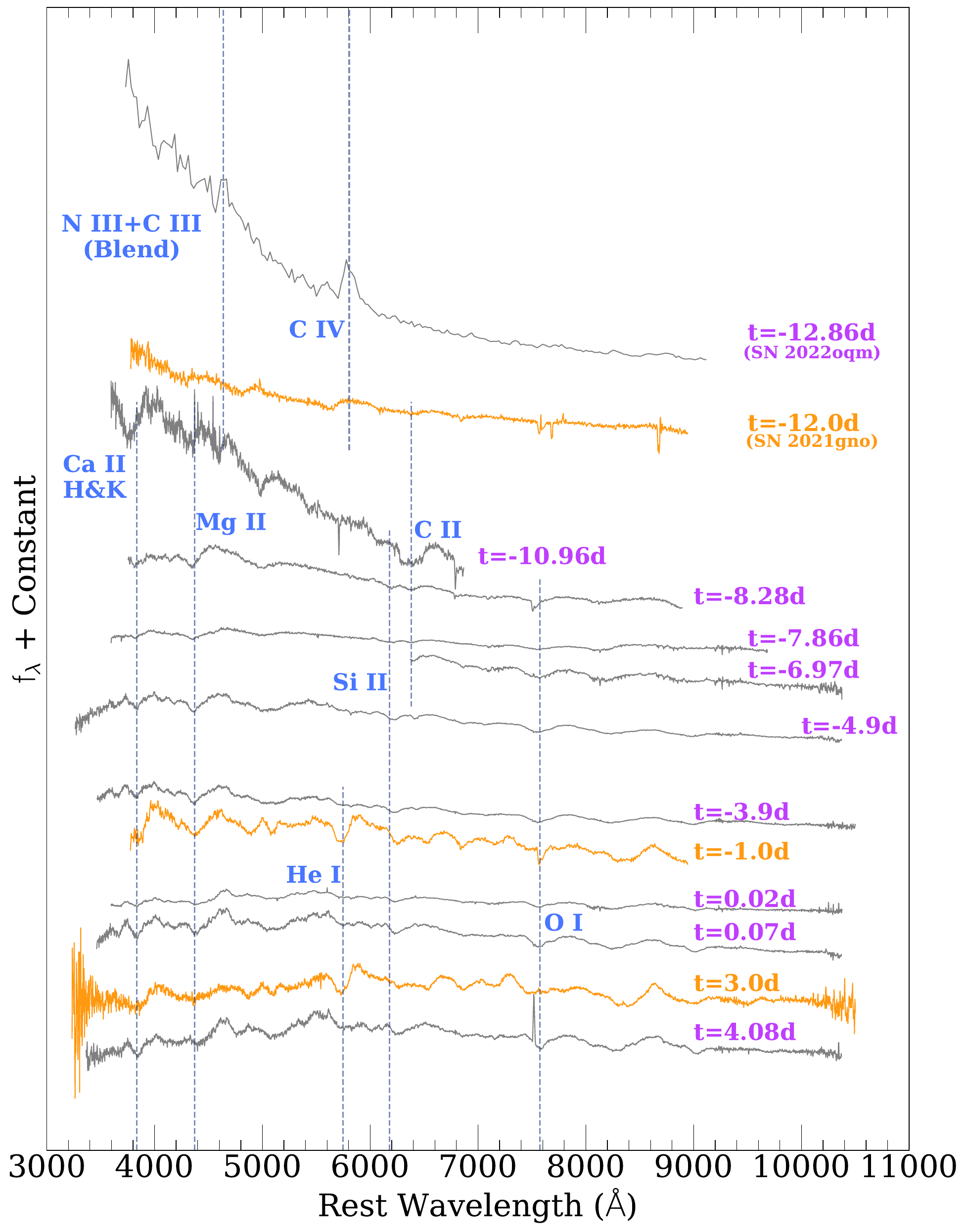}
    \caption{Spectral sequence of SN~2022oqm during $-13d\leq t \leq5d$ (grey) overlaid with the same-phase spectra of SN~2021gno, a He-rich \cart{} \citep{Wynn2022}. We identify \ion{Ca}{2} $\lambda\lambda3933$ and $3968$, \ion{Mg}{2} $\lambda4481$, \ion{N}{3}$\lambda4638$, \ion{C}{3} $\lambda4658$, \ion{C}{4} $\lambda 5801$, \ion{He}{1} $\lambda 5876$, \ion{Si}{2} $\lambda 6347$, \ion{C}{2} $\lambda 6580$, and \ion{O}{1} $\lambda 7775$ with blue vertical lines. Note that \ion{He}{1} $\lambda5876$ is only seen in the spectra of SN~2021gno. All spectra are in the rest frame. The vertical axis is scaled linearly. The earliest spectrum is presented in \cite{2022oqm_original_classification}. }
    \label{fig:spec_sequence_1}
\end{figure*}

\begin{figure*}
    \centering
    \includegraphics[width = 0.95\linewidth]{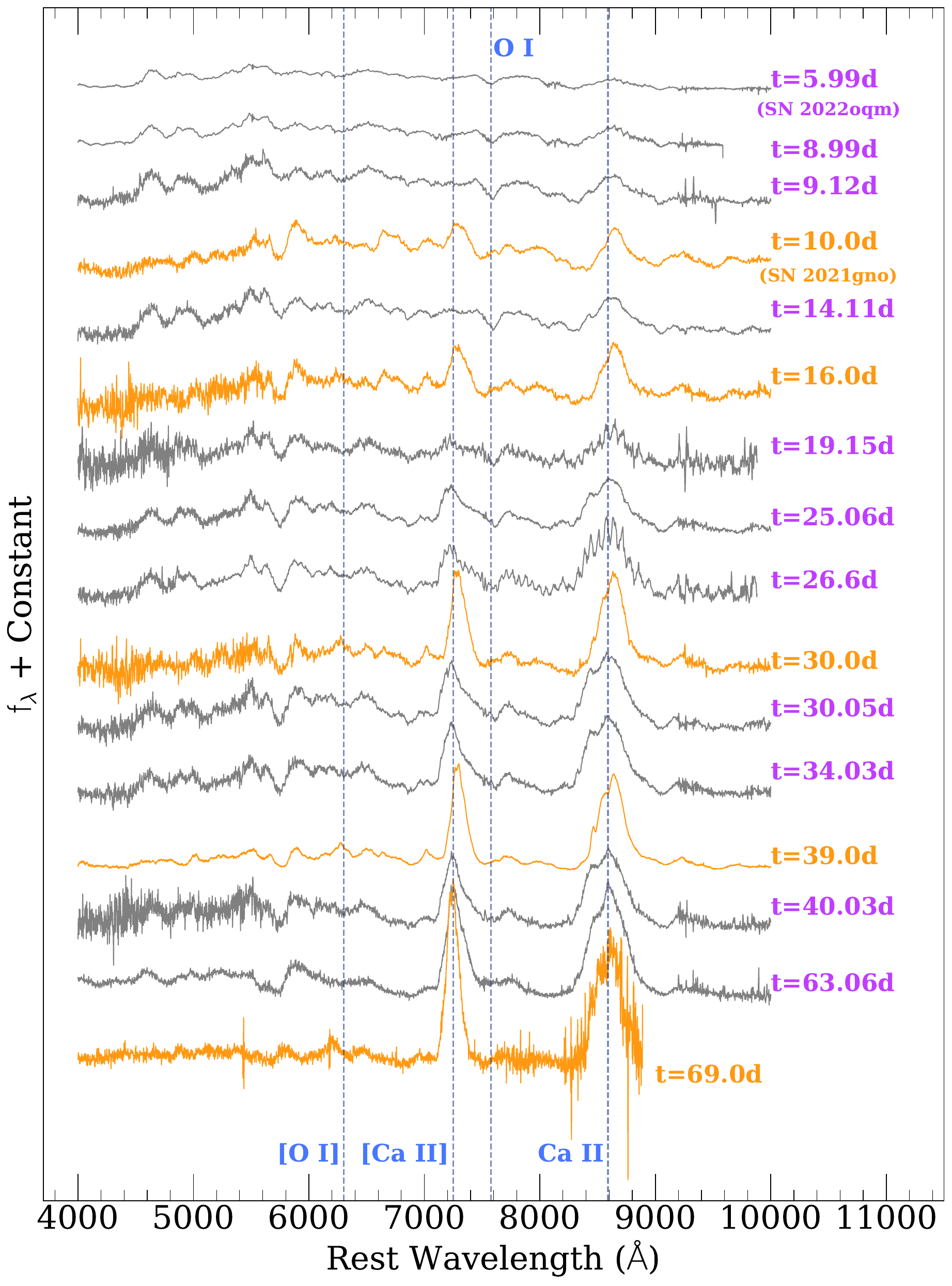}
    \caption{Spectral sequence of SN~2022oqm during $5d\leq t \leq65d$, overlaid with the same-phase spectra of the He-rich \cart{} SN~2021gno \citep{Wynn2022}. The features \ion{O}{1} $\lambda7775$, [\ion{Ca}{2}] $\lambda\lambda7291,7324$, and \ion{Ca}{2} $\lambda\lambda8498$, $8542$, and $8662~\mathrm{\AA}$ are clearly detected and marked. We also show the location of [\ion{O}{1}] $\lambda\lambda6300,6364$, but this line is not clearly detected in our spectral sequence. The strong [\ion{Ca}{2}] $\lambda\lambda7291,7324$ feature is characteristic of \carts. The Ca complex found redward of $8000~\mathrm{\AA}$ is a permitted feature and is not used to qualify a transient as a CaRT. All spectra are in the rest frame. The vertical axis is scaled linearly.}
    \label{fig:spec_sequence_2}
\end{figure*}

\begin{figure*}
    \centering
    \includegraphics[width = 0.95\linewidth]{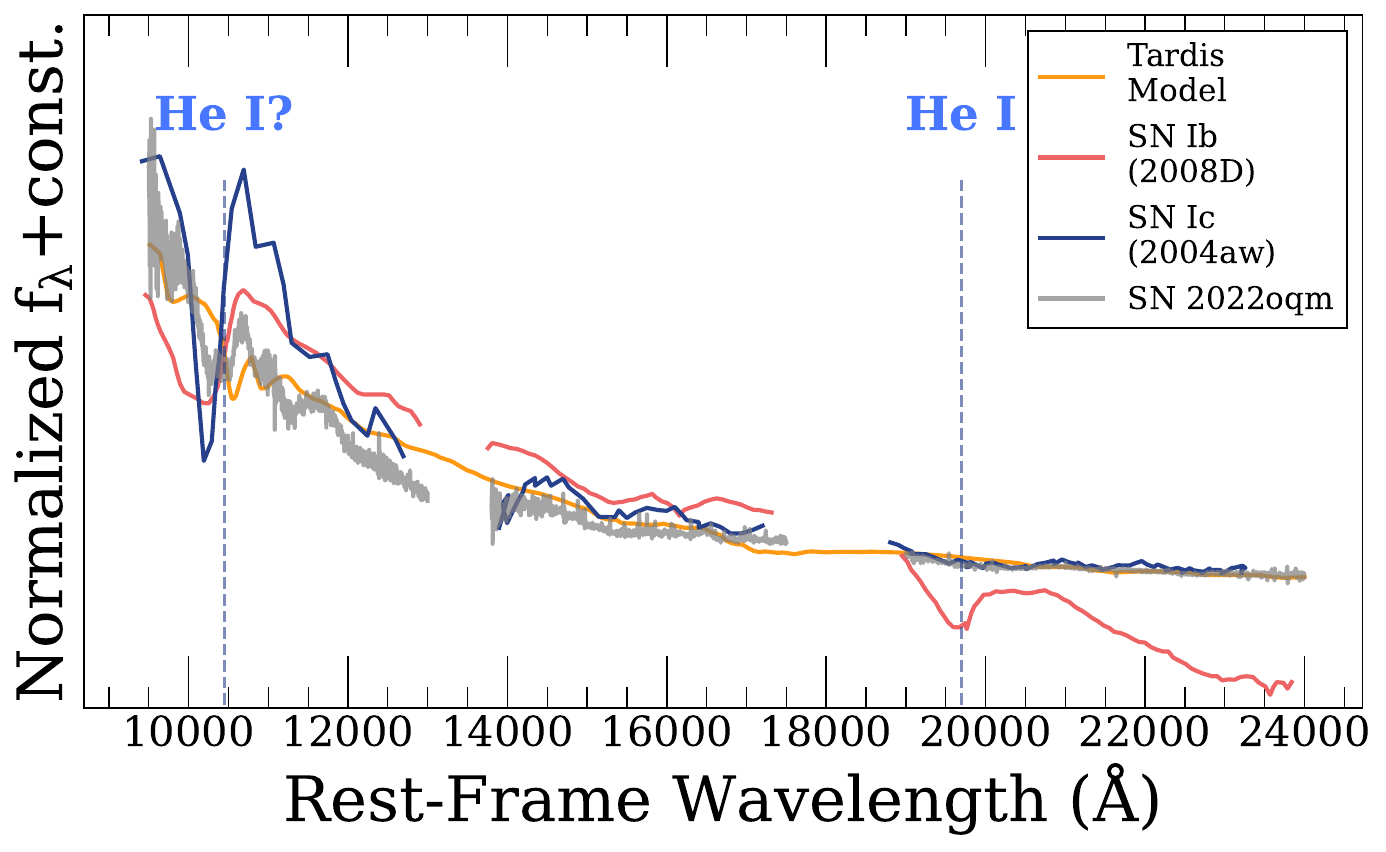}
\caption{NIR Spectrum of SN~2022oqm taken at $t = -4.9$d in grey. Also overlaid are the corresponding phase \texttt{TARDIS} radiative transfer simulation (see Figure~\ref{fig:TARDIS}) in orange, the NIR spectrum of the SN~Ib SN~2008D in red, and the NIR spectrum of the SN~Ic SN~2004aw in dark blue. In contrast to SN~2022oqm, SN~2008D has clear \ion{He}{1} $\lambda10830$ and \ion{He}{1} $\lambda20581$ absorption features. Similar to SN~2022oqm, SN~2004aw has a potential \ion{He}{1} $\lambda10830$ absorption feature, but lacks the \ion{He}{1} $\lambda20581$ absorption feature. The spectrum of SN~2004aw is from \cite{Taubenberger2004aw} and the spectrum of SN~2008D is from \cite{Modjaz+09}. The vertical axis is scaled linearly.}
    \label{fig:nir_series}
\end{figure*}

We present 22 spectra that were taken between $t=-14$d and $t=62$d in 
figures~\ref{fig:spec_sequence_1}, \ref{fig:spec_sequence_2}, and \ref{fig:nir_series}. The 
earliest spectrum ($-12.86$d) is publicly available and was presented by \cite{2022oqm_original_classification} to obtain the earliest classification of SN~2022oqm. This spectrum is dominated by a blue continuum with a \CIV{} emission feature at 
$5805~\mathrm{\AA}$ and \ion{N}{3} at $4638~\mathrm{\AA}$. \cite{irani2022} present multiple spectra with a higher signal-to-noise ratio at this phase, identifying the same emission features in addition to strong \ion{O}{3}, \ion{O}{4}, and \ion{O}{5} features. The presence of such high ionization levels suggests the presence of an intense UV radiation field \citep{leloudas2019, Quimby2007}. Indeed fitting a blackbody curve to the earliest spectrum suggests a temperature in excess of $3\times10^4$ K, in agreement with our photometrically inferred earliest temperature (see Figure~\ref{fig:extrabol_output}). The observed narrow highly ionized species during the earliest phase of SN~2022oqm points to the presence of circumstellar material (CSM) in the system, where non-thermal photons are being created by the interaction between the SN ejecta and CSM. After shock breakout, the excited CSM cools through shock cooling, powering the earliest phase of the light curve. The earliest spectrum suggests that the CSM lacks H and He, although C and O are observed.

We observe the \ion{Ca}{2} H\&K complex at $3933~\mathrm{\AA}$, the \ion{Mg}{2} $\lambda 4481$ feature, a weak \ion{S}{2} $\lambda 6355$ absorption line, a weak \ion{C}{2} $\lambda 6580$ absorption line, and an \ion{O}{1} $\lambda 7775$ feature before peak 3. The NIR spectrum taken on $t=-6$d (Fig~\ref{fig:nir_series}) shows no clear evidence of helium, unlike many \carts{} (see section \ref{sec:ca_rich_comparisons} for a detailed comparison between SN~2022oqm and CaRT spectroscopic signatures).

By $t\approx 30d$, we find the SN has begun transitioning to its nebular phase, revealing significant forbidden line emission. By this time, a clear [\ion{Ca}{2}] $\lambda\lambda7291, 7324$ emission feature becomes apparent. Using the most recently available spectrum ($t = 62d$), in which SN~2022oqm approaches the nebular phase, we attempt to measure an emission flux ratio [\ion{Ca}{2}] $\lambda\lambda7291,7324$/[\ion{O}{1}] $\lambda\lambda6300,6364$. Because [\ion{O}{1}] emission is not clearly detected, we measure the total flux within the equivalent width of the [\ion{Ca}{2}] $\lambda\lambda7291,7324$ line centered at the [\ion{O}{1}] $\lambda\lambda6300,6364$ line location. We place a lower limit of $\approx4.4$, greater than the defined cutoff of $2$ to be considered a \cart{}~\citep{Perets10, Wynn2022, de2020zwicky}. 

We also plot the spectra of the He-rich \cart{} SN~2021gno studied in detail by \cite{Wynn2022} and \cite{2023Ertini} in Figures~\ref{fig:spec_sequence_1} and \ref{fig:spec_sequence_2}. We show the phase of the SN~2021gno spectra with respect to the epoch of $r$-band peak of SN~2021gno, facilitating easy comparison between the phases of SN~2021gno and SN~2022oqm. We find that with the exception of the highly ionized line transitions found in the earliest spectrum of SN~2022oqm and the \ion{He}{1} $\lambda5876$ feature in SN~2021gno, all lines found in the spectra of SN~2022oqm are also found in those of SN~2021gno. This broad agreement with both spectral sequences further supports the \cart{} classification of SN~2022oqm. We compare photospheric and nebular spectra of SN~2022oqm to those of SN~2021gno in more detail in Subsection~\ref{subsec:phot_neb_overlay}.

\begin{figure}
    \centering
    \includegraphics[width = \linewidth]{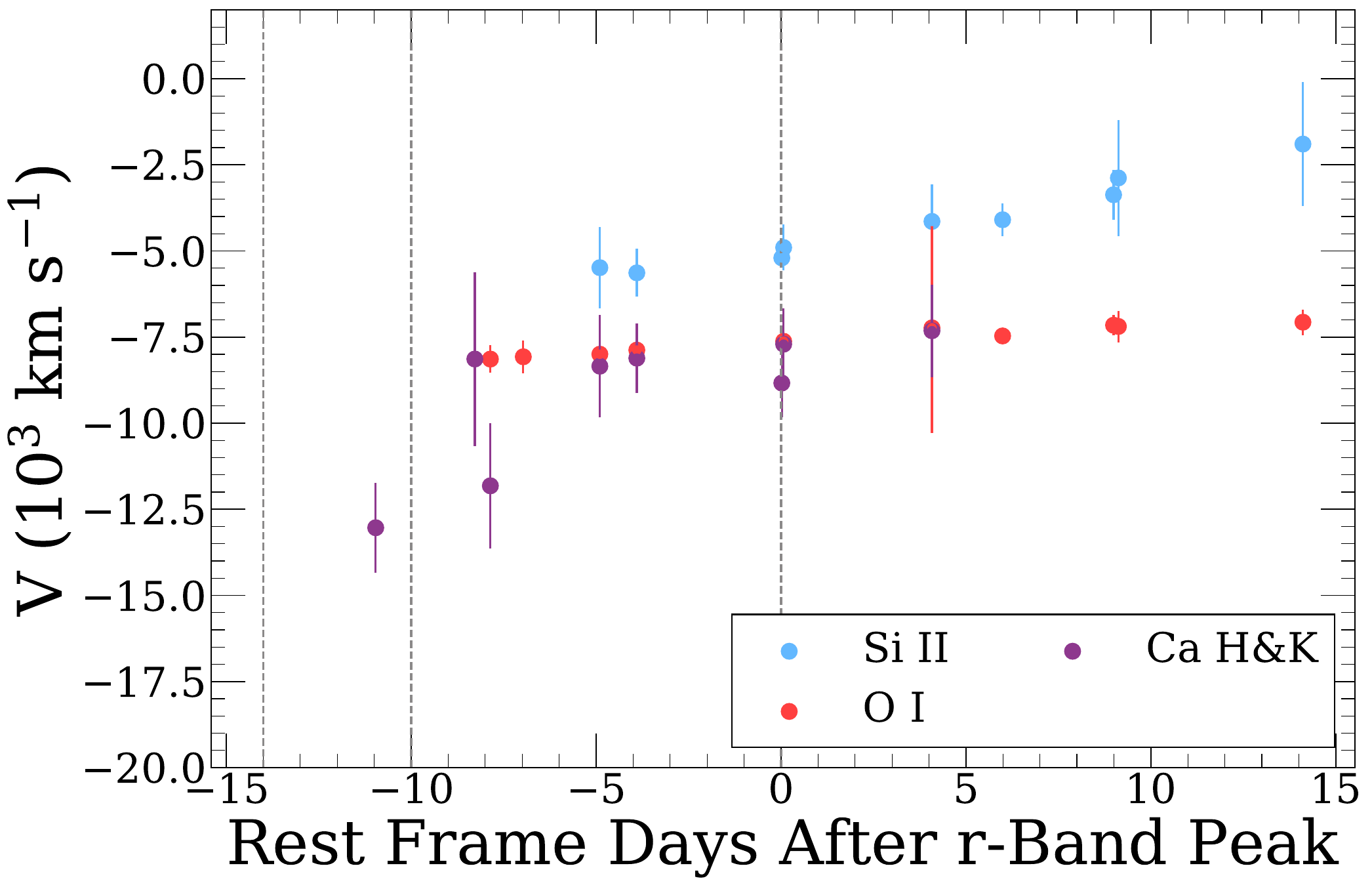}
    \caption{Velocities of the \ion{Si}{2}, \ion{O}{1}, and Ca H\&K absorption lines measured during the photospheric phase. Silicon is consistently at a smaller velocity (and therefore radius) than oxygen and calcium, which seem to be at similar radius throughout the early evolution. Times of peaks 1, 2, and 3 are shown as dotted grey lines.}
    \label{fig:photosphere_vel}
\end{figure}

Spectroscopic observations enable a census of elemental abundance, density, and velocity structures within the SN ejecta. As the ejecta expand, optical depth decreases with time, and the photosphere recedes into deeper layers of the ejecta. By measuring the line velocity of several elements over time, one can probe the velocity gradient of the ejecta. We present the velocities of \ion{O}{1} $\lambda 7775$, \ion{Si}{2} $\lambda 6355$, and \ion{Ca}{2} H\&K absorption lines during the photospheric phase of the SN in Figure~\ref{fig:photosphere_vel}. \ion{O}{1} $\lambda 7775$ velocity is roughly constant at $\sim 8000$~km~s$^{-1}$ during the $\sim 20$ days when this line is visible. \ion{Si}{2} $\lambda 6355$ declines in velocity, starting from $\sim 5000$~km~s$^{-1}$ and reaching $3000$~km~s$^{-1}$. At early times, we observe \ion{Ca}{2} H\&K at $\sim 13000$~km~s$^{-1}$ and then approaches the \ion{O}{1} $\lambda 7775$ velocity from the time of peak 2 until immediately after peak 3. The times of peaks 1, 2, and 3 are highlighted with grey vertical lines. These velocities are typical for \carts{} and SNe~Ic \citep{modjaz2016spectral, de2020zwicky}.

\subsection{Spectral synthesis analysis}
\label{subsec:Tardis_modeling}
We model the observed spectral series of SN~2022oqm using \texttt{TARDIS} \citep{Kerzendorf2014}, a one-dimensional Monte-Carlo radiative transfer code that numerically solves the radiative transport equation and models the spectral emission of SN ejecta. \texttt{TARDIS} treats photon-atom line interactions using a macroatom formalism, where the ``macroatom'' is the combination of the photon and atom. The macroatom formalism prescribes how the excited atom eventually de-excites and emits a photon and is explained in detail by \cite{lucy2002}. \texttt{TARDIS} also assumes local thermodynamic equilibrium to simplify modeling the photon-atom interactions. In order to model helium, we have used the \texttt{recomb-nlte} treatment, which considers \ion{He}{1} excited states as if they are in local thermodynamic equilibrium with higher ionization \ion{He}{2} \citep[see e.g.][]{Boyle2017}.

By assuming homologous expansion of ejecta, \texttt{TARDIS} can parameterize ejecta layers by velocity. \texttt{TARDIS} takes ejecta density distribution, elemental abundance distribution, bolometric luminosity, and time since explosion as input to simulate an instantaneous SN spectrum produced by a prescribed range of velocities within the ejecta.

\begin{figure}
    \centering
    \includegraphics[width = \linewidth]{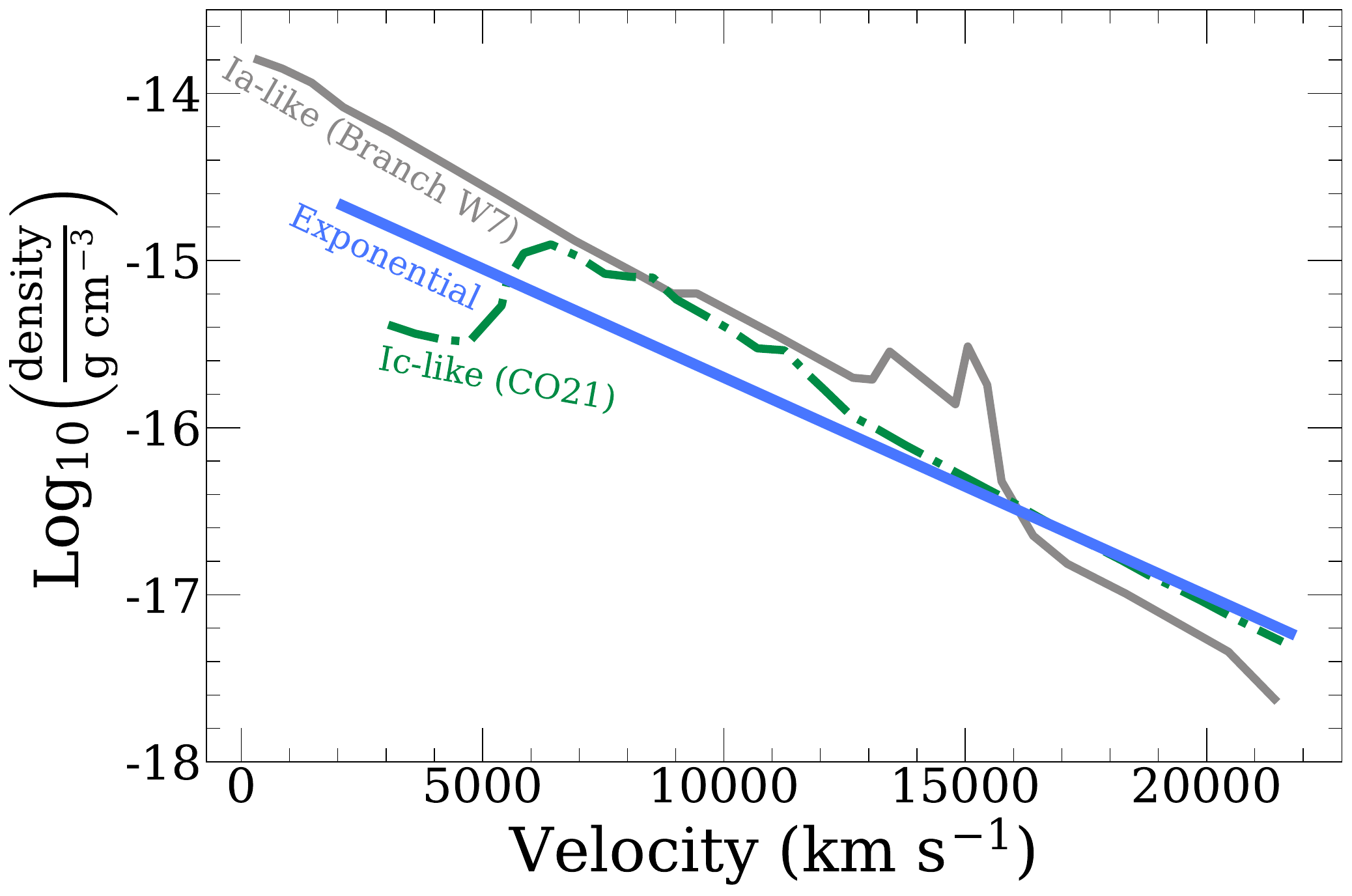}
    \caption{We use the exponential ejecta mass density profile to model SN~2022oqm ejecta using \texttt{TARDIS}. This is shown in relation to two commonly used density profiles for SN~Ia and SN~Ic modeling. The exponential density profile, which is shallower than both the Ia-like and Ic-like profiles, provides the best fit to the SN~2022oqm ejecta spectra. The fitted spectra are shown in Figure~\ref{fig:TARDIS}}
    \label{fig:density_profiles}
\end{figure}

The CO21 model \citep{CO21ModelOriginal} and the W7 Branch model \citep{Branch_moddel_1984, BranchModel1985} are the canonical density profiles in \texttt{TARDIS} for stripped-envelope and thermonuclear explosions respectively. We find that neither of these models reproduces the SN~2022oqm spectra well. As such, we utilize an exponential ejecta density profile to adequately fit the spectral sequence of SN~2022oqm. These three density profiles are reproduced in Figure~\ref{fig:density_profiles}. The exponential density profile mimics the Ic-like outer ejecta profile throughout the ejecta. The exponential density profile is constructed as $\rho \propto \exp\left(\frac{-3v}{10^4~\rm{km~s^{-1}}}\right)$ and has an integrated mass of $M_{ej} = 0.57$ M$_{\odot}$. The agreement to an exponential density profile is reminiscent of a disk-like ejecta structure (i.e. an alpha-disk model), but here we are limited to a one-dimensional geometry and therefore do not truly probe the angular distribution of the ejecta.

\begin{figure}
    \centering
    \includegraphics[width = \linewidth]{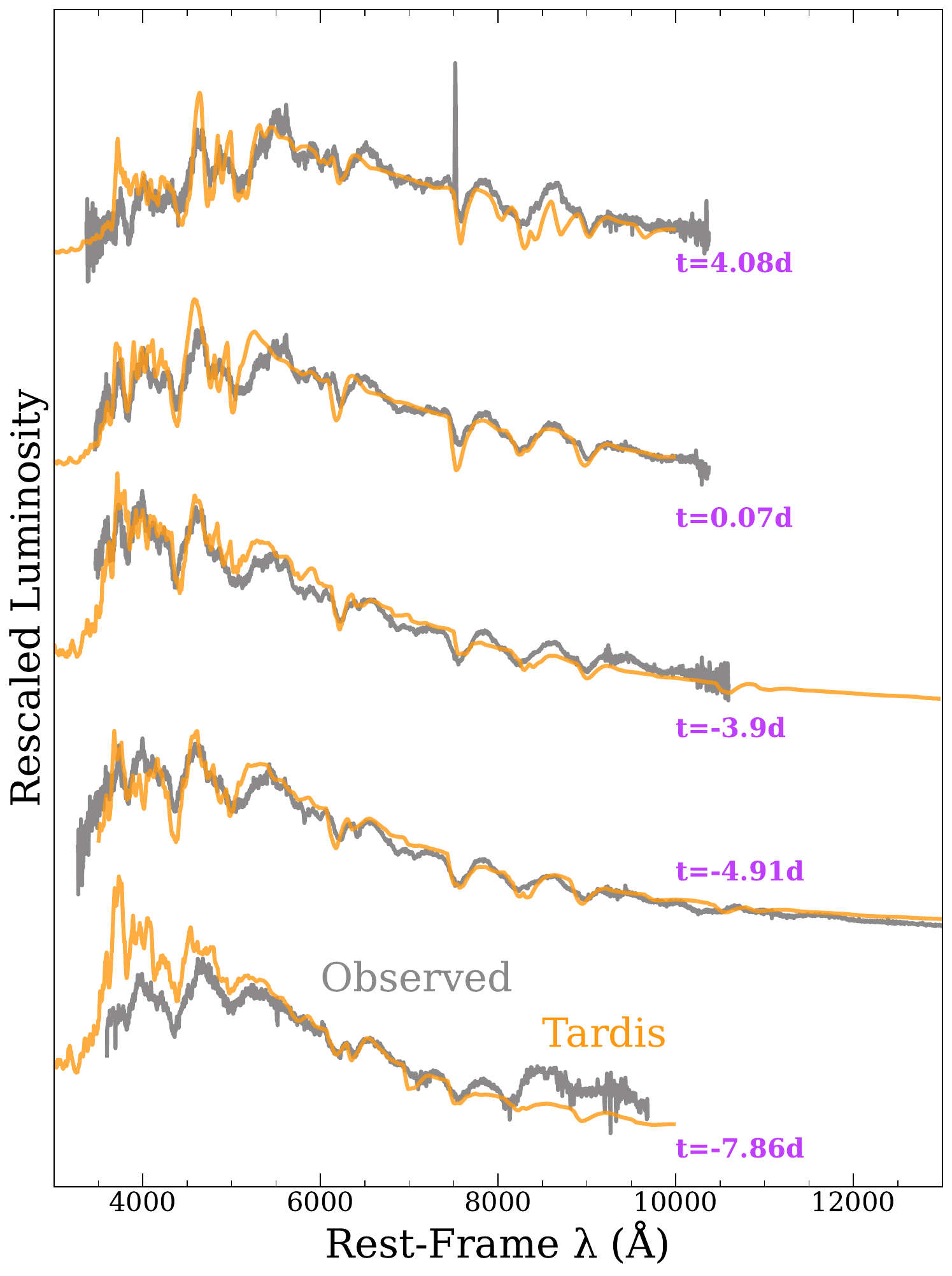}
    \caption{Spectral series of SN 2022oqm (grey curves) and the corresponding \texttt{TARDIS} radiative transfer simulation (orange curves). We find good agreement between the \texttt{TARDIS} models and our observed spectral sequence. The vertical axis is scaled linearly. The inferred ejecta density profile and the ejecta abundances, as inferred from \texttt{TARDIS} are shown in Figures~\ref{fig:density_profiles} and \ref{fig:abundance}, respectively.}
    \label{fig:TARDIS}
\end{figure}

\begin{figure}
    \centering
    \includegraphics[width = \linewidth]{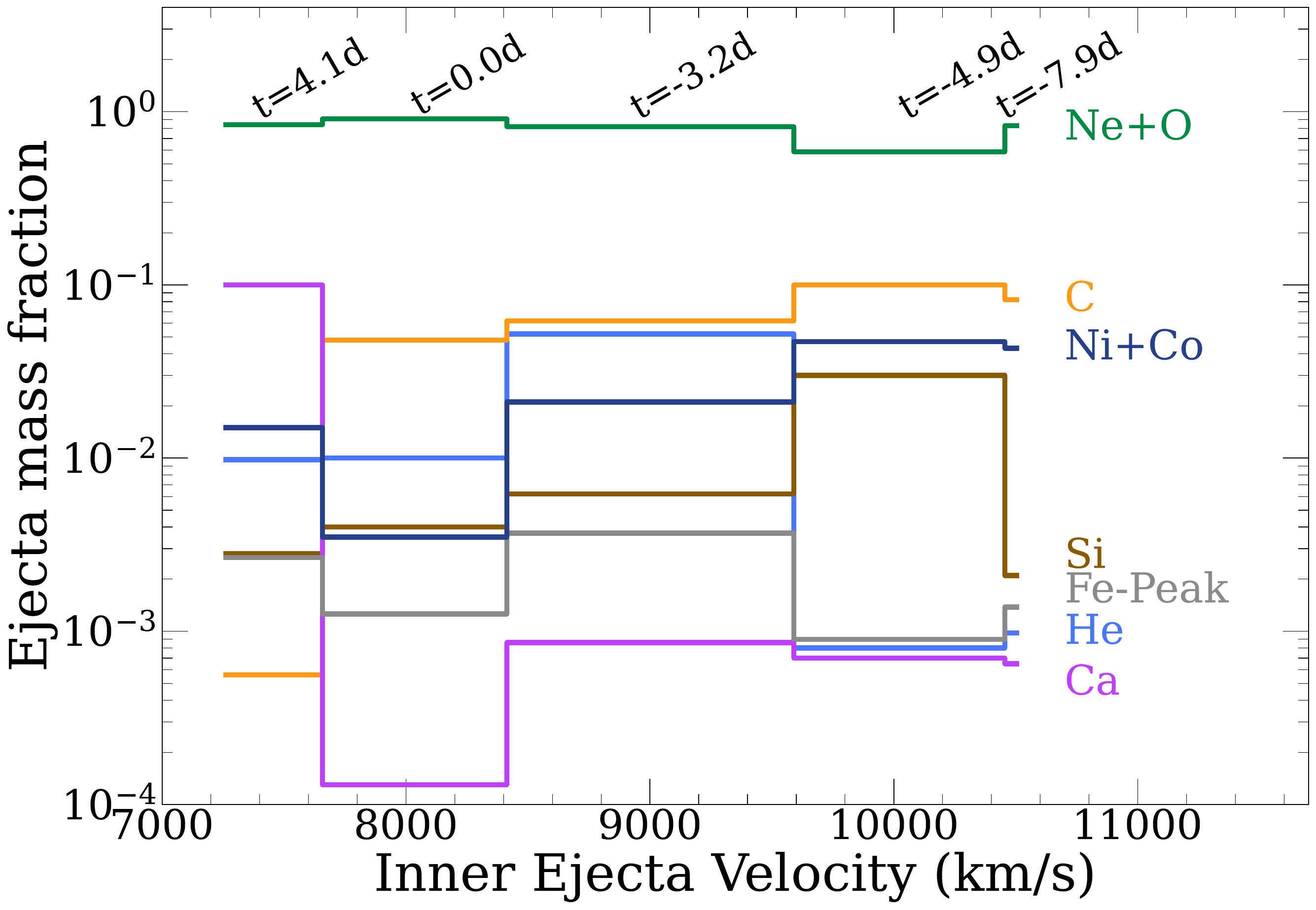}
    \caption{The abundance pattern derived from \texttt{TARDIS} fitting to spectra at $t=-7.86, -3.22, 0.07, 4.08$~d. Atomic abundances which are degenerate via \texttt{TARDIS} modeling are combined into one category (i.e. Ne+O, Ni+Co, and Fe-Peak, which is Fe, Ti, and Cr). The distributions of Fe-peak elements, radioactive elements, and lower-mass elements are flat throughout the ejecta.}
    \label{fig:abundance}
\end{figure}

We fit four optical spectra at $t=-7.86, -3.22, 0.07, 4.08$~d and one NIR spectrum at $t = -4.9$~d. Each simulation fits the velocity range and abundance of He, C, O, Ne, Si, Ca, and Fe-peak elements (including Ni and Co). We present atomic abundances which are denegerate in our modeling in the same category (e.g. Ne and O as ``Ne+O'' and Ni and Co as ``Ni+Co''). We assume these elements are uniformly distributed through the velocity range fitted by each spectral fit. We show the fitted spectra in Figure~\ref{fig:TARDIS} and the abundance distributions for each element in Figure~\ref{fig:abundance}. We find a general agreement between the simulated and the observed spectra, where the majority of spectral features are well reproduced. Some specific features, such as the broad absorption around $5200~\mathrm{\AA}$ at peak 3 are not correctly reproduced, likely due to a different ionization level of the line responsible for that absorption. We find a total ejecta mass of $\sim0.6~M_{\rm{\sun}}$.

The resulting abundance pattern is characterized by: 1) a predominant ejecta composition with a homogeneous distribution of both light elements (He, C, O, and Ne) and Fe-peak elements (Fe, Co, and Ni); and 2) a low Ca abundance at velocities $v_{ej} \gtrsim 7900$~km~s$^{-1}$, rapidly increasing when the most inner regions of the ejecta become visible at late epochs, e.g. days after peak brightness. The roughly uniform distribution of both Fe-peak and Ni+Co abundances throughout the ejecta shows agreement with the $\rm{^{56}Ni}$ mixing and double-decay model suggested in subsections \ref{subsec:mosfit_reflections}. A more detailed analysis of the iron features in the spectra could elucidate this, but such analysis is beyond the scope of this article. The enhancement of Ca during later spectra is consistent with the broadening of the later Ca lines observed $\sim$30 days after peak. A lack of distinct O and Ne features in our optical spectra mean that we are unable to constrain the individual abundance of either O or Ne.

Detailed spectroscopic modeling suggests a low He abundance in the ejecta of SN~2022oqm. Through \texttt{TARDIS} modeling, we find a helium mass upper limit of $8.43\times10^{-3}~M_{\mathrm{\sun}}$, corresponding to a mass fraction of $\sim1.5\%$ (see Figure~\ref{fig:abundance}). In addition, the commonly used \ion{He}{1} $\lambda10830$ and \ion{He}{1} $\lambda20581$ features are nonexistent in the NIR spectrum of SN~2022oqm (see Figure~\ref{fig:nir_series}). A possible absorption feature is present near the \ion{He}{1} $\lambda10830$ line, but it has been shown that this could be a blend of Mg \citep{Filippenko1995, williamson2021Tardis}. \cite{williamson2021Tardis} show that the \ion{He}{1} $\lambda20581$ can be used to unambiguously constrain the presence of He because it is relatively separate from nearby lines. Moreover, \cite{williamson2021Tardis} show that if He is present in considerable amounts in the system, the \ion{He}{1} $\lambda20581$ will be visible. This feature is not detected in the NIR spectrum of SN~2022oqm, strongly suggesting a low abundance of He (see Figure~\ref{fig:nir_series}). In Figure~\ref{fig:nir_series} we compare the NIR spectrum of SN~2022oqm to that of the He-poor SN~Ic SN~2004aw \citep{Taubenberger2004aw} and the He-rich SN~Ib SN~2008D \citep{Modjaz+09}. SN~2008D has a clear \ion{He}{1} $\lambda20581$ absorption line, revealing the presence of He in SN~2008D, whereas SN~2004aw and SN~2022oqm both do not have a clear detection of the \ion{He}{1} $\lambda20581$ absorption line, suggesting a paucity of He (see \cite{Taubenberger2004aw} for a more detailed exploration of the presence of He in the spectrum of SN~2004aw).

\subsection{Comparing SN~2022oqm to Other Transients}
\label{subsec:phot_neb_overlay}
\begin{figure}
    \centering
    \includegraphics[width = \linewidth]{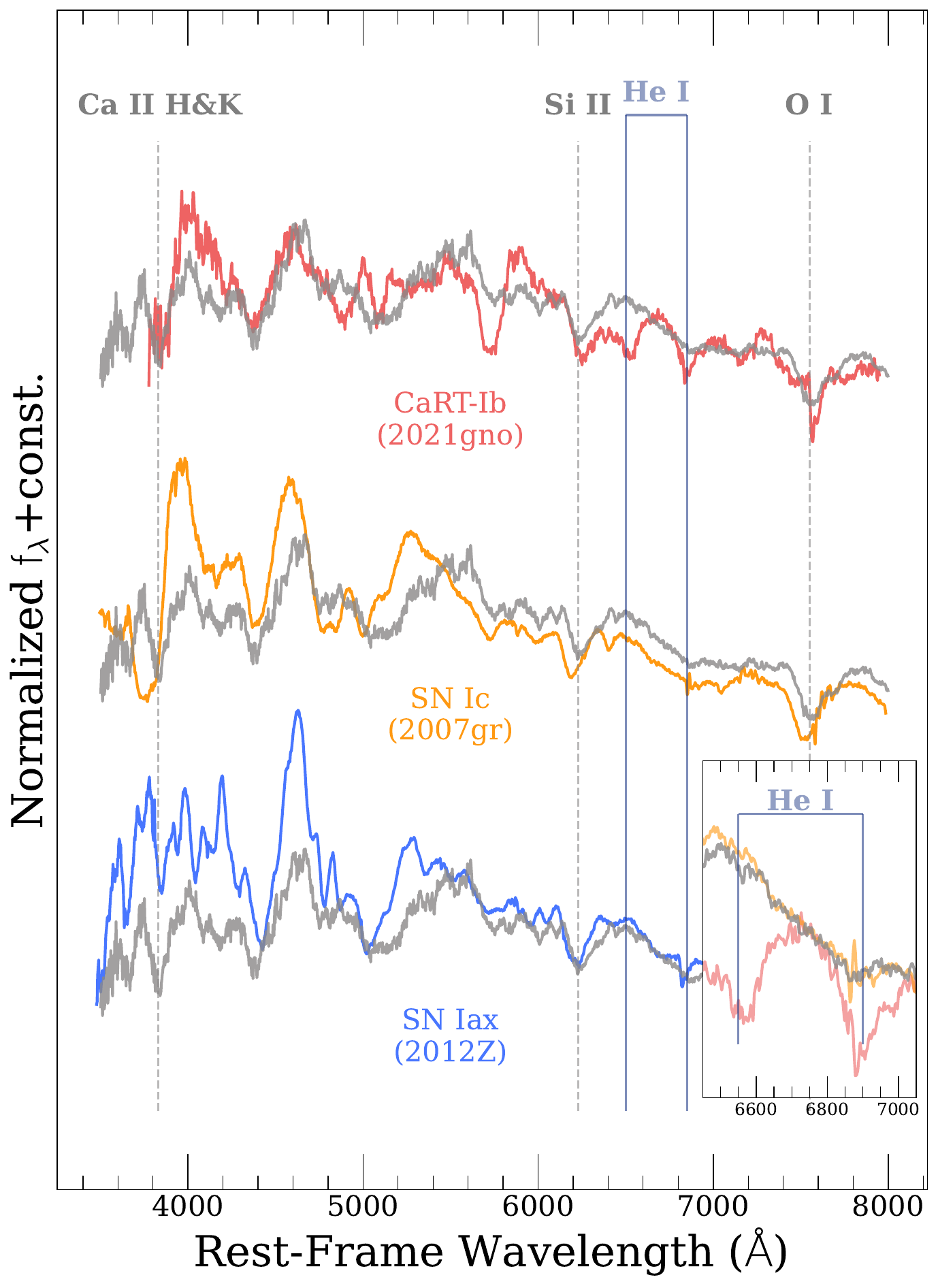}
    \caption{Peak-time spectrum of SN~2022oqm (grey) overlaid with peak-time spectra of SN 2012Z (SN~Iax), SN 2007gr (SN~Ic), and SN 2021gno (\cart{}). Absorption lines visible in the spectrum of SN~2022oqm are marked. SN~2022oqm appears quite similar to SN~Iax, SN~Ic, and the \cart{}. We show \ion{Ca}{2} $\lambda 3933~\mathrm{\AA}$, \ion{Si}{2} $\lambda 6347$, \ion{He}{1} $\lambda6678$, \ion{He}{1} $\lambda7065$, and \ion{O}{1} $\lambda 7775$, where the helium lines are only clearly discernable in the CaRT-Ib spectrum.}
    \label{fig:phot_spec_overlay}
\end{figure}

\begin{figure}
    \centering
    \includegraphics[width = \linewidth]{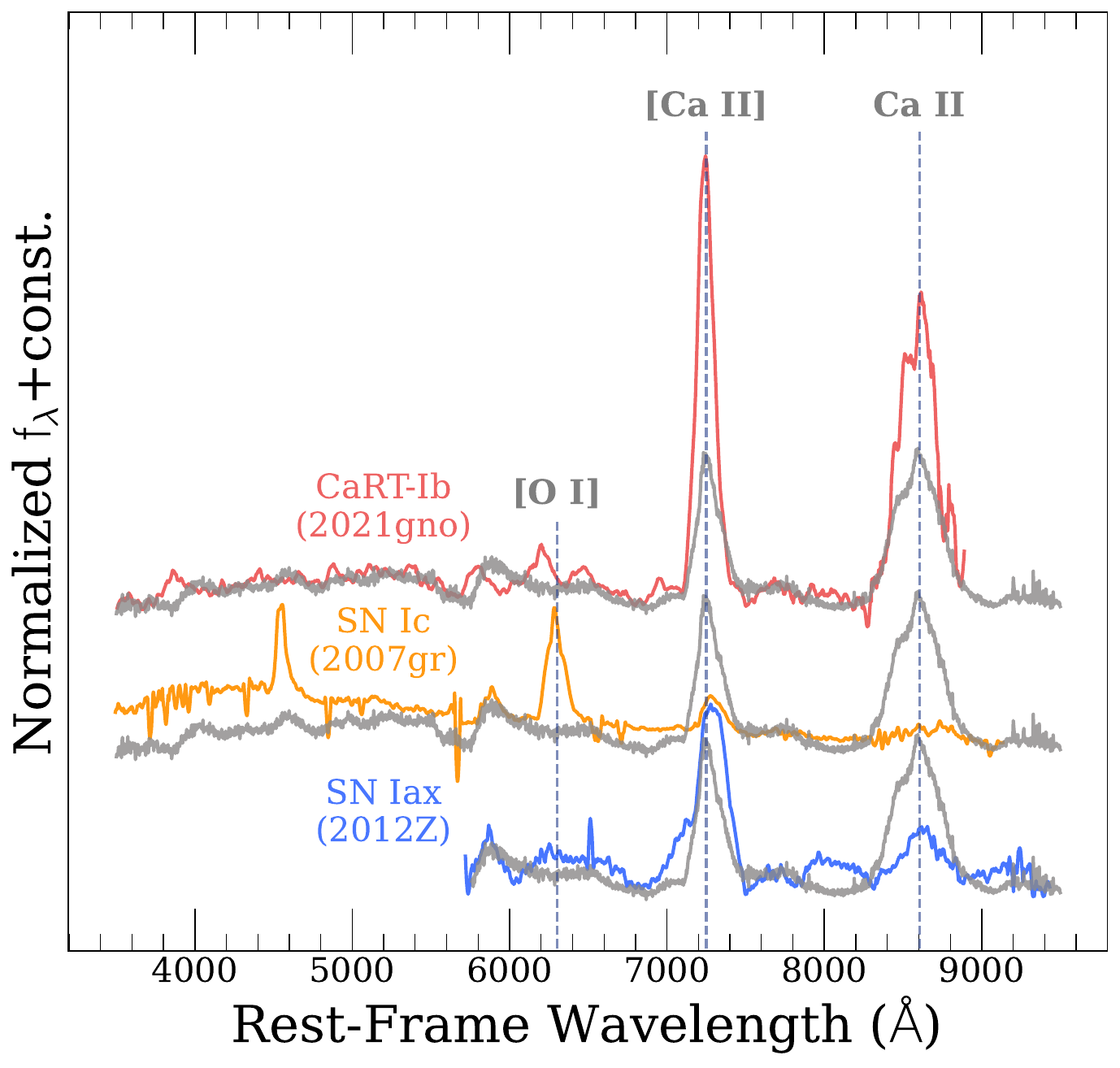}
    \caption{Nebular spectrum of SN~2022oqm (grey) overlaid with nebular spectra of SN 2012Z (SN~Iax), SN 2007gr (SN~Ic), and SN~2021gno (\cart{}). The [\ion{Ca}{2}] $\lambda\lambda7291,7324$ forbidden emission line is present in all spectra but is strongest in the \carts{}. The [\ion{O}{1}] $\lambda\lambda6300,6364$ feature is completely absent in the spectrum of SN~2022oqm. Note that this [\ion{O}{1}] line is stronger in the SN~Ic, a characteristic difference between SNe~Ic and Ca-Ic.}
    \label{fig:nebular_spec_overlay}
\end{figure}

We overlay the spectrum of SN~2022oqm with those of SN~2007gr (a SN~Ic, \citealt{Hunter2007gr, valenti2007gr}), SN~2021gno (a \cart{}, \citealt{Wynn2022, 2023Ertini}), and SN~2012Z (a SN~Iax, \citealt{Stritzinger2012z, mccully2022still}) at peak (Figure~\ref{fig:phot_spec_overlay}) and during their nebular phases (Figure~\ref{fig:nebular_spec_overlay}). At peak, we show \ion{Ca}{2} $\lambda\lambda 3933$ and $3968$, \ion{Si}{2} $\lambda 6347$, and \ion{O}{1} $\lambda7775$ with vertical lines, while for the nebular phase (Fig.~\ref{fig:nebular_spec_overlay}), we show the [\ion{O}{1}] $\lambda\lambda6300,6364$, [\ion{Ca}{2}] $\lambda\lambda7291,7324$, and \ion{Ca}{2} $\lambda\lambda8498$, $8542$, and $8662$ with vertical lines. During the photospheric phase of SN~2022oqm, the spectrum closely resembles that of a SN~Ic, where the strengths of calcium, silicon, and oxygen appear to be almost identical. The only inconsistency appears to be at shorter wavelengths, where line blanketing reduces flux more strongly in \carts{} than in SNe~Ic. We also note that helium is clearly detected in at least some \carts{} but not in SN~2022oqm or the SN~Ic. In the nebular phase, however, SN~2022oqm has a very different spectrum from that of the SN~Ic. During the nebular phase, SNe~Ic spectra feature a strong [\ion{O}{1}] $\lambda\lambda6300,6364$ emission line \citep{1985F_Oxygen_line, shivvers2019berkeley}. This line is not detected in the spectrum of SN 2022oqm, which instead has a very strong [\ion{Ca}{2}] $\lambda\lambda7291,7324$ emission line, characteristic of \carts{}. 

\section{Detailed Light Curve Modeling}
\label{sec:mosfit}

\begin{deluxetable*}{cccccc}
\tabletypesize{\footnotesize}
\tablecolumns{6}
\tablecaption{Best-fit parameters of each model produced in \texttt{MOSFiT} \label{table:mosfit}. The goodness of fit is parameterized by the Watanabe Akaike Information Criterion (WAIC) measure and by the reduced chi-squared ($\chi^2_\mathrm{red}$) measure. }
\tablehead{
\colhead{Module} &\colhead{Parameter}  & \colhead{\begin{tabular}{@{}c@{}}Power Law \\ + Single $\mathrm{^{56}Ni}$ Decay \\ (Section ~\ref{subsec:pow_arnett})\end{tabular}} & \colhead{\begin{tabular}{@{}c@{}} Single $\mathrm{^{56}Ni}$ Decay \\ (Section~\ref{subsec:just_arnett})\end{tabular}} & \colhead{\begin{tabular}{@{}c@{}} Double $\mathrm{^{56}Ni}$ Decay \\ (Section~\ref{subsec:just_arnett})\end{tabular} } & \colhead{\begin{tabular}{@{}c@{}} Power Law \\ + Double $\mathrm{^{56}Ni}$ Decay \\ (Section~\ref{subsec:pow_double_arnett})\end{tabular}}}
 \startdata
 \\
     Power Law & $log~\frac{L_\mathrm{pow}}{\mathrm{erg~s^{-1}}}$  & $43.43^{0.89}_{1.17}$  & -- & -- & $41.43^{1.79}_{1.15}$ \\
     \\
      & $\alpha$   & $1.10^{0.03}_{0.03} $ & -- & -- & $4.55^{0.78}_{0.95}$\\
      \\
      & $log~\frac{t_\mathrm{pow}}{\mathrm{days}}$ & $-0.5^{1.11}_{0.83}$ & -- & -- & $0.71^{0.3}_{0.35}$\\
      \\
      & $log~\frac{v_\mathrm{ej, pow}}{\mathrm{km~s^{-1}}}$ & $4.14^{0.03}_{0.05}$ & -- & -- & $3.82^{0.09}_{0.1}$\\
      \\
      & $\kappa_\mathrm{pow}$ & $0.11^{0.05}_{0.03}$ & -- & -- & $0.11^{0.05}_{0.04}$\\
      \\
      & $log~\frac{T_\mathrm{pow}}{\mathrm{K}}$ & $3.79^{0.01}_{0.01}$ & -- & -- & $3.43^{0.41}_{0.27}$\\
     \\
     \hline
     \\
     $\mathrm{^{56}Ni}$ Decay 1 & $log~\frac{m_\mathrm{1, ej}}{\mathrm{M_\odot}}$ & $-0.41^{0.01}_{0.01}$ & $-0.26^{0.02}_{0.02}$ & $0.07^{0.16}_{0.21}$ & $-0.35^{0.07}_{0.06}$ \\
     \\
      & $log~f_\mathrm{1, Ni}$ & $-0.70^{0.01}_{0.01}$ & $-0.79^{0.03}_{0.02}$ & $-1.30^{0.21}_{0.16}$ & $-0.90^{0.04}_{0.06}$ \\
      \\
      & $M_\mathrm{1, Ni}$ & $0.074_{0.001}^{0.001}$  & $0.09_{0.002}^{0.003}$  & $0.055_{0.002}^{0.002}$  & $0.06_{0.002}^{0.002}$   \\
      \\
      & $log~\frac{v_\mathrm{1, ej}}{\mathrm{km~s^{-1}}}$ & $4.0^{0.02}_{0.02}$ & $4.31^{0.01}_{0.01}$ & $3.97^{0.01}_{0.02}$ & $3.92^{0.02}_{0.01}$ \\
      \\
      & $log~\frac{T_\mathrm{1}}{\mathrm{K}}$ & $2.7^{0.39}_{0.42}$ &$3.68^{0.01}_{0.0}$ & $2.82^{0.28}_{0.62}$ & $2.57^{0.54}_{0.38}$ \\
      \\
      & $log~\kappa_\mathrm{\gamma,12}$ & $-0.98^{0.03}_{0.02}$ & $-0.99^{0.02}_{0.01}$ & $-0.40^{0.21}_{0.15}$ & $-0.99^{0.02}_{0.01}$ \\
      \\
      & $\kappa_\mathrm{1,2}$ & $0.18^{0.01}_{0.01}$ & $0.19^{0.0}_{0.01}$ & $0.10^{0.05}_{0.04}$ & $0.19^{0.00}_{0.01}$ \\
      \\
      & $\log~\frac{t_\mathrm{d}}{\mathrm{days}}$ & $1.09^{0.08}_{0.9}$ & $0.78_{0.01}^{0.01}$ & $1.1_{0.02}^{0.02}$ & $1.1_{0.03}^{0.01}$ \\
     \\
     \hline
     \\
     $\mathrm{^{56}Ni}$ Decay 2 & $log~\frac{m_\mathrm{2, ej}}{\mathrm{M_\odot}}$ & -- & -- & $-0.96^{0.15}_{0.27}$ & $-0.73^{0.08}_{0.11}$ \\
     \\
      & $log~f_\mathrm{2, Ni}$ & -- & -- & $-0.49^{0.25}_{0.17}$ & $-0.67^{0.11}_{0.08}$ \\
      \\
      & $M_\mathrm{2, Ni}$ & -- & -- & $0.04_{0.002}^{0.002}$  & $0.04_{0.003}^{0.002}$    \\
      \\
      & $log~\frac{v_\mathrm{2, ej}}{\mathrm{km~s^{-1}}}$ & -- & -- & $4.21^{0.03}_{0.03}$ & $4.06^{0.03}_{0.02}$ \\
      \\
      & $log~\frac{T_\mathrm{2}}{\mathrm{K}}$ & -- & -- & $3.78^{0.01}_{0.01}$ & $3.76^{0.01}_{0.01}$ \\
      \\
      & $log~\kappa_\mathrm{\gamma,12}$ & -- & -- & $-0.40^{0.21}_{0.15}$ & $-0.99^{0.02}_{0.01}$ \\
      \\
      & $\kappa_\mathrm{1,2}$ & -- & -- & $0.10^{0.05}_{0.04}$ & $0.19^{0.00}_{0.01}$ \\
      \\
      & $\log~\frac{t_\mathrm{d}}{\mathrm{days}}$ & -- & -- & $0.46_{0.06}^{0.06}$ & $0.81_{0.05}^{0.04}$ \\
     \\
     \hline
     \\
     General & $log~\sigma$ & $-0.65^{0.01}_{0.02}$ & $-0.43^{0.04}_{0.02}$ & $-0.71^{0.01}_{0.01}$ & $-0.73^{0.01}_{0.01}$ \\
     \\
      & WAIC & 705.1 & 550.4 & 727.1 & 822.9\\
     \\
      & $\chi^2_\mathrm{red}$ & 1.29 & 3.53 & 1.52 & 0.85\\
     \\
 \enddata
\end{deluxetable*}

In this section, we model the light curve of SN 2022oqm using several potential underlying power sources through the Modular Open-source Fitter for Transients (\texttt{MOSFiT}, \citealt{guillochon2018}). \texttt{MOSFiT} is an open-source software which self-consistently models the time-variable SED following the framework originally presented in \cite{arnett1982}. For each fit, we present goodness of fit through the Watanabe-Akaike Information Criterion (WAIC) score, where a higher score represents a better fit to the data. The WAIC score accounts for parameter size by punishing models with more parameters (see \citealt{guillochon2018}). Therefore, the WAIC score presents a method of comparing models with different parameter numbers and avoids the problem of overfitting. A WAIC score increase of at least $10$ implies a significantly better model fit \citep{watanabe2013, gelman2014WAIC}. We also measure goodness of fit for each model with the reduced chi-squared ($\chi^2_\mathrm{red}$) term.

The early excess in the light curve seen in peak 1 (see Figure~\ref{fig:reduced_photometry}) is reminiscent of shock cooling following  breakout of a SN shockwave out of surrounding circumstellar material (CSM, see e.g, \citealt{Smith2017, Piro2021}). Following this peak, the behavior of the SN near peak 3 is similar to the more classical SN light curve, which is powered by the radioactive decay of $\rm{^{56}Ni}$ diffused through optically thick ejecta \citep{arnett1982,chatzopoulos2012}. Within this framework, the ejecta mass ($m_\mathrm{ej}$), ejecta velocity ($v_\mathrm{ej}$), and greybody opacity ($\kappa$) are fully degenerate, as all three primarily impact the overall diffusion timescale of the transient: $t_d\propto \sqrt{\kappa m_\mathrm{ej}/ v_\mathrm{ej}}$. As such, we present $t_d$ in addition to these three parameters. 

\subsection{Joint Shock Cooling and Radioactive Decay Model}
\label{subsec:pow_arnett}

First, we explore a toy model in which the emission can be modeled by the combination of a power-law model (a reasonable description for shock cooling) and a radioactive decay model. The first peak would be captured by the power-law model and the subsequent light curve evolution would be captured by the radioactive decay. For the power law model, we utilize a power-law source of bolometric luminosity, described by:
\begin{equation}
    L_1 = L_{\rm pow} {\left(t/t_{\rm pow}\right)}^{-\alpha},
\end{equation}
where $L_{\rm pow}$ is a luminosity scaling factor, $t_{\rm pow}$ is the time-scale of decay, and $\alpha$ is the power-law index. This emission is treated as a single blackbody, with a characteristic photosphere velocity $v_{\rm ej, pow}$, opacity $\kappa_{\rm pow}$, and minimum photosphere temperature $T_{\rm pow}$ (see \citealt{villar2017theoretical, guillochon2018} for more details). 

We note that the power-law model is a well-known model to describe the luminosity of shock-heating material over time (see e.g., \citealt{Nakar&Sari10,Piro2021}). In this case, the time-dependence is directly linked to the outer ejecta profile, $L_1\propto t^{-4/(n-2)}$, where $n$ is the index of the power law that describes the outer ejecta density profile. We also note that we do not allow for photons emitted via this power-law model to diffuse through any material after emission. This is equivalent to assuming there is no material beyond the region where these power-law photons are emitted which can diffuse photons. Allowing diffusion of the power-law photons results in a much poorer fit (see Appendix~\ref{appx:piro}, where we attempt this fit).

The radioactive decay of $^{56}$Ni produces gamma-rays that are thermalized in optically thick ejecta, which then emit as a blackbody, driving peaks 2 and 3 \citep{arnett1982,chatzopoulos2012, villar2017theoretical}. The free parameters corresponding to this are the $\mathrm{^{56}Ni}$ mass ($m_\mathrm{Ni}$) and diffusion time $t_d \propto \sqrt{\kappa m_\mathrm{ej}/ v_\mathrm{ej}}$ (see Power Law + Single $\mathrm{^{56}Ni}$ Decay column in Table~\ref{table:mosfit}). 

\begin{figure*}
    \centering
    $\vcenter{\hbox{\includegraphics[width = 0.8\linewidth]{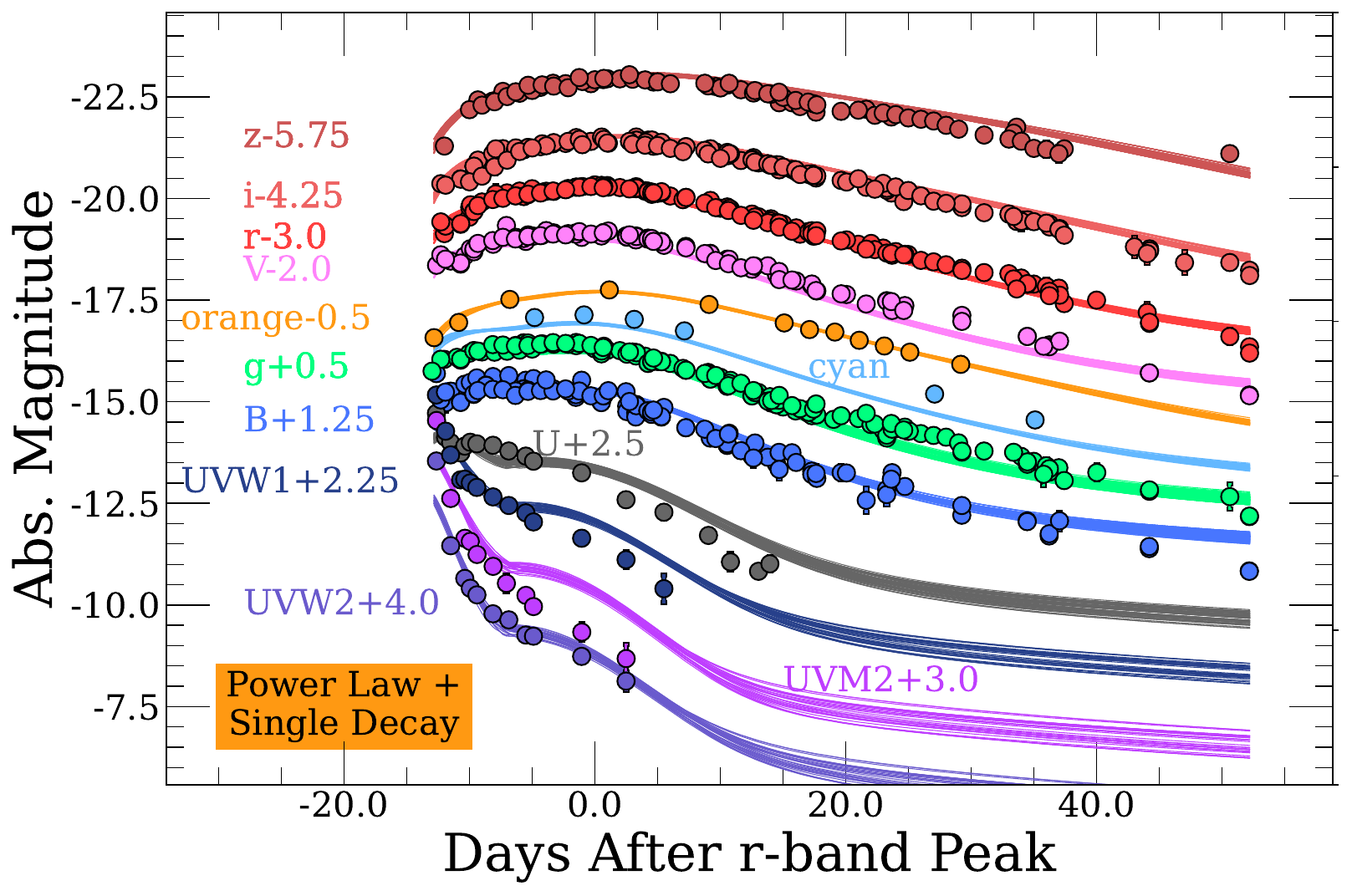}}}$
    \caption{Best-fit model to the UVONIR light curves for the shock cooling-$\mathrm{^{56}Ni}$ model. Points represent observed photometry, while colored lines are random draws from the resulting posterior. Fitting is done using the \texttt{MOSFiT} package.}
    \label{fig:pow_law_arnett}
\end{figure*}

We present the fitted light curve in Figure~\ref{fig:pow_law_arnett}, and show the fitted parameters in Table~\ref{table:mosfit}. We overall find agreement between the model and data ($\chi^2_\mathrm{red}=1.41$), but the fit is poor in the $U$, $UVW1$, and $UVM2$ bands. Most notably, this model overpredicts the emission in $UVM2$ by $\sim1$~mag after peak 1. 

\subsection{Capturing Individual Peaks with Radioactive Decay Models}
\label{subsec:just_arnett}
To find a better fit than was shown in Subsection~\ref{subsec:pow_arnett}, we focus on only peak 3. We cut peaks 1 and 2 from the light curve by including only data points from $t>-6$~d and fit a single radioactive decay model. We find an ejecta mass of $0.55~\mathrm{M_\odot}$, with a $^{56}\rm{Ni}$ fraction $f_{\rm Ni} \approx 0.16$, corresponding to $^{56}\rm{Ni}$ mass $m_{\rm Ni} = 0.09~\mathrm{M_\odot}$. Even though the first two peaks have been removed for this fit, the best fit still requires all $0.09~M_\sun$ to power peak 3, hinting that a single radioactive decay model of $^{56}\mathrm{Ni}$ does not fully account for the photons emitted during peak 3. 

We attempt to fit peaks 2 and 3 with a toy ``two-zone" model, by considering two discrete $^{56}\rm{Ni}$ central sources. This is similar to \citet{maeda2003two}, who presented a similar formalism of a two-zone model by constructing two concentric sources of photons. Photons emitted by $^{56}\rm{Ni}$ located at larger radii diffuse through a total mass $m_{2, \mathrm{ej}}$, before escaping from the system and driving peak 2. Photons emitted from the $^{56}\rm{Ni}$ decay that are located deeper inside the ejecta would diffuse through a larger amount of mass ($m_{1, \mathrm{ej}}+m_{2, \mathrm{ej}}$) before escaping, resulting in the delayed, third peak. 

We create a custom model in \texttt{MOSFiT} to perform this fit. We find an inner ejecta layer with mass of $log_{10}~[m_{1, \mathrm{ej}}/\mathrm{M_\sun}] = 0.07^{0.16}_{0.21}$ and a $^{56}\rm{Ni}$ fraction of $log_{10}~f_{\rm 1, Ni}=-1.30^{0.21}_{0.16}$, corresponding to $log_{10}~[M_{1, \mathrm{Ni}}/\mathrm{M_\sun}] = 0.055^{0.002}_{0.002}$ of $^{56}\rm{Ni}$. We find an outer ejecta layer with mass of $log_{10}~[m_{2, \mathrm{ej}}/\mathrm{M_\sun}] = -0.96^{0.15}_{0.27}$ and a $^{56}\rm{Ni}$ fraction of $log_{10}~f_{\rm 2, Ni}=-0.49^{0.25}_{0.17}$, corresponding to $log_{10}~[M_{2, \mathrm{Ni}}/\mathrm{M_\sun}] = 0.04^{0.002}_{0.002}$ of $^{56}\rm{Ni}$.

\subsection{Joint Shock Cooling and Double Radioactive Decay Model}
\label{subsec:pow_double_arnett}
\begin{figure}
    \centering
    \includegraphics[width = 0.9\linewidth]{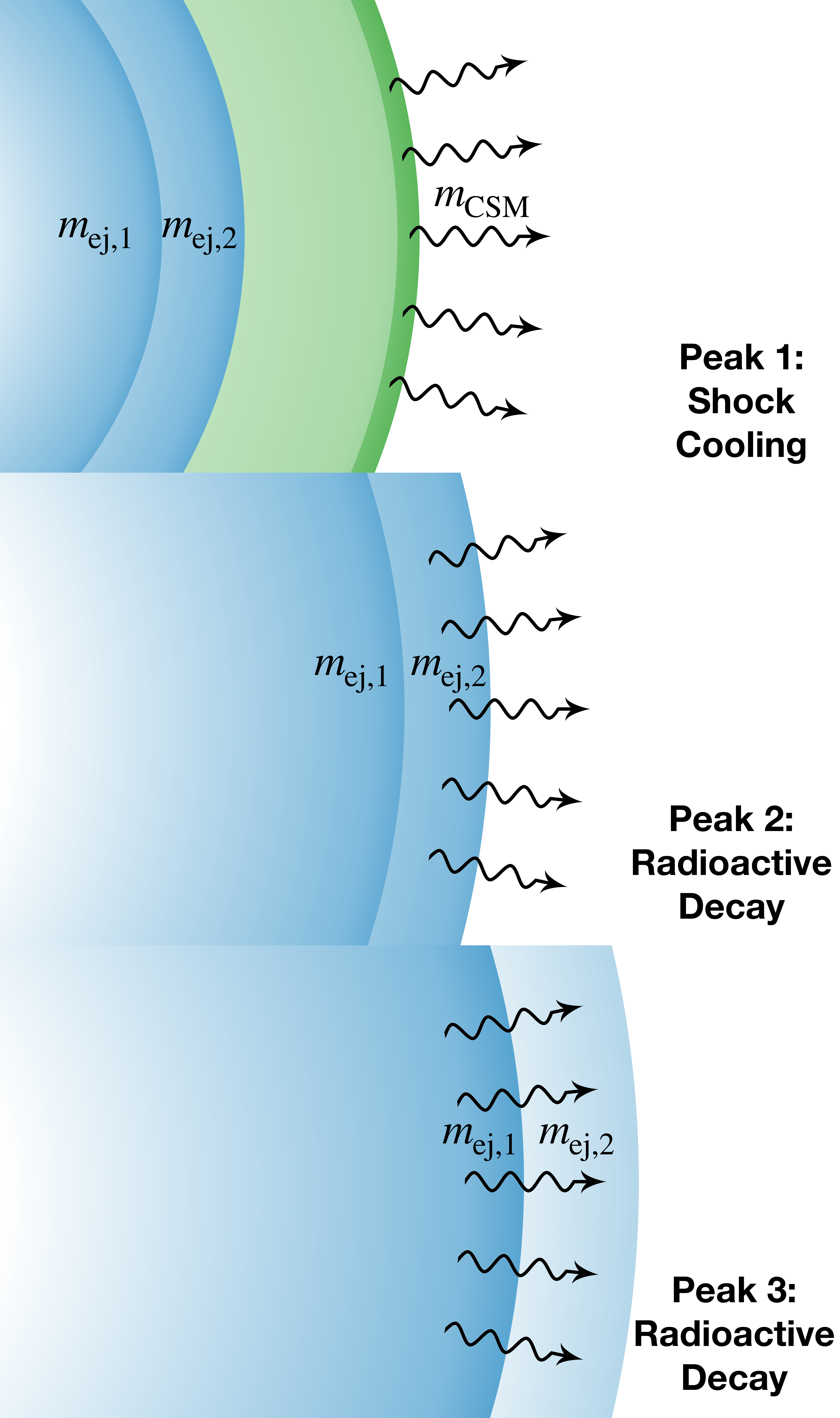}
    \caption{Schematic to highlight the sources of photons during peaks 1, 2, and 3 in our shock cooling-double decay model (Section~\ref{subsec:pow_double_arnett}). Fits to this model are shown in Figure~\ref{fig:pure_pow_arnett_arnett} and the fitted parameters are shown in the far-right column of Table~\ref{table:mosfit}.}
    \label{fig:pure_pow_arnett_arnett_schematic}
\end{figure}

Finally, we combine the power-law model of peak 1 to the Double $^{56}\rm{Ni}$ Decay model of peaks 2 and 3 to fit the entire light curve. In such a model, peak 1 is powered by shock cooling of the CSM following shock breakout. Peak 2 is powered by radioactive decay of $^{56}\rm{Ni}$ distributed near the outer regions of the ejecta. Peak 3 is powered by radioactive decay of $^{56}\rm{Ni}$ distributed near the inner regions of the ejecta. Photons  emitted from the inner regions of the ejecta diffuse through more mass and take longer before they are emitted. We show a schematic of this model in Figure~\ref{fig:pure_pow_arnett_arnett_schematic}. We find a power-law index of $\alpha = 4.55^{0.78}_{0.95}$, $\log_{10}~[L_{\rm pow}/\mathrm{erg~s^{-1}}] = 41.43^{1.79}_{1.15}$, $\log~[t_{\rm pow}/\mathrm{days}] = 0.71^{0.30}_{0.35}$~days. The inner layer of ejecta has $\log_{10}~[m_{1, \mathrm{ej}}/\mathrm{M_\sun}] = -0.35^{0.07}_{0.06}$ and  $^{56}\rm{Ni}$ mass of $m_{1, \mathrm{Ni}} = 0.055^{0.002}_{0.001}~\mathrm{M_\odot}$. The outer layer of ejecta has $\log_{10}~[m_{2, \mathrm{ej}}/\mathrm{M_\sun}] = -0.73^{0.08}_{0.11}$ and $^{56}\rm{Ni}$ mass of $m_{Ni, 2} = 0.040^{0.002}_{0.003}~\mathrm{M_\odot}$. The total ejecta mass from this model is $\log\left[(m_1+m_2)/M_\sun\right] = -0.2 \pm 0.11$, and $m_{1, \mathrm{Ni}}+m_{2, \mathrm{Ni}} \approx 0.09~\mathrm{M_\odot}$ of it is $^{56}\rm{Ni}$.

Compared to the model presented in Subsection~\ref{subsec:pow_arnett}, our joint Power Law + Double $\rm{^{56}Ni}$ decay provides a better fit to the SN~2022oqm photometry. We obtain a WAIC score of 758 and reduced $\chi^2_\mathrm{red}\approx1$.  This suggests that the SN~2022oqm light curve is powered by three distinct sources, leading to the triple peaked behavior seen in Figure~\ref{fig:reduced_photometry}.

\begin{figure*}
    \centering
    \includegraphics[width = 0.9\linewidth]{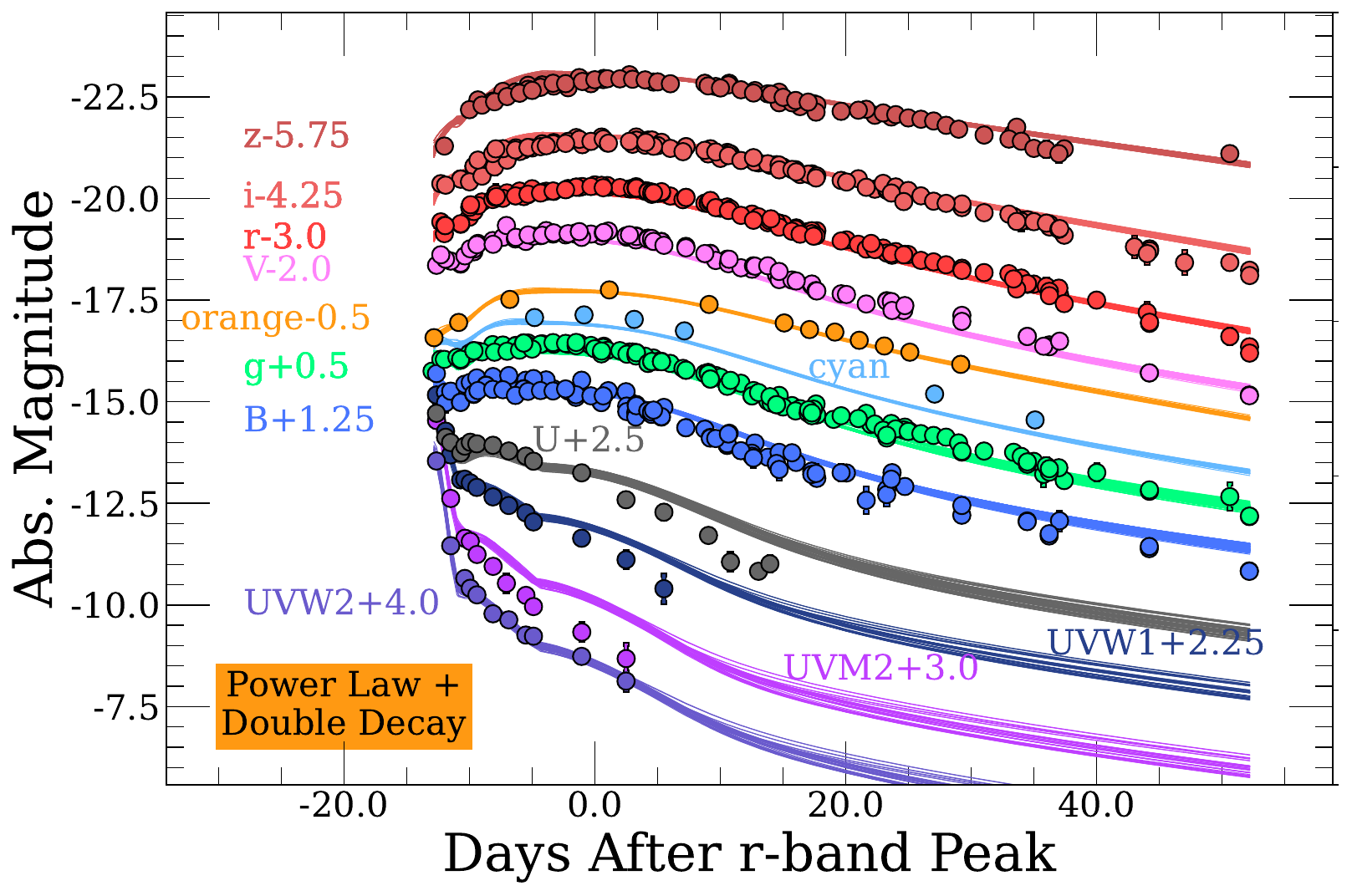}
    \includegraphics[width = 0.9\linewidth]{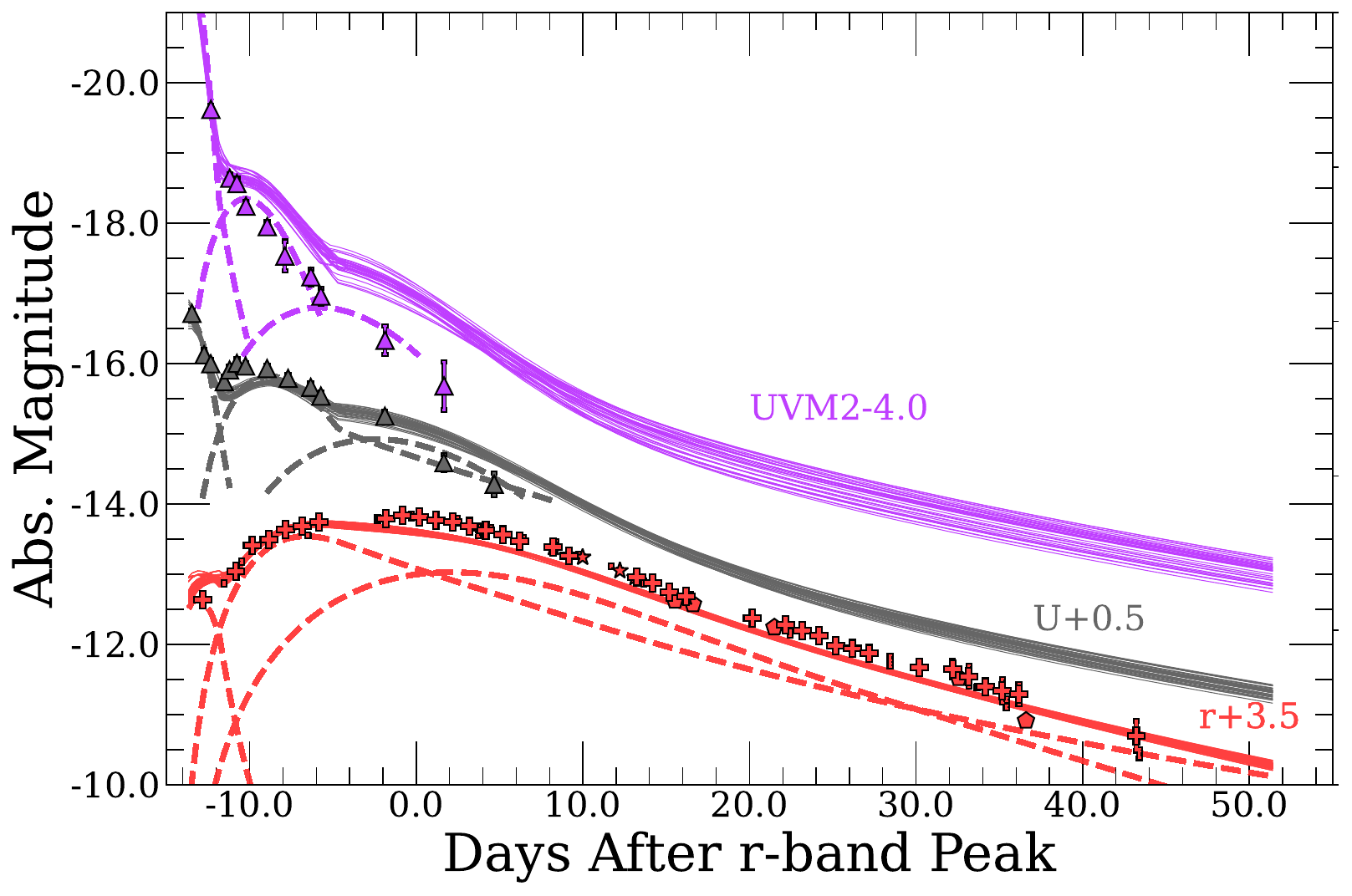}
    \caption{\textit{Left}: The result of fitting shock cooling and the double Arnett model to the photometry of peaks 1, 2, and 3. In agreement with the double Arnett model, this fit finds a total $^{56}\rm{Ni}$ mass of $\sim 0.09~\mathrm{M_\odot}$, with total ejecta mass $\sim 1 \mathrm{M_\odot}$. \textit{Right}: Contribution to r-, U-, and UVM2-band light curves from the three sources of photons. Peak 1 is powered by shock interactions with CSM, and is most clearly visible in UV bands. Peak 2 is powered by the radioactive decay of $^{56}\rm{Ni}$ in the outer ejecta, and is most clearly visible in the U-band. Because photons emitted by this overdensity of $^{56}\rm{Ni}$ have less mass to diffuse through, they will be observed before photons emitted by $^{56}\rm{Ni}$ in more inner regions of the ejecta. Peak 3, which is most clear in the r-band, is powered by the radioactive decay of $^{56}\rm{Ni}$ in the inner ejecta. Fitting is done using the \texttt{MOSFiT} package.}
    \label{fig:pure_pow_arnett_arnett}
\end{figure*}

We note here that all three peaks are not clearly visible in all bands. Rather, peak 1 is most clearly visible in UV bands, e.g., \textit{Swift}; peaks 1 and 2 are visible in slightly longer wavelengths, e.g., $u$-band; and peak 3 is strongest in longer wavelengths, e.g., $r$-band. To clearly visualize these peaks, we overlay the data in these bands with our \texttt{MOSFiT} models in these three regimes in the right panel of Figure~\ref{fig:pure_pow_arnett_arnett}.

\subsection{Reflections on Photometric Modeling}
\label{subsec:mosfit_reflections}

Through Sections~\ref{subsec:pow_arnett}, \ref{subsec:just_arnett}, and \ref{subsec:pow_double_arnett}, we find that
\begin{enumerate}
    \item Peaks 2 and 3 can be modeled as two distinct radioactive peaks
    \item The total nickel mass is $\sim 0.09~\mathrm{M_\odot}$
    \item Peak 1 can be modeled with a power-law model of emission in time, suggesting a shock cooling peak
\end{enumerate}
 The presence of narrow, highly ionized emission lines in the earliest spectrum also suggests emission from shocked material during the first $\sim2$~d of this SN (see Section~\ref{sec:spectroscopy}). We capture peak 1 with a power law index $\alpha~=4.55$. \cite{Piro2021} show that shock cooling emission following an astrophysical explosion can be captured by a power law emission in time $t^{-4/(n-2)}$, where $n$ is the radial dependence of density in the outer region of the ejecta upon radius. \cite{Piro2021} then connect $n$ to $s$, the polytropic index of the exploding star: $n=\left(1+1/s+3\beta\right)/\beta$, where $\beta\approx0.19$. Following this, our fitted value of $\alpha=4.55$ results in a polytropic index of $s\approx-0.9$, an unphysical value. Likely, more precise prescriptions of early emission are necessary to describe the early light-curve behavior. We attempt fitting the model of \cite{morag2023shock} (hereafter, M23) to peak 1, and find that this fit does not converge to the photometry of SN~2022oqm. Furthermore, this model finds total ejecta mass ($0.63~M_\odot$) is within $1\sigma$ of the mass found by spectroscopic modeling ($\approx0.6~M_\sun$, see Section~\ref{sec:spectroscopy}), which shows independent support to the best-fit light-curve model we use here.  As such, we consider inferred physical properties from the best self-consistent model of the light curve explored: the combination of a power law shock cooling emission model and a double \niVI~decay model.

We emphasize that our double $\mathrm{^{56}Ni}$ decay model is a toy approximation for a multi-zone model. Our model suggests that the light curve is powered by a centrally located radioactive material, with a fraction of Fe-peak elements mixed into the outer ejecta layers. In particular, the fact that the best fit model requires that at least some radioactive material is not centrally located hints at the mixing of $^{56}\rm{Ni}$ throughout the ejecta. The mixed distribution of $^{56}\rm{Ni}$ throughout the ejecta is consistent with our spectroscopic modeling (see Figure~\ref{fig:abundance}). The outer ejecta layer is less massive, more nickel-rich, and faster moving than the inner ejecta layer. In addition, such agreement with such a two-zone model of SN emission highlights the need for more robust multidimensional analytical models of SNe emission \citep{maeda2003two}. 

It is worth noting that \cite{irani2022} also find that fitting the first peak to the model of \cite{Piro2021} results in an unphysical fit. \cite{irani2022} then instead fit the photometry during only peak 2 to a combined \niVI~radioative decay + M23 model and find agreement to the photometry. After peak 2, they find agreement with an independent \niVI~radioactive decay model. Therefore, the \cite{irani2022} interpretation of the multiple peaks of SN~2022oqm is one of CSM interaction during peak 1, shock cooling and \niVI~radioactive decay during peak 2, and only \niVI~radioactive decay during peak 3.

\section{Comparing to the \cart{} Population}
\label{sec:ca_rich_comparisons}

\begin{figure*}
    \centering
    \includegraphics[width = 0.9\linewidth]{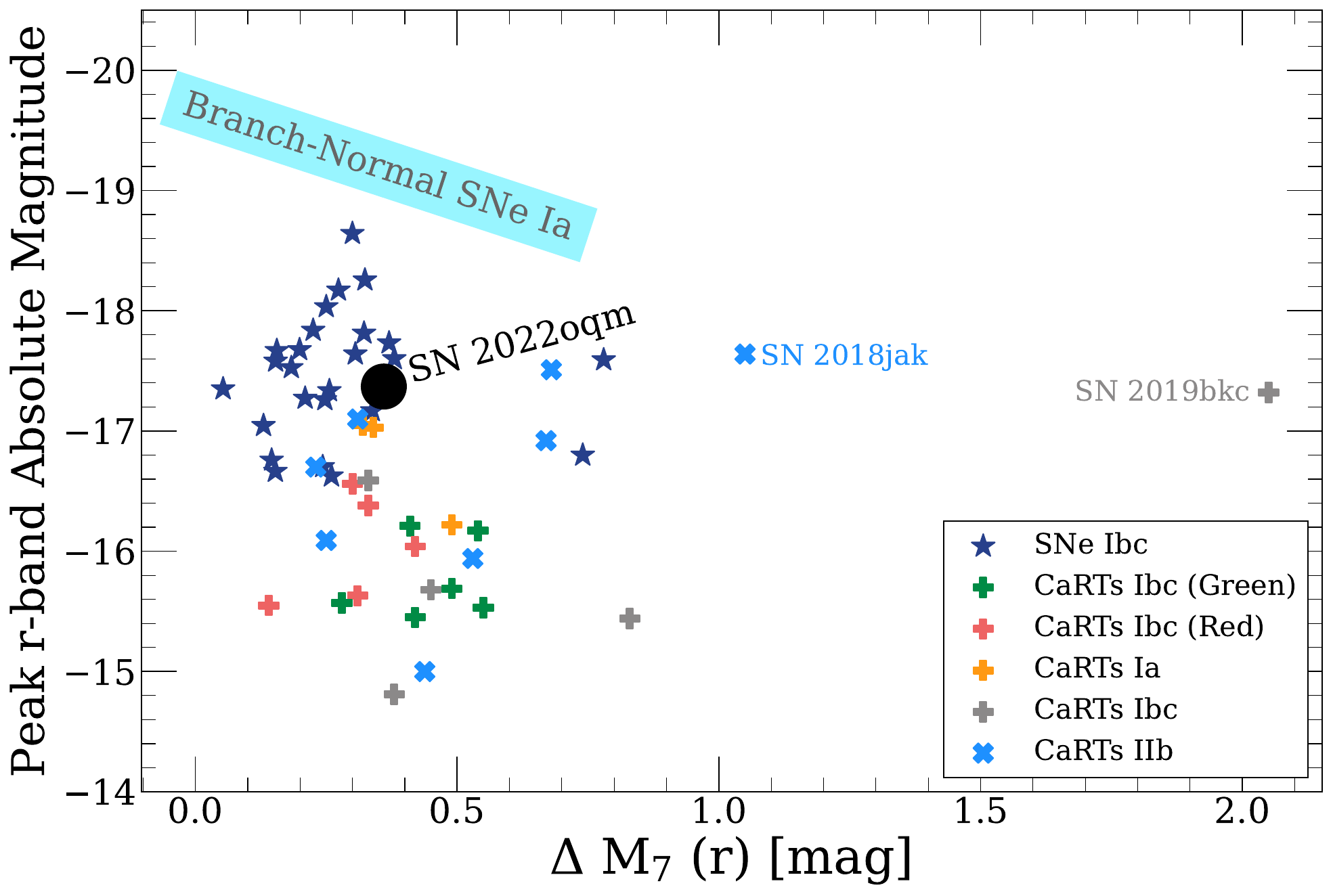}
    \caption{Comparison of the decline rate and absolute magnitude for \carts{}, SNe~Ibc, and SNe~Ia. These SNe~Ibc are presented in \cite{Drout2011}. We generate the sample of Branch-Normal SNe~Ia using the methodology described in Equations~\ref{eq:t_dur} and \ref{eq:s}. \carts{}~Ia, green \carts{}~Ibc, and red \carts{}~Ibc are presented in \citet{de2020zwicky}. SN~2022oqm is among the brightest \carts{} detected. Its peak $r$-band absolute magnitude is typical of SNe~Ibc. \carts{}~Ibc shown as gray pluses are unclassified as Green or Red either because the \citet{de2020zwicky} sample pre-dates their discovery or because \citet{de2020zwicky} did not classify them as Green or Red. These are SN~2007ke \citep{2007keDiscovery, Perets10}, PTF11bij \citep{Kasliwal+12}, SN~2019bkc \citep{Chen2019bkc, bkc_extraordinary2020, bkc_notCart2021}, SN~2021gno \citep{Wynn2022, 2023Ertini}, and SN~2021inl \citep{Wynn2022}. The sample of \carts{}~IIb is presented in \cite{Das2022}.  }
    \label{fig:delta_7_r}
\end{figure*}

SN~2022oqm is the 39th transient to be classified as a \cart{}. We present an aggregate list of all known \carts{} at the time of writing in Table~\ref{table:all_CaRTs}, along with the original references and basic observational properties. Although the prototypical \cart{} is SN~2005E \citep{Perets10}, several \carts{} were detected before 2005 (see Table~\ref{table:all_CaRTs}). The most comprehensive study of \carts{} observed before 2020 is presented in \cite{de2020zwicky}. Since then, $11$ \carts{} have been presented in detail by \cite{Das2022}, \cite{Jacobson-Galan2019ehk}, and \cite{Wynn2022}\footnote{Note though SNe~Iax have CaRT-like nebular spectra, and their late-time Ca emission strength would place them in the realm of \carts{}, however we do not consider them CaRTs for this analysis. See \citet{Foley13_Iax} for a detailed analysis of the $\gtrapprox 25$ members of this class.}.

In Figure~\ref{fig:delta_7_r}, we compare the r-band peak absolute magnitude and the Philips-like measure $\Delta M_7$ (change in $r$-band magnitude during the $7$~days after $r$-band peak) of SN~2022oqm to those of the populations of \carts-Ibc, the \cite{Drout2011} population of SNe~Ibc, and a sample of SN~Ia light curves. We generate our SN~Ia light curve sample using an empirical relation described by \cite{tripp1999} and \cite{Betoule2014}:
    \begin{equation}
        \label{eq:t_dur}
        t_{dur} = 35s~\mathrm{days},
    \end{equation}
    \begin{equation}
        \label{eq:s}
        M_{peak, \rm{r}} = -19.2 - 1.4(s-1.0),
    \end{equation}
where $s$ parameterizes the light-curve stretch, for which we provide a range of 0.6 to 4. We use the canonical $s=1$ model from \cite{Nugent2002} to find a template $R$-band light curve to stretch using $s$.  \cite{de2020zwicky} presented $\Delta M_7$ instead of the canonical $\Delta M_{15}$ \citep{Phillips1993} because \carts{} often evolve too quickly to have a complete enough light curve to measure $\Delta M_{15}$. We adopt the measure of $\Delta M_7$ for the full population of \carts{} from the samples of \cite{de2020zwicky}, \cite{Das2022}, and from detailed analyses of these objects \citep{2007keDiscovery, Perets10, Kasliwal+12, bkc_extraordinary2020, bkc_notCart2021, Wynn2022}.
SN~2022oqm is among the brightest of all known \carts. It has a peak light magnitude that is characteristic of most SNe~Ibc, but evolves faster than most SNe~Ibc.

\cite{de2020zwicky} classify the population of \carts{} observed before 2020 into Ia-like \carts{} (\carts-Ia) and Ibc-like \carts{} (\carts-Ibc). SN~2022oqm shows many similarities with the \carts-Ibc class (see Table~\ref{table:ca_ibc_red_green_table}). They further classify \carts-Ibc into ``Red'' \carts-Ibc and ``Green'' \carts-Ibc. Red \carts-Ibc have somewhat redder spectra, due to line-blanketing of flux in shorter wavelengths, lower ejecta velocity, somewhat brighter peak magnitude, and smaller [\ion{Ca}{2}]/[{\ion{O}{1}}] ratio. Green \carts-Ibc have flatter spectra, with no blanketing and equal flux in longer and shorter wavelengths, higher ejecta velocities, somewhat dimmer peak magnitudes, and larger [\ion{Ca}{2}]/[{\ion{O}{1}}] emission ratio. 

Each population is characterized by properties listed in Table~\ref{table:ca_ibc_red_green_table}, where we compare SN~2022oqm to the Red and Green \carts-Ibc, the only two \cart{} populations with which SN~2022oqm has similarities. SN~2022oqm has many characteristics consistent with the Red Ca-Ibc population: weak signatures of \ion{Si}{2} and \ion{He}{1}; suppressed blue flux; has an r-band Philips-like measure $\Delta~M_7\approx 0.3$, and [\ion{Ca}{2}]/[\ion{O}{1}]$\approx 4.3$. In contrast, SN~2022oqm has a high photospheric velocity ($\sim 8000$~km~s$^{-1}$) and bluer $g-r$ color of $\sim0.5$~mag, both of which are characteristic of the Green \carts-Ibc population. Furthermore, the peak r-band magnitude near $-17$~mag is unlike both populations. As such, we suggest that SN~2022oqm may be an intermediate object between the Green and Red \carts-Ibc populations, with multiple characteristics of each, and a peak absolute magnitude that is uncharacteristic of either classes. This may also indicate that there is a continuum between Red and Green \carts-Ibc, with the two subpopulations resulting from a relatively small sample size of \carts-Ibc.

SN~2022oqm is not the only event to fall outside of the classification scheme presented by \cite{de2020zwicky}. The sample of \carts{} presented in \cite{Das2022} includes several IIb-like \carts{}, which are brighter than most \carts{} and are shown in light blue in Figure~\ref{fig:delta_7_r}. Though they have brighter peak magnitudes, they appear to have a similar light curve decay timescale to most \carts{}. \carts{}-IIb are likely not directly comparable to the rest of the population of \carts{} as they likely have a different progenitor system \citep{Das2022}, given their relatively bright peak magnitudes. Indeed, \cite{Das2022} suggest the progenitor of these \carts{}-IIb is a strongly stripped-envelope star with ZAMS mass of $8-12~\mathrm{M_\odot}$. 

In addition to these, \cite{Chen2019bkc}, and \cite{bkc_extraordinary2020} present the ``extraordinary'' supernova SN~2019bkc, a fast and bright \cart{}-Ic. Similar to SN~2022oqm, SN~2019bkc has a Ic-like peak magnitude and Ic-like photospheric spectrum. As such, SN~2019bkc would have been classified as a peculiar SN~Ic without a nebular spectrum, just as SN~2022oqm would have \citep{bkc_extraordinary2020}. SN~2019bkc reaches a similar peak magnitude ($-17.32\pm0.04$, \citealt{bkc_extraordinary2020}) as SN~2022oqm ($-17.37\pm0.051$) but decays more than 2 magnitudes in 7 days, faster than any other known \cart{}. This object is an outlier within the overall class of \carts-Ibc.

\begin{deluxetable}{cccc}
\tabletypesize{\footnotesize}
\tablecolumns{4}
\tablecaption{Comparison of SN~2022oqm to parameters of Red and Green \carts-Ibc from \cite{de2020zwicky}. These comparisons suggest SN~2022oqm is a red Ca-Ibc, with the exceptions that the $g-r$ color of SN~2022oqm suggests that SN~2022oqm is more like a green Ca-Ibc and that the peak magnitude for SN~2022oqm is much brighter than that of both classes. We find that SN~2022oqm is very likely a Ca-Ibc, and propose that SN~2022oqm may be a transition object between red and green \carts-Ibc. \label{table:ca_ibc_red_green_table}}
\tablehead{
\colhead{Observable} & \colhead{Ca-Ibc Red} & \colhead{Ca-Ibc Green} & \colhead{SN~2022oqm}}
 \startdata
     \ion{Si}{2}? & strong to weak & weak & weak\\
     \ion{He}{1}? & weak to strong & strong & weak\\\
     V($10^3~\mathrm{km}~\mathrm{s}^{-1}$) & 4-10 & 8-12 & $\sim 8$\\
     Blanketed? & yes & no & yes\\
     $M_{peak}$ & $-15.3$- $16.7$ & $-15.5$ - $16.2$ & $\sim -17$\\
     $\Delta m_7$ & 0.3 & 0.5 & 0.3\\
     $g-r$ & 1.5 & 0.4 & 0.5\\
     $\mathrm{[Ca~II]/[O~I]}$ & 2.5-10& 7-13 & 4.3\\
 \enddata
\end{deluxetable}

\section{Progenitor Models for SN~2022oqm}
\label{sec:allmodels}

\subsection{Massive Star Progenitors}

Given the early classification as a SN~Ic (\citealt{2022oqm_class_Icpec, 2022oqm_class_Ic_2, 2022oqm_class_Ic_1, 2022oqm_original_discovery}; see Section~\ref{sec:observations}), we consider: does SN~2022oqm have a massive-star progenitor?  During the photospheric phase, the spectra of SN~2022oqm are similar to those of normal SNe~Ic, displaying \ion{Ca}{2}, \ion{Si}{2}, and \ion{O}{1} absorption features (see Figure~\ref{fig:phot_spec_overlay}). SN~2022oqm also peaks at M$_r=-17.37$~mag, which is typical of SNe~Ic (see Figure~\ref{fig:delta_7_r}). 

Furthermore, the inferred ejecta mass for SN~2022oqm of $0.6~\rm{M_\sun}$ is only 1.2$\sigma$ less than the mean ejecta for SNe~Ic and 1.5$\sigma$ and 2.5$\sigma$ less than the mean ejecta mass for SNe~Ib and broad-lined SNe~Ic, respectively \citep{Drout2011}.  The 84th percentile fitted ejecta mass ($1\sigma$ greater than the mean) is $10^{-0.2+0.12}~\rm{M_\odot}= 0.81~\rm{M_\odot}$; this is $\sim1\sigma$ less than the mean ejecta mass for SNe~Ic, and $1.5\sigma-2\sigma$ less than the mean ejecta mass for SNe~Ib and broad-lined SNe~Ic \citep{Drout2011}. The ejecta mass is therefore also consistent with that of the SN~Ic population. 

Given the short lifetimes of high-mass progenitor stars, SNe~Ic are typically found in their host galaxy's disk, often near star-forming regions \citep[e.g.,][]{blanchard2016offset}. SN~2022oqm is found offset by $13.1$~kpc from the center of its host galaxy. This offset, while notable, is consistent with the population of SESNe. Comparing to the \citet{schulze2021palomar} sample of 151 SN~Ibc host-SN separations, the offset of SN~2022oqm is at the 97th percentile of all SNe~Ibc. SN~2022oqm is located $2.4$ half-light radii from the host center, which is greater than all but two SNe Ibc (121 events) in the \citet{kelly2012core} sample: SN~2002ap ($2.7$ half-light radii) and SN~2007bg ($3.6$ half-light radii), both broad-lined SNe~Ic.

An important observable of SN~2022oqm is the inferred C-, N-, and O-enhanced CSM that surrounds the exploding SN. If SN~2022oqm were a CCSN, the stripping mechanism would have needed to strip not only the hydrogen and helium layers of the progenitor star, but also sufficient carbon, nitrogen, and oxygen to excite clear signatures of them in the earliest spectrum. 

A population of SESNe with markedly high stripping and low ejecta masses are ultra-stripped supernovae (USSNe). These are SESNe which have been stripped dramatically by a compact binary companion, resulting in very small ejecta masses (e.g., \citealt{de2018_ultrastripped, yao2020_sn2019dge}). However, USSNe are expected to have smaller ejecta masses ($m_{\rm{ej}}\leq~0.2\rm{M_\sun}$; \citealt{Tauris2015}), although it is possible to have higher ejecta masses in USSNe (see e.g., \citealt{Sawada2022}), the mass of \niVI~produced by these explosions ($<0.05~\rm{M_\sun}$) is still smaller than what is seen in SN~2022oqm. Furthermore, these progenitor systems are not expected to travel $\gtrsim800$~pc during their lifetime \citep{Tauris2015}, considerably shorter than the $\sim3$~kpc offset of SN~2022oqm from its nearest visible star-forming region. In short, a USSN origin is possible for SN~2022oqm, but not strongly favored by our observations.

There exists a small subclass of SNe~Ic with ionized narrow C emission lines: SNe~Icn \citep{Icn_intro}. Only five SNe~Icn have been identified in literature: SN~2019jc, SN~2019hgp, SN~2021ckj, SN~2021csp, and SN~2022ann \citep{2021ckj_cite, 19hgp_cite, Pellegrino_Icn, Perley2022, davis2022sn}. \citet{Pellegrino_Icn} present a fit to the light curves of four SNe~Icn, combining interaction- and \niVI{}-powered components into a single model. For all SNe, they find that circumstellar interaction dominates the observed luminosity throughout the evolution of each SN and place upper limits of $\sim0.04\rm{M_\sun}$ of \niVI{} for each SN. \cite{davis2022sn} extensively studied SN~2022ann, placing an upper limit of $\sim0.04\rm{M_\odot}$ of \niVI{}, and finding its emission is dominated by circumstellar interaction. Indeed all SNe~Icn have narrow and ionized features at all observed phases, suggesting interactions with CSM throughout their observed spectroscopic evolution. 

Of the five known SNe~Icn, SN~2019jc is the most similar to SN~2022oqm. It has a relatively large host-galaxy separation ($11.2~\rm{kpc}$, compared to $13.1~\rm{kpc}$ for SN~2022oqm), similar $g$-band absolute peak magnitude ($-17.2$~Mag, compared to $-17.3$~Mag for SN~2022oqm), and similar derived ejecta mass ($\sim0.6~\rm{M_\sun}$ for both SN~2019jc and SN~2022oqm). Moreover, SN~2019jc is also the SN~Icn with the largest fitted contribution to the light curve from \niVI{} radioactive decay. However, \cite{Pellegrino_Icn} infer an upper limit of $0.04~\rm{M_\odot}$ for the \niVI{} mass of SN~2019jc, lower than the $\sim0.1~\rm{M_\odot}$ we infer for SN~2022oqm. It is possible that SN~2022oqm could be a SN~2019jc-like SN~Icn with much less CSM and significantly more \niVI{} than SN~2019jc.

As such, we find that if SN~2022oqm were the result of a massive stellar explosion, it could have been a SN~Icn-like event, but one that has shorter-lived (only the first $\sim2$~days post explosion) CSM interaction and a relatively large  \niVI{} mass compared to the population. Owing to the very small observed population, the progenitor system of these SNe is not well constrained, and we are unable to consider whether a system similar to the progenitor of SNe~Icn could result in a SN~2022oqm-like event. 

\subsection{Systems Involving a NS}
\label{subsubsec:WD_NS_merger}

A potential progenitor for SN~2022oqm is the merger of a WD-NS binary. In such a merger, the WD is tidally disrupted and sheared into an accretion disk \citep{Metzger+08, Zenati19_NSWD, Kaltenborn+22}. Unlike the CSM commonly found around massive stars, the WD CSM is rich in helium, carbon, and oxygen. Therefore, this model could provide the necessary CO enrichment of CSM to support the detection of these lines in the spectrum during peak 1 for SN~2022oqm. Moreover, the expected magnitudes of the most massive ONe WD-NS models in principle approach those of the known \carts{}. However, these models \textit{likely} are not bright enough to produce a 2022oqm-like SN due to low masses of radioactive material produced  \citep{Fernandez+19, Zenati+20_NSCOWD, Bobrick+21}.

Another possible progenitor is one in which a short-lived magnetar with a strong surface dipole magnetic field $B_{\rm NS} \sim 10^{13}-10^{15} G$ is born. Given that the spin down of a magnetar has been proposed as a possible source of energy in some SNe \citep{Ostriker&Gunn1971, Kasen&Bildsten10, Woosley2010}, we explore this possibility in the case of SN~2022oqm. Specifically, could the combination of spin down of a newly born magnetar and radioactive decay of the $^{56}\rm{Ni}$ produced in the SN power SN~2022oqm? To model such a scenario, we use \texttt{MOSFiT} to combine the magnetar spin-down model and $^{56}\rm{Ni}$ decay to model peaks 2 and 3. We find this model is an extremely poor fit to the data and does not converge to the photometry of SN~2022oqm. It cannot reduce the scatter of the fit to less than $\sim~5$~mag of the observed photometry. Models involving the spindown of a magnetar are ordinarily invoked to obtain peak magnitudes near $-20$ and $-21$ mag, where $^{56}\rm{Ni}$ decay cannot reasonably explain the light curve of such events e.g., in the case of Type I superluminous SNe  \citep{nicholl2017magnetar}. As such, attempting to explain the relatively low luminosity of SN~2022oqm requires a relatively weak magnetar and a small ejecta mass which are, in general, inconsistent with masses derived from spectroscopic modeling (see Section~\ref{sec:spectroscopy}) and photometric modeling (see Section~\ref{sec:mosfit}). Moreover, a magnetar is the remnant of a CCSN, which we disfavor earlier in this section. Therefore, we rule out this scenario as the progenitor system of SN~2022oqm.

Another scenario involving a NS is accretion-induced collapse: the direct collapse of a Chandrasekhar WD into a NS caused by electron capture or other processes \citep[see][]{MiyajiNomoto+1980, Dessart+07, AICMetzger2009, Piro&Kulkarni13, Schwab21}. Generally, such a collapse is unlikely to produce a SN, but \cite{AICMetzger2009} consider the possibility that such a NS could be supported by centrifugal forces owing to the fast expected spin rate, and interactions with the debris disk might create a SN-like event. In their model, \cite{AICMetzger2009} find that when the WD collapses into a proto-NS, if an accretion disk is formed around the NS, the disk's evolution would naturally eject $\sim10^{-2}\mathrm{M_\odot}$ of $^{56}\rm{Ni}$, and total mass $\sim 2 \times 10^{-2}\mathrm{M_\odot}$. Photometric and spectroscopic modeling of SN~2022oqm both find larger ejecta mass and $^{56}\rm{Ni}$ mass, with a lower $^{56}\rm{Ni}$ fraction. Therefore, we rule out this system as the progenitor of SN~2022oqm.

\subsection{WD Binaries}
\label{subsec:CO_HeCO_model_detailed}
Several models of \cart{} progenitors invoke the detonation of a helium shell on a WD \citep{Perets10, Fernandez&Metzger13,  Zenati19_NSWD, Zenati+19a, de2020zwicky, Zenati+23_carich}. \cite{Perets10} presented the first such explanation proposed for SN~2005E: the detonation of a $0.3~\mathrm{M_\sun}$ He envelope. Such a detonation would result in the overproduction of radioactive materials \citep{Shen2010, woosley2011sub}. However, if such a detonation happens in the lower densities expected in a merging WD binary \citep{Dessart2015}, an overproduction of $^{56}\rm{Ni}$ can be avoided. In addition, the He shell detonation should not trigger a core detonation, as that would result in a SN~Ia \citep{nomoto1982, Fink2007, Fink2010, Shen2019}. According to \citet{Shen2019}, the core detonation can be avoided if the WD mass is $\leq~0.8~M_\sun$, or if it has an O/Ne composition. \cite{de2020zwicky} suggest that \cart{}s can be produced by the detonation of a helium shell on a WD, where the helium shell mass, the WD mass, and WD composition can influence whether the resulting \cart{} is Ia-like or Ibc-like. 

Recently \cite{Zenati+23_carich} showed that the disruption of a low-mass C/O WD by a binary companion hybrid He+C/O WD during merger could explain the origin and properties of thermonuclear SNe with strong [\ion{Ca}{2}] emission in their nebular spectra, including SNe~Ia, peculiar SNe~Ia, and \carts{} \citep{Perets+19, Pakmor+21, Burmester+23_HVS, Wynn2022}. The accretion of C/O material onto a He+C/O-WD heats its He-shell, leading to a weak detonation and ejection of the shell. Liberated C/O from the disrupted WD also is sent into the surroundings. This detonation results in a Ca-rich SN while leaving the C/O core of the He+C/O-WD intact as a hot remnant WD. This model predicts many of the observational features (e.g., strong nebular Ca, $B-V$ color evolution, C- and O-enhanced CSM) of SN~2022oqm, but also fails to predict others (e.g., peak luminosity, $^{56}\rm{Ni}$ mass, ejecta mass, ejecta composition, and $r-i$ color evolution).

However, it is difficult to reproduce our inferred ejecta mass in this progenitor model. The range of fitted ejecta masses ($0.48~M_\sun<m_{\rm 1, ej}+m_{\rm 2, ej}<0.81~M_\sun$) from photometric modeling (see Section~\ref{sec:mosfit}) and the spectroscopically fitted ejecta mass of $\sim 0.6~M_\sun$ (see Section~\ref{sec:spectroscopy}) would imply an unexpectedly massive He+C/O WD. In particular, WDs with mass larger than $\sim1~\rm{M_\sun}$ are associated with an O/Ne composition \citep{Wu2022}. Our spectroscopic modeling suggests that the ejecta mass fraction of C is only $\sim 10\%$ and that the ejecta mass fraction of Ne could be up to $\sim 50\%$ (although O/Ne are not distinguished; see Figure~\ref{fig:abundance}). The inferred mass and abundances may suggest instead an O/Ne WD progenitor. A more massive WD could also explain the relatively bright light curve of SN~2022oqm, which is $\sim1.5-2$~mag more luminous than most \carts. However, \cite{Zenati+19a} find that in general, more massive WDs have progressively smaller helium mass fractions. Whether WDs that are massive enough to potentially explain SN~2022oqm-like explosions can have enough helium to detonate as a \cart{} is still unclear. More three-dimensional, high-resolution simulations exploring the higher WD mass region parameter space are required to help further understand whether WD systems can lead to SN~2022oqm-like events.

\section{Discussion}
\label{sec:discussion}
The collective observational picture of SN~2022oqm presented here showcases a highly unusual SN. SN~2022oqm was detected offset from the center of its host galaxy, NGC~5875, by $50.6''$ (13.1~kpc). At the time of detection (MJD=59771), SN~2022oqm presented a strong UV excess (peak 1). This UV excess (see Figure~\ref{fig:reduced_photometry}), the contemporaneous hot continuum with narrow emission lines from highly ionized atomic species (see Figure~\ref{fig:spec_sequence_1}), the fit to the light curve with a shock cooling model (see Figure~\ref{fig:pure_pow_arnett_arnett} and Section~\ref{sec:mosfit}), all point to CSM interaction at early times \citep[e.g., see][]{Tinyanont2021}.

After peak 1, the light curve shows a weak local maximum near MJD $59774$, followed by a broader peak 3 ($\sim$~MJD $59785$). The fact that the spectral sequence does not abruptly change during peaks 2 and 3 suggests that peaks 2 and 3 are caused by the same emission mechanism, rather than a result of interactions with more CSM. We find that the best fit to the photometry is given by a model that has three power sources corresponding to three photometric peaks. In this model, peak 1 is driven by shock cooling, and peaks 2 and 3 are driven by the radioactive decay of radially separated $^{56}\rm{Ni}$ throughout the SN ejecta. Throughout its photospheric phase, SN~2022oqm appears both spectroscopically (see Figure~\ref{fig:phot_spec_overlay}) and photometrically (see Figure~\ref{fig:delta_7_r}) as a SN~Ic. As the SN approaches its nebular phase, a strong [\ion{Ca}{2}]$\lambda\lambda7291,7324$ emission feature emerges, only then revealing the ``calcium-rich'' nature of SN~2022oqm. At a peak $r$-band magnitude of $-17.37$, it is among the brightest \carts\ detected, complicating the progenitor nature of SN~2022oqm. 

In  Section~\ref{sec:allmodels}, we explore possible progenitors  for SN~2022oqm.  Here we suggest that BWD systems with primary WD mass $M \gtrapprox 0.6\mathrm{M_{\rm{\odot}}}$ could possibly produce 1) the C/O-enriched CSM surrounding the binary WD required to drive peak 1 and 2) the relatively bright light curve within the class of \carts{}. However, such a progenitor system has difficulties. In particular, the mass we infer here is likely too large to form the He+C/O Hybrid WD required by the models of \citet{Zenati+23_carich}. The ejecta abundances we find in Figure~\ref{fig:abundance} would also be inconsistent with the composition of a He+C/O WD. We find a C mass fraction of $10\%$, which is less than is required for He+C/O WD at these masses \citep{Zenati+19a}. 

Instead, the detonation of a He shell on an O/Ne WD has been suggested as a potential \carts{} progenitor \citep{Shen2019}. Indeed, the inferred progenitor WD mass of $M \gtrapprox 0.6M_{\rm{\odot}}$ (as implied from the fitted ejecta mass) could be consistent with a WD with an O/Ne composition, the canonical composition for WDs with mass $M\geq 1 M_{\rm{\odot}}$\citep{Wu2022}. However, our modeling of abundances (Figure~\ref{fig:abundance}), which cannot distinguish between O and Ne, does not strongly constrain this possibility. 

In all, the SN~2022oqm's inconsistencies with the typical model of \carts{} (e.g., peak magnitude) and with the proposed progenitor systems of \carts{} (e.g., ejecta mass and abundances) strongly point to a gap in the theoretical understanding of \carts{}. Whether a \cart can result from a WD with mass $\gtrapprox 0.6\mathrm{M_\sun}$ is an open question which requires dedicated theoretical study.

We also find that that SN~2022oqm could result from a massive stellar progenitor. We find that such a model would be most consistent with SNe~Icn given the signs of interaction with a H- and He-free CSM and low ejecta mass. Unlike SNe~Icn, SN~2022oqm does \textit{not} show spectroscopic signatures of CSM interaction throughout the light curve and has a higher inferred \niVI{} mass than others in the SN~Icn class. Because only five SNe~Icn are known \citep{Pellegrino_Icn, davis2022sn}, the progenitor of SNe~Icn is poorly constrained. Of the five known SNe~Icn, four are thought to be the result of a Wolf-Rayet progenitor explosion and one (SN~2019jc, the SN~Icn most similar to SN~2022oqm) the result of an USSN. Given the relatively poor constraints on the progenitor system of SNe~Icn and the inconsistency in the inferred light curve power source, it is unclear whether the progenitor system of SNe~Icn can also give rise to an SN~2022oqm-like transient.

As such, of the progenitor systems we explore in this article, we find that SN~2022oqm might have been the result of either 1) the detonation of a He shell on a WD with mass $M \gtrapprox 0.6\rm{M_\odot}$ or 2) the core collapse of a highly stripped, massive star resulting in a SN~Icn-like event, but with less CSM and more \niVI{} than all SNe~Icn observed today. We emphasize that neither proposed progenitor case explains \textit{all} observed properties of SN~2022oqm, and both progenitors are still poorly understood. However, these are the only two cases we find are able to reasonably explain many of the observed properties of SN~2022oqm.

Finally, given the photometric and spectroscopic similarity between SN~2022oqm and the SNe~Ic population, we address the concern that some SN~2022oqm-like \carts{} are misclassified as SNe~Ic. The relatively high peak brightness of SN~2022oqm suggests that the population of \carts-Ibc may have a broader range of peak magnitudes than is currently known and that, due to the absence of later-time spectroscopic follow-up observations (where a \cart-like spectrum would have been detected), some \carts-Ibc have been incorrectly classified as SNe~Ibc. To quantify the probability of such misclassifications, we calculate the relative rates of \carts{}-Ibc and SNe~Ibc: the observed rate of SNe~Ibc is $\approx 35\%$ that of SNe~Ia (the ZTF magnitude $m<18.5$ limited survey \citet{perley2020zwicky}), and the observed rate of \carts{} is $\approx 5\%$ the SNe~Ia observed rate \citep{Perets10}. Therefore, the observed rate of \carts{} is $\approx 5/35 \approx 14\%$ the observed rate of SNe~Ibc. Of the $28$ observed \carts{} with spectroscopic subtype labels, we find that $15$ are \carts-Ibc. Therefore, if $15/28 \approx 53\%$ of \carts{} are \carts-Ibc, only $\approx 53\% \times 14\% \approx 7.7\%$ of observed SNe~Ibc might actually be CaRTs-Ibc. Therefore, the sample of SNe~Ibc is likely not strongly contaminated by CaRTS-Ibc. However, the population of \carts-Ibc is possibly undercounted because many have been classified as other classes of SNe. In addition, as the Rubin Observatory is planned to begin observing in the coming months, many more transients like SN~2022oqm will inevitably be discovered. However, the upcoming Vera C.\ Rubin Observatory's planned cadence of $\sim3$ days presents an interesting observational challenge for the two successive radioactive decay-like peaks in SN~2022oqm. Without deliberate photometric follow-up to interweave between the planned cadence of the Rubin Observatory, such short-timescale light curves could be missed. Photometric characteristics which distinguish \carts\ from normal Type Ibc SNe in surveys such as YSE or the Rubin Observatory to enable spectroscopic followup are yet unexplored.

\section{Conclusions}
\label{sec:conclusions}

In this article, we have presented the photometric and spectroscopic observations of the recent Ic-like \cart{} SN~2022oqm. 

SN~2022oqm is a multi-peaked \cart{} with observational similarities to multiple disparate classes of SNe. It is among the brightest \carts{} known. We find that a model that combines shock cooling with the radioactive decay of two separate sources of $^{56}\rm{Ni}$ describes the three peaks well (see Subsection~\ref{subsec:pow_double_arnett}). The first peak is well-captured by a power-law model, which we ascribe to shock cooling of the CSM; and the second two peaks are captured by a ``double radioactive decay'' model, a rudimentary model of $^{56}\rm{Ni}$ mixing throughout the ejecta. We find that a mass of $0.06~\rm{M_\sun}$ of $^{56}\rm{Ni}$ in the inner portion of the ejecta, along with a mass of $0.04~\rm{M_\sun}$ of $^{56}\rm{Ni}$ in the outer portion of the ejecta may reproduce peaks 2 and 3. The potential for $^{56}\rm{Ni}$ mixing is further supported by detailed spectroscopic modeling of the ejecta, which predicts a flat ejecta abundance profile of iron-peak elements (see Figure~\ref{fig:abundance}).

We summarize our key conclusions here:

\begin{enumerate}
    \item SN~2022oqm is a Ic-like \cart{} (\cart-Ic), with [\ion{Ca}{2}] $\lambda\lambda7291,7324$/[\ion{O}{1}] $\lambda\lambda6300,6364\approx4.4$. Photometric evolution suggests it is most similar to the population of SNe~Ibc (see Figures~\ref{fig:color_evolution} and \ref{fig:delta_7_r})

    \item We find spectroscopic similarities with SNe~Iax, SNe~Ic, and other \carts{} at early photospheric phases, and similarities only with SNe~Iax and \carts{} at later times. Detailed spectroscopic modeling fits SN~2022oqm with an exponential ejecta density profile, rather than the standard Ia-like (Branch W7) or the Ic-like (CO21) density profile; with a flat ejecta abundance profile for both low-mass elements and  for Fe-peak elements.

    \item We identify three peaks in the light curves of SN~2022oqm. Two are clearly visible in the observed light curves while the third peak emerges from modeling of the light curve. The peaks are: 1) an early ($\approx 0$ d post explosion), blue ($B-V \approx -0.6$~mag) peak, 2) a weaker, less blue ($B-V \approx 0$~mag) peak $\approx 4$d post explosion, and 3) a third, broader peak ($B-V \approx 0.2$~mag) about $\approx 14$d post explosion. Using a combination of a power-law and two radioactive decay models, we successfully model the complete UV/O/NIR light curve. We infer a total $^{56}\rm{Ni}$ mass of $\sim 0.09\mathrm{M_\odot}$ and a total ejecta mass of $0.6~\mathrm{M_\odot}$. Ejecta velocities are $\leq 10^4$~km~s$^{-1}$, in agreement with line-velocities as measured from our spectral sequence.

    \item SN~2022oqm is located in the spiral host galaxy NGC 5875 at an angular separation of $50.6''$, corresponding to a projected physical distance of $13.1$~kpc from the center of the galaxy and $\sim2.4$ half-light radii of the galaxy. We do not observe any obvious ongoing star formation at the location of the transient, with the nearest detected star-forming region $\approx 3$~kpc away. Any progenitor system born in this star-forming region would require an average velocity $v = 147\frac{\rm km}{\rm s}\left[\frac{d}{3.0~\rm{kpc}}\right]\left[\frac{20 \rm Myr}{\tau}\right]$ over its entire lifetime $\tau$.

    \item While SN~2022oqm has similarities to SNe~Ibc in terms of photometry (peak absolute magnitude) and spectroscopy (photospheric phase similarity to SNe~Ic), the location of the SN in its host galaxy, inferred ejecta mass, and strong [\ion{Ca}{2}]~$\lambda\lambda7291,7324$ feature are somewhat inconsistent with SNe~Ibc. We  find that of the possible stripped envelope SN populations, only Type Icn supernovae (SNe~Icn) share some similarities with the observed properties of SN~2022oqm. SN~2019jc is the most similar SN~Icn to SN~2022oqm, but significant differences persist between the two objects. Importantly, SN~2019jc is dominated by interactions between the SN ejecta and circumstellar material (CSM) throughout its evolution and much less \niVI{} than SN~2022oqm. As such, we suggest that if SN~2022oqm were a SESN, it may have resulted from a similar progenitor as SNe~Icn, but with significantly less CSM and with a surprisingly large mass of \niVI{} produced. 

    \item We find that the detonation of a He layer on a white dwarf (WD) with mass $\gtrapprox 0.6\mathrm{M_\sun}$ could also be consistent with the observed characteristics of SN~2022oqm. This WD progenitor of SN~2022oqm is surrounded by CSM that contains C and O, consistent with the disruption of a C/O WD binary companion progenitor. Ejecta mass and abundance constraints suggest that the WD could be an O/Ne WD, or perhaps a massive C/O WD. Regardless, whether such massive WDs can have enough He to detonate in this method is not well-constrained.

\end{enumerate}

More detailed theoretical modeling of \cart{} progenitors, focusing on massive hybrid WDs, could help further understand luminous \carts{} like SN~2022oqm. Alternatively, SN~2022oqm could present a link between SNe~Ic and SNe~Icn. Exploration of such a continuum, especially predictions on the correlations between CSM mass, SN ejecta mass and \niVI{} mass, could further explain the progenitor system of SN~2022oqm-like events. Such studies will also shed light on the full range of possible photometric and spectroscopic properties of \carts{}, allowing for much more comprehensive constraints on the space of possible \carts{}.

\section*{Software and Facilities}
\software{ \texttt{YSEPZ} \citep{Coulter22, Coulter23}, \texttt{MOSFiT} \citep{guillochon2018}, \texttt{TARDIS} \citep{Kerzendorf2014}, \texttt{extrabol} \citep{extrabol_cite}, \texttt{numpy} \citep{numpy_cite}, \texttt{astropy}, \citep{astropy:2013, astropy:2018, astropy:2022}, \texttt{matplotlib} \citep{matplotlib}}, \texttt{scipy} \citep{2020SciPy}, 

\facilities{YSE/PS1, Shane Telescope (Kast), Nickel Telescope, Las Cumbres Observatory, Lulin Telescope, Hobby Eberly Telescope (LRS2), Apache Point Observatory (KOSMOS), ATLAS, Swift, NIRES, ZTF}

\section*{Acknowledgements}
We thank Ido Irani and Avishay Gal-Yam for insightful feedback on modeling this object and interpreting the progenitor system.
We thank Greg Zeimann for help with HET data reduction and Dan Weisz for Kast observations.
S. K. Yadavalli and V. A. Villar acknowledge support by the NSF through grant AST-2108676. 
Much of the photometry and spectroscopy presented here were acquired as part of the ongoing Young Supernova Experiment \citep{yse_ref_3, Aleo22}
The Young Supernova Experiment (YSE) and its research infrastructure is supported by the European Research Council under the European Union's Horizon 2020 research and innovation programme (ERC Grant Agreement 101002652, PI K.\ Mandel), the Heising-Simons Foundation (2018-0913, PI R.\ Foley; 2018-0911, PI R.\ Margutti), NASA (NNG17PX03C, PI R.\ Foley), NSF (AST-1720756, AST-1815935, PI R.\ Foley; AST-1909796, AST-1944985, PI R.\ Margutti), the David \& Lucille Packard Foundation (PI R.\ Foley), VILLUM FONDEN (project 16599, PI J.\ Hjorth), and the Center for AstroPhysical Surveys (CAPS) at the National Center for Supercomputing Applications (NCSA) and the University of Illinois Urbana-Champaign. J.C.W. and J.V. acknowledge the support from the NSF grant AST-1813825. W.J-G is supported by the National Science Foundation Graduate Research Fellowship Program under Grant No.~DGE-1842165. W.J-G acknowledges support through NASA grants in support of {\it Hubble Space Telescope} program GO-16075 and 16500. C.D.K. is partly supported by a CIERA postdoctoral fellowship. D. A. Coulter acknowledges support from the National Science Foundation Graduate Research Fellowship under Grant DGE1339067. CRA was supported by a grant from VILLUM FONDEN (project number 16599). D.F is supported by a VILLUM FONDEN Young Investigator Grant (project number 25501).

This project has been supported by the GINOP-2-3-2-15-2016-00033 project and the NKFIH/OTKA grants FK-134432 and K-142534 of the National Research, Development and Innovation (NRDI) Office of Hungary, partly funded by the European Union.
T.S. is supported by the J{\'a}nos Bolyai Research Scholarship of the 
Hungarian Academy of Sciences, and by the New National Excellence 
Program (UNKP-22-5) of the Ministry for Culture and Innovation from the 
source of the NRDI Fund, Hungary.
This project was supported by the KKP-137523 `SeismoLab' \'Elvonal grant of the Hungarian Research, Development and Innovation Office (NKFIH) and by the Lend\"ulet Program  of the Hungarian Academy of Sciences under project No. LP2018-7.

This publication has made use of data collected at Lulin Observatory, partly supported by MoST grant 108-2112-M-008-001. We additionally acknowledge the use of public data from the \textit{Swift} data archive. This work has made use of data from the Asteroid Terrestrial-impact Last Alert System (ATLAS) project. The Asteroid Terrestrial-impact Last Alert System (ATLAS) project is primarily funded to search for near-earth asteroids through NASA grants NN12AR55G, 80NSSC18K0284, and 80NSSC18K1575; byproducts of the NEO search include images and catalogs from the survey area. This work was partially funded by Kepler/K2 grant J1944/80NSSC19K0112 and HST GO-15889, and STFC grants ST/T000198/1 and ST/S006109/1. The ATLAS science products have been made possible through the contributions of the University of Hawaii Institute for Astronomy, the Queen's University Belfast, the Space Telescope Science Institute, the South African Astronomical Observatory, and The Millennium Institute of Astrophysics (MAS), Chile.
Y.Z was partially supported by NASA TCAN and Grant Number NNH17ZDA001N and TCAN-80NSSC18K1488.
We acknowledge observations made with the Nordic Optical Telescope, owned in collaboration by the University of Turku and Aarhus University, and operated jointly by Aarhus University, the University of Turku and the University of Oslo, representing Denmark, Finland and Norway, the University of Iceland and Stockholm University at the Observatorio del Roque de los Muchachos, La Palma, Spain, of the Instituto de Astrofisica de Canarias.
This work makes use of observations from the Las Cumbres Observatory global telescope network. The LCO group is supported by NSF grants AST-1911225 and AST0-1911151.
This work makes use of observations from the Las Cumbres Observatory network. The LCO team is supported by NSF grants AST-1911225 and AST-1911151, and NASA Swift grant 80NSSC19K1639.
This research has made use of the Spanish Virtual Observatory (https://svo.cab.inta-csic.es) project funded by MCIN/AEI/10.13039/501100011033/ through grant PID2020-112949GB-I00. The work of the Spanish Virtual Observatory's Filter Profile Service has proven to be the backbone of using the proper filter bandpasses in proper photometric modeling. We cite them here \citep{SVO2, SVO1}.

\appendix
\section{All Known \carts{}}
\begin{deluxetable}{ccccccc}
\tabletypesize{\footnotesize}
\tablecolumns{7}
\tablecaption{Table of every known \cart. We present the right ascension and declination in J2000 units, any known aliases, peak r-band absolute magnitude, $\Delta M_7$ in $r$-band, and references for every \cart{}. Objects with insufficient photometric coverage do not have $r$-band absolute magnitude and $\Delta M_7$ \label{table:all_CaRTs}. We present references to detailed studies of objects, where available. Otherwise, we provide discovery and classification announcements of objects.}
\tablehead{ \colhead{Name} & \colhead{Ra} & \colhead{Dec} & \colhead{$r$-peak} & \colhead{$\mathrm \Delta m_7~(r)$} & \colhead{Aliases} & \colhead{Reference} }
\startdata
SN~2000ds & $09^h11^m36.24^s$ & $+60^\circ01'42''.2$ & --  & -- & -- & \cite{IAUC7507, Perets10} \\
SN~2001co & $14^h19^m11.80^s$ & $+24^\circ47'42''.9$ & --  & -- & -- & \cite{aazami2001supernova, Filippenko2003} \\
SN~2003H & $6^h16^m25.68^s$ & $-21^\circ22'23''.8$ & --  & -- & -- & \cite{hamuy2003supernovae, Filippenko2003} \\
SN~2003dg & $11^h57^m31.97^s$ & $-01^\circ15'13''.6$ & --  & -- & -- & \cite{Filippenko2003, 2003dgDiscovery} \\
SN2003dr & $14^h38^m11.13^s$ & $46^\circ38'03''.4$ & --  & -- & -- & \cite{Filippenko2003, 2003drDiscovery} \\
SN~2005cz & $12^h37^m27.85^s$ & $74^\circ11'24''.5$ & --  & -- & -- & \cite{2005czDiscovery}\\
SN~2005E & $02^h39^m14.34^s$ & $+01^\circ05'55''.0$ & $-15.53$ & $0.55$ & -- & \cite{2005EDiscovery, Perets10} \\
SN~2007ke & $02^h54^m23.90^s$ & $+41^\circ24'16''.3$ & $-16.59$ & $0.33$ & -- & \cite{2007keDiscovery, Perets10} \\
& & & & &  & \cite{Kasliwal+12}\\
PTF09dav & $22^h46^m55.15^s$ & $+21^\circ37'34''.1$ & $-16.22$ & $0.49$ & -- & \cite{sullivan2011, Kasliwal+12} \\
SN~2010et & $17^h16^m54.27^s$ & $+31^\circ33'51''.7$ & $-15.69$ & $0.49$ & PTF10iuv & \cite{Kasliwal+12} \\
PTF10hcw & $08^h43^m36.22^s$ & $+50^\circ12'38''.5$ & --  & -- & -- & \cite{lunnan2017} \\
PTF11bij & $12^h58^m58.39^s$ & $+37^\circ23'12''.0$ & $-15.68$ & $0.45$ & -- & \cite{Kasliwal+12} \\
PTF11kmb & $22^h22^m53.61^s$ & $+36^\circ17'36''.5$ & $-15.57$ & $0.28$ & -- & \cite{Foley+15,lunnan2017} \\
SN~2012hn & $06^h42^m42.55^s$ & $-27^\circ26'49''.8$ & $-15.55$ & $0.14$ & -- & \cite{Valenti2012hn} \\
PTF12bho & $13^h01^m16.65^s$ & $+28^\circ01'18''.1$ & $-16.04$ & $0.42$ & -- & \cite{lunnan2017} \\
iPTF15eqv & $10^h52^m11.40^s$ & $+32^\circ57'01''$ & --  & -- & -- & \cite{iPTF15eqvDiscovery} \\
SN~2016hgs & $00^h50^m51.39^s$ & $+27^\circ22'48''.0$ & $-15.45$ & $0.42$ & iPTF16hgs & \cite{De2018} \\
SN~2016hnk & $02^h13^m16.63^s$ & $-07^\circ39'40''.80$ & $-17.05$ & $0.32$ & ATLAS16dpc & \cite{jacobson-galan2016hnk} \\
SN~2018ckd & $14^h06^m11.95^s$ & $-09^\circ20'39''.30$ & $-16.17$ & $0.54$ & ZTF18aayhylv & \cite{de2020zwicky} \\
SN~2018gwo & $12^h08^m38.83^s$ & $+68^\circ46'44''.40$ & --  & -- & Gaia18dfp, PS19lf  & \cite{de2020zwicky} \\
& & & & &  ZTF18acbwazl & \\
SN~2018gjx & $02^h16^m15.58^s$ & $+28^\circ35'28''.64$ & $-17.51$ & $0.68$ & ATLAS18vis, Gaia18csc & \cite{Das2022} \\
& & & & &  kait-18ao, PS10do& \\
& & & & &  PSP18c, ZTF18abwkrbl& \\
SN~2018jak & $09^h59^m18.20^s$ & $+34^\circ53'43''.78$ & $-17.64$ & $1.05$ & ATLAS18zqa, PS18clq & \cite{Das2022} \\
& & & & &  ZTF18acqxyiq & \\
SN~2018kjy & $06^h47^m17.96^s$ & $+74^\circ14'05''.90$ & $-15.63$ & $0.31$ & PS~18cfh, ZTF18acsodbf& \cite{de2020zwicky} \\
SN~2018lqo & $16^h28^m43.26^s$ & $+41^\circ07'58''.70$ & $-16.21$ & $0.41$ & ZTF18abmxelh & \cite{de2020zwicky} \\
SN~2018lqu & $15^h54^m11.48^s$ & $+13^\circ30'50''.90$ & --  & -- & ZTF18abttsrb & \cite{de2020zwicky} \\
SN~2019bkc & $10^h00^m22.54^s$ & $-03^\circ01'12''.64$ & $-17.32$ & $2.05$ & ATLAS19dqr & \cite{2019bkcDiscovery, Chen2019bkc} \\
& & & & &  & \cite{bkc_extraordinary2020, bkc_notCart2021}\\
SN~2019ehk & $12^h22^m56.15^s$ & $+15^\circ49'34''.06$ & $-16.09$ & $0.25$ & ATLAS19ibr, Gaia19bqn & \cite{Jacobson-Galan2019ehk, De19ehk}\\
& & & & &  PS19ayq, ZTF19aatesgp&  \cite{Jacobson-Galan2019ehk_later}\\
& & & & &  &  \cite{nakaoka2021, Das2022}\\
SN~2019hvg & $14^h06^m01.58^s$ & $+12^\circ46'50''.38$ & $-16.7$ & $0.23$ & ATLAS19bftc, ZTF19abacxod& \cite{Das2022} \\
SN~2019hty & $12^h55^m33.03^s$ & $+32^\circ12'21''.70$ & $-16.38$ & $0.33$ & ATLAS19nhp, PS~19bhn& \cite{de2020zwicky} \\
& & & & &  ZTF19aaznwze & \\
SN~2019ofm & $14^h50^m54.65^s$ & $+27^\circ34'57''.60$ & $-17.03$ & $0.34$ & ATLAS~19tjf, ZTF19abrdxbh & \cite{de2020zwicky} \\
SN~2019pof & $01^h12^m37.88^s$ & $+33^\circ02'05''.75$ & $-15.94$ & $0.53$ & ATLAS19uvh, PS19fbm & \cite{Das2022} \\
& & & & &  ZTF19abxtcio & \\
SN~2019pxu & $05^h10^m12.61^s$ & $-00^\circ46'38''.60$ & $-16.56$ & $0.30$ & ATLAS19uvg, PS19fwq  & \cite{de2020zwicky} \\
& & & & & ZTF19abwtqsk & \\
SN~2020sbw & $02^h46^m03.32^s$ & $+03^\circ19'47''.67$ & $-17.1$ & $0.31$ & ATLAS20yle, PS20hhn & \cite{Das2022} \\
& & & & & ZTF20abwzqzo & \\
SN~2021M & $14^h14^m14.73^s$ & $+35^\circ25'23''.14$ & --  & -- & ATLAS21cum, ZTF21aaabwfu& \cite{Das2022} \\
SN~2021gno & $12^h12^m10.29^s$ & $+13^\circ14'57''.04$ & $-15.44$ & $0.83$ & ATLAS21iro, Gaia21cdw& \cite{Wynn2022, 2023Ertini} \\
& & & & &  PS21cjz, ZTF21aaqhhfu & \\
SN~2021inl & $13^h01^m33.24^s$ & $+27^\circ49'55''.10$ & $-14.81$ & $0.38$ & ATLAS21lqc, PS21dal & \cite{Wynn2022} \\
& & & & &  ZTF21aasuego& \\
SN~2021pb & $09^h44^m46.80^s$ & $+51^\circ41'14''.64$ & $-16.92$ & $0.67$ & ZTF21aabxjqr & \cite{Das2022} \\
SN~2021sjt & $20^h37^m19.20^s$ & $+66^\circ06'23''.13$ & $-15.00$ & $0.44$ & ZTF21abjyiiw & \cite{Das2022} \\
SN~2022oqm & $15^h09^m12.09^s$ & $+52^\circ32'05''.14$ & $-17.37$ & $0.36$ & ZTF22aasxgjp &\cite{irani2022}; This Work \\
\enddata
\end{deluxetable}

\begin{deluxetable}{ccccc}
\tabletypesize{\footnotesize}
\tablecolumns{5}
\tablecaption{Log of SN~2022oqm spectra presented in this article. Phase is measured in days from time of r-band peak magnitude (MJD 59785). Note that the very first spectrum was presented on the Transient Name Server \citep{2022oqm_original_classification} and we are presenting that spectrum as is. \label{table:spec_log}}
\tablehead{ \colhead{Time (MJD)} & \colhead{Phase (days)} & \colhead{Telescope} & \colhead{Instrument} & \colhead{Wavelength Range ($\mathrm{\AA}$)} }
\startdata
59771.31 & -12.86 & Palomar 60~inch Telescope & SED Machine & 3776-9223 \\
59773.21 & -10.96 & Hobby Eberly Telescope & LRS2 & 3640-6950 \\
59775.89 & -8.28 & Nordic Optical Telescope & ALFOSC & 3800-8999 \\
59776.31 & -7.86 & Hobby Eberly Telescope & LRS2 & 3640-9799 \\
59777.20 & -6.97 & Hobby Eberly Telescope & LRS2 & 6450-10500 \\
59779.26 & -4.91 & Keck II Telescope& NIRES & 9538-24377 \\
59779.27 & -4.90 & Shane 3m Telescope& Kast & 3303-10496 \\
59780.27 & -3.90 & Shane 3m Telescope & Kast & 3508-10722 \\
59784.19 & 0.02 & Hobby Eberly Telescope & LRS2 & 3640-10500 \\
59784.24 & 0.07 & Shane 3m Telescope & Kast & 3503-10494 \\
59788.25 & 4.08 & Shane 3m Telescope & Kast & 3405-10495 \\
59790.16 & 5.99 & Hobby Eberly Telescope & LRS2 & 3640-10500 \\
59793.16 & 8.99 & Hobby Eberly Telescope & LRS2 & 3640-9700 \\
59793.29 & 9.12 & Shane 3m Telescope & Kast & 3253-10896 \\
59798.28 & 14.11 & Shane 3m Telescope & Kast & 3253-10894 \\
59803.32 & 19.15 & Faulkes Telescope North & FLOYDS & 3459-9882 \\
59809.23 & 25.06 & Shane 3m Telescope & Kast & 3508-10730 \\
59810.77 & 26.60 & Faulkes Telescope North & FLOYDS & 3459-9880 \\
59814.22 & 30.05 & Shane 3m Telescope & Kast & 3503-10893 \\
59818.20 & 34.03 & Shane 3m Telescope & Kast & 3253-10893 \\
59824.20 & 40.03 & Shane 3m Telescope & Kast & 3509-10703 \\
59847.23 & 63.06 & Keck I Telescope & LRIS & 3120-10282 \\
\enddata
\end{deluxetable}

\begin{deluxetable}{ccccccc}
\tabletypesize{\footnotesize}
\tablecolumns{7}
\tablecaption{Partial log of SN~2022oqm photometry presented in this article. Phase is measured in days from time of $r$-band peak magnitude (MJD 59784.24). \label{table:phot_log}}
\tablehead{ \colhead{Time (MJD)} & \colhead{Phase (d)} & \colhead{AB magnitude} & \colhead{AB magnitude Uncertainty} & \colhead{Telescope} & \colhead{Instrument} & \colhead{Band} }
\startdata
59771.19 & -13.81 & 17.323 & 0.038 & P48 & ZTF-Cam &  \textit{g} \\
59771.351 & -13.649 & 17.499 & 0.048 & Atlas & Atlas & \textit{orange} \\
59771.54 & -13.46 & 16.159 & 0.045 & \textit{Swift} & UVOT & \textit{UVW1} \\
59771.54 & -13.46 & 16.033 & 0.04 & \textit{Swift} & UVOT & \textit{UVW2} \\
59771.54 & -13.46 & 16.633 & 0.074 & \textit{Swift} & UVOT & \textit{B} \\
59771.89 & -13.11 & 17.28 & 0.1 & Baja & Baja & \textit{B} \\
59772.173 & -12.827 & 17.033 & 0.052 & Thacher & ACP & \textit{g} \\
59772.205 & -12.795 & 17.235 & 0.024 & P48 & ZTF-Cam &  \textit{r} \\
59772.241 & -12.759 & 17.243 & 0.025 & P48 & ZTF-Cam &  \textit{r} \\
59772.259 & -12.741 & 17.089 & 0.065 & Nickel & Nickel & \textit{B} \\
59772.27 & -12.73 & 17.351 & 0.206 & \textit{Swift} & UVOT & \textit{B} \\
59772.27 & -12.73 & 16.95 & 0.076 & \textit{Swift} & UVOT & \textit{U} \\
\enddata
\end{deluxetable}

\section{Piro Model of Shock Cooling in \texttt{MOSFiT}}
\label{appx:piro}

In addition to the power-law model of shock interaction presented in \ref{sec:mosfit}, we present a custom \texttt{MOSFiT} model of Shock Cooling-Double Radiocative decay, where we allow for diffusion of photons emitted by the shock interaction engine. Analogous to the presentation in subsection~\ref{subsec:pow_double_arnett}, peak 1 would be modeled by the Piro model while peaks 2 and 3 are modeled with a $\rm{^{56}Ni}$ radioactive decay model \citep{arnett1982, chatzopoulos2012}. We find that allowing for photon diffusion results in poor agreement with peak 1, with reduced chi-squared $29.1$.

\bibliography{sample631}{}
\bibliographystyle{aasjournal}

\end{document}